\documentclass[longauth]{aa}

\usepackage{graphicx}
\usepackage{natbib}
\usepackage{scalerel}
\AtBeginDocument{\let\linenumbers\relax}
\usepackage[table]{xcolor}
\usepackage{listings}  

\usepackage{txfonts}
\usepackage[pdfencoding=auto,psdextra]{hyperref}
\hypersetup{
    colorlinks=true,
    linkcolor=blue,
    filecolor=magenta,      
    urlcolor=blue,
    citecolor=blue
}
\makeatletter
\renewcommand*\aa@pageof{, page \thepage{} of \pageref*{LastPage}}
\makeatother
\usepackage[utf8]{inputenc}

\usepackage{euclid}

\newcommand{\ariii}[1]{[\ion{Ar}{iii}]#1}
\newcommand{\oiii}[1]{[\ion{O}{iii}]#1}
\newcommand{\oii}[1]{[\ion{O}{ii}]#1}
\newcommand{\oi}[1]{[\ion{O}{i}]#1}

\newcommand{\neiii}[1]{[\ion{Ne}{iii}]#1}
\newcommand{\nii}[1]{[\ion{N}{ii}]#1}
\newcommand{\sii}[1]{[\ion{S}{ii}]#1}
\newcommand{\siii}[1]{[\ion{S}{iii}]#1}
\newcommand{\hi}{\ion{H}{i}}
\newcommand{\hei}{\ion{He}{i}}

\newcommand{\Pb}{\ensuremath{\mathrm{Pa}\,\beta}}

\newcommand{\ecgs}
{\ensuremath{\mathrm{erg\,s}^{-1}\,\mathrm{cm}^{-2}}}

\definecolor{burgundy}{rgb}{0.5, 0.0, 0.13}

\defcitealias{Q1-TP007}{SPE-Q1}

\begin{document}

\title{Euclid Data Release 1 (DR1)}
\subtitle{
Performance of the redshift measurement and spectroscopic analysis 
} 

\newcommand{\orcid}[1]{} 
\author{Euclid Collaboration: V.~Le~Brun\orcid{0000-0002-5027-1939}\thanks{\email{Vincent.Lebrun@lam.fr}}\inst{\ref{aff1}}
\and M.~Moresco\orcid{0000-0002-7616-7136}\inst{\ref{aff2},\ref{aff3}}
\and M.~Bethermin\orcid{0000-0002-3915-2015}\inst{\ref{aff4}}
\and S.~Quai\orcid{0000-0002-0449-8163}\inst{\ref{aff2},\ref{aff3}}
\and N.~Borghi\orcid{0000-0002-2889-8997}\inst{\ref{aff2},\ref{aff3},\ref{aff5}}
\and Z.~Mao\orcid{0000-0003-4016-845X}\inst{\ref{aff3},\ref{aff6}}
\and P.~A.~C.~Cunha\orcid{0000-0002-9454-859X}\inst{\ref{aff7},\ref{aff3}}
\and X.~Lopez~Lopez\orcid{0009-0008-5194-5908}\inst{\ref{aff3}}
\and D.~Vibert\orcid{0009-0008-0607-631X}\inst{\ref{aff1}}
\and P.-Y.~Chabaud\inst{\ref{aff1}}
\and G.~Daste\inst{\ref{aff1}}
\and F.~Dufresne\inst{\ref{aff1}}
\and T.~Bedrine\inst{\ref{aff1}}
\and G.~Zamorani\orcid{0000-0002-2318-301X}\inst{\ref{aff3}}
\and L.~Pozzetti\orcid{0000-0001-7085-0412}\inst{\ref{aff3}}
\and Y.~Copin\orcid{0000-0002-5317-7518}\inst{\ref{aff8}}
\and D.~Vergani\orcid{0000-0003-0898-2216}\inst{\ref{aff3}}
\and A.~Lan\c{c}on\orcid{0000-0002-7214-8296}\inst{\ref{aff4}}
\and L.~Guzzo\orcid{0000-0001-8264-5192}\inst{\ref{aff9},\ref{aff10},\ref{aff11}}
\and C.~Scarlata\orcid{0000-0002-9136-8876}\inst{\ref{aff12}}
\and M.~Talia\orcid{0000-0003-4352-2063}\inst{\ref{aff2},\ref{aff3}}
\and T.~Contini\orcid{0000-0003-0275-938X}\inst{\ref{aff13}}
\and P.~Cassata\orcid{0000-0002-6716-4400}\inst{\ref{aff14},\ref{aff15}}
\and D.~Sluse\orcid{0000-0001-6116-2095}\inst{\ref{aff16}}
\and S.~Bisogni\orcid{0000-0003-3746-4565}\inst{\ref{aff17}}
\and B.~Altieri\orcid{0000-0003-3936-0284}\inst{\ref{aff18}}
\and S.~Andreon\orcid{0000-0002-2041-8784}\inst{\ref{aff10}}
\and N.~Auricchio\orcid{0000-0003-4444-8651}\inst{\ref{aff3}}
\and H.~Aussel\orcid{0000-0002-1371-5705}\inst{\ref{aff19}}
\and C.~Baccigalupi\orcid{0000-0002-8211-1630}\inst{\ref{aff20},\ref{aff21},\ref{aff22},\ref{aff23}}
\and M.~Baldi\orcid{0000-0003-4145-1943}\inst{\ref{aff7},\ref{aff3},\ref{aff5}}
\and A.~Balestra\orcid{0000-0002-6967-261X}\inst{\ref{aff15}}
\and S.~Bardelli\orcid{0000-0002-8900-0298}\inst{\ref{aff3}}
\and A.~Basset\orcid{0009-0001-5490-3139}\inst{\ref{aff24}}
\and P.~Battaglia\orcid{0000-0002-7337-5909}\inst{\ref{aff3}}
\and A.~Biviano\orcid{0000-0002-0857-0732}\inst{\ref{aff21},\ref{aff20}}
\and M.~Bolzonella\orcid{0000-0003-3278-4607}\inst{\ref{aff3}}
\and E.~Branchini\orcid{0000-0002-0808-6908}\inst{\ref{aff25},\ref{aff26},\ref{aff10}}
\and M.~Brescia\orcid{0000-0001-9506-5680}\inst{\ref{aff27},\ref{aff28}}
\and J.~Brinchmann\orcid{0000-0003-4359-8797}\inst{\ref{aff29},\ref{aff30},\ref{aff31}}
\and S.~Camera\orcid{0000-0003-3399-3574}\inst{\ref{aff32},\ref{aff33},\ref{aff34}}
\and G.~Ca\~nas-Herrera\orcid{0000-0003-2796-2149}\inst{\ref{aff35}}
\and V.~Capobianco\orcid{0000-0002-3309-7692}\inst{\ref{aff34}}
\and C.~Carbone\orcid{0000-0003-0125-3563}\inst{\ref{aff17}}
\and J.~Carretero\orcid{0000-0002-3130-0204}\inst{\ref{aff36},\ref{aff37}}
\and F.~J.~Castander\orcid{0000-0001-7316-4573}\inst{\ref{aff38},\ref{aff39}}
\and M.~Castellano\orcid{0000-0001-9875-8263}\inst{\ref{aff40}}
\and G.~Castignani\orcid{0000-0001-6831-0687}\inst{\ref{aff3}}
\and S.~Cavuoti\orcid{0000-0002-3787-4196}\inst{\ref{aff28},\ref{aff41}}
\and K.~C.~Chambers\orcid{0000-0001-6965-7789}\inst{\ref{aff42}}
\and A.~Cimatti\inst{\ref{aff43}}
\and C.~Colodro-Conde\inst{\ref{aff44}}
\and G.~Congedo\orcid{0000-0003-2508-0046}\inst{\ref{aff45}}
\and C.~J.~Conselice\orcid{0000-0003-1949-7638}\inst{\ref{aff46}}
\and L.~Conversi\orcid{0000-0002-6710-8476}\inst{\ref{aff47},\ref{aff18}}
\and A.~Costille\inst{\ref{aff1}}
\and F.~Courbin\orcid{0000-0003-0758-6510}\inst{\ref{aff48},\ref{aff49},\ref{aff50}}
\and H.~M.~Courtois\orcid{0000-0003-0509-1776}\inst{\ref{aff51}}
\and M.~Cropper\orcid{0000-0003-4571-9468}\inst{\ref{aff52}}
\and J.-G.~Cuby\orcid{0000-0002-8767-1442}\inst{\ref{aff53},\ref{aff1}}
\and A.~Da~Silva\orcid{0000-0002-6385-1609}\inst{\ref{aff54},\ref{aff55}}
\and H.~Degaudenzi\orcid{0000-0002-5887-6799}\inst{\ref{aff56}}
\and S.~de~la~Torre\inst{\ref{aff1}}
\and G.~De~Lucia\orcid{0000-0002-6220-9104}\inst{\ref{aff21}}
\and H.~Dole\orcid{0000-0002-9767-3839}\inst{\ref{aff57},\ref{aff58}}
\and F.~Dubath\orcid{0000-0002-6533-2810}\inst{\ref{aff56}}
\and X.~Dupac\inst{\ref{aff18}}
\and S.~Dusini\orcid{0000-0002-1128-0664}\inst{\ref{aff59}}
\and S.~Escoffier\orcid{0000-0002-2847-7498}\inst{\ref{aff60}}
\and M.~Fabricius\orcid{0000-0002-7025-6058}\inst{\ref{aff61},\ref{aff62}}
\and M.~Farina\orcid{0000-0002-3089-7846}\inst{\ref{aff63}}
\and R.~Farinelli\inst{\ref{aff3}}
\and S.~Ferriol\inst{\ref{aff8}}
\and F.~Finelli\orcid{0000-0002-6694-3269}\inst{\ref{aff3},\ref{aff64}}
\and S.~Fotopoulou\orcid{0000-0002-9686-254X}\inst{\ref{aff65}}
\and N.~Fourmanoit\orcid{0009-0005-6816-6925}\inst{\ref{aff60}}
\and M.~Frailis\orcid{0000-0002-7400-2135}\inst{\ref{aff21}}
\and E.~Franceschi\orcid{0000-0002-0585-6591}\inst{\ref{aff3}}
\and M.~Fumana\orcid{0000-0001-6787-5950}\inst{\ref{aff17}}
\and L.~Gabarra\orcid{0000-0002-8486-8856}\inst{\ref{aff66}}
\and S.~Galeotta\orcid{0000-0002-3748-5115}\inst{\ref{aff21}}
\and K.~George\orcid{0000-0002-1734-8455}\inst{\ref{aff67}}
\and W.~Gillard\orcid{0000-0003-4744-9748}\inst{\ref{aff60}}
\and C.~Giocoli\orcid{0000-0002-9590-7961}\inst{\ref{aff3},\ref{aff5}}
\and J.~Gracia-Carpio\orcid{0000-0003-4689-3134}\inst{\ref{aff61}}
\and B.~R.~Granett\orcid{0000-0003-2694-9284}\inst{\ref{aff10}}
\and A.~Grazian\orcid{0000-0002-5688-0663}\inst{\ref{aff15}}
\and F.~Grupp\inst{\ref{aff61},\ref{aff62}}
\and W.~G.~Hartley\orcid{0000-0002-4563-5101}\inst{\ref{aff56}}
\and S.~V.~H.~Haugan\orcid{0000-0001-9648-7260}\inst{\ref{aff68}}
\and S.~Hemmati\orcid{0000-0003-2226-5395}\inst{\ref{aff69}}
\and H.~Hoekstra\orcid{0000-0002-0641-3231}\inst{\ref{aff35}}
\and W.~Holmes\orcid{0009-0007-8554-4646}\inst{\ref{aff70}}
\and I.~M.~Hook\orcid{0000-0002-2960-978X}\inst{\ref{aff71}}
\and F.~Hormuth\inst{\ref{aff72}}
\and A.~Hornstrup\orcid{0000-0002-3363-0936}\inst{\ref{aff73},\ref{aff74}}
\and M.~Huertas-Company\orcid{0000-0002-1416-8483}\inst{\ref{aff44},\ref{aff75},\ref{aff76}}
\and K.~Jahnke\orcid{0000-0003-3804-2137}\inst{\ref{aff77}}
\and M.~Jhabvala\inst{\ref{aff78}}
\and S.~Kermiche\orcid{0000-0002-0302-5735}\inst{\ref{aff60}}
\and A.~Kiessling\orcid{0000-0002-2590-1273}\inst{\ref{aff70}}
\and B.~Kubik\orcid{0009-0006-5823-4880}\inst{\ref{aff8}}
\and M.~K\"ummel\orcid{0000-0003-2791-2117}\inst{\ref{aff62}}
\and M.~Kunz\orcid{0000-0002-3052-7394}\inst{\ref{aff79}}
\and H.~Kurki-Suonio\orcid{0000-0002-4618-3063}\inst{\ref{aff80},\ref{aff81}}
\and A.~M.~C.~Le~Brun\orcid{0000-0002-0936-4594}\inst{\ref{aff82}}
\and D.~Le~Mignant\orcid{0000-0002-5339-5515}\inst{\ref{aff1}}
\and P.~Liebing\orcid{0009-0005-8043-7150}\inst{\ref{aff52}}
\and S.~Ligori\orcid{0000-0003-4172-4606}\inst{\ref{aff34}}
\and P.~B.~Lilje\orcid{0000-0003-4324-7794}\inst{\ref{aff68}}
\and V.~Lindholm\orcid{0000-0003-2317-5471}\inst{\ref{aff80},\ref{aff81}}
\and I.~Lloro\orcid{0000-0001-5966-1434}\inst{\ref{aff83}}
\and M.~Magliocchetti\orcid{0000-0001-9158-4838}\inst{\ref{aff63}}
\and G.~Mainetti\orcid{0000-0003-2384-2377}\inst{\ref{aff84}}
\and E.~Maiorano\orcid{0000-0003-2593-4355}\inst{\ref{aff3}}
\and O.~Mansutti\orcid{0000-0001-5758-4658}\inst{\ref{aff21}}
\and O.~Marggraf\orcid{0000-0001-7242-3852}\inst{\ref{aff85}}
\and M.~Martinelli\orcid{0000-0002-6943-7732}\inst{\ref{aff40},\ref{aff86}}
\and N.~Martinet\orcid{0000-0003-2786-7790}\inst{\ref{aff1}}
\and F.~Marulli\orcid{0000-0002-8850-0303}\inst{\ref{aff2},\ref{aff3},\ref{aff5}}
\and R.~J.~Massey\orcid{0000-0002-6085-3780}\inst{\ref{aff87}}
\and N.~Mauri\orcid{0000-0001-8196-1548}\inst{\ref{aff43},\ref{aff5}}
\and E.~Medinaceli\orcid{0000-0002-4040-7783}\inst{\ref{aff3}}
\and S.~Mei\orcid{0000-0002-2849-559X}\inst{\ref{aff88},\ref{aff89}}
\and M.~Melchior\inst{\ref{aff90}}
\and M.~Meneghetti\orcid{0000-0003-1225-7084}\inst{\ref{aff3},\ref{aff5}}
\and E.~Merlin\orcid{0000-0001-6870-8900}\inst{\ref{aff15}}
\and G.~Meylan\orcid{0000-0001-6503-0209}\inst{\ref{aff91}}
\and P.~Monaco\orcid{0000-0003-2083-7564}\inst{\ref{aff92},\ref{aff21},\ref{aff22},\ref{aff20}}
\and A.~Mora\orcid{0000-0002-1922-8529}\inst{\ref{aff93}}
\and C.~Moretti\orcid{0000-0003-3314-8936}\inst{\ref{aff21},\ref{aff20},\ref{aff22}}
\and L.~Moscardini\orcid{0000-0002-3473-6716}\inst{\ref{aff2},\ref{aff3},\ref{aff5}}
\and R.~Nakajima\orcid{0009-0009-1213-7040}\inst{\ref{aff85}}
\and C.~Neissner\orcid{0000-0001-8524-4968}\inst{\ref{aff94},\ref{aff37}}
\and R.~C.~Nichol\orcid{0000-0003-0939-6518}\inst{\ref{aff95}}
\and S.-M.~Niemi\orcid{0009-0005-0247-0086}\inst{\ref{aff96}}
\and C.~Padilla\orcid{0000-0001-7951-0166}\inst{\ref{aff94}}
\and S.~Paltani\orcid{0000-0002-8108-9179}\inst{\ref{aff56}}
\and F.~Pasian\orcid{0000-0002-4869-3227}\inst{\ref{aff21}}
\and W.~J.~Percival\orcid{0000-0002-0644-5727}\inst{\ref{aff97},\ref{aff98},\ref{aff99}}
\and V.~Pettorino\orcid{0000-0002-4203-9320}\inst{\ref{aff96}}
\and S.~Pires\orcid{0000-0002-0249-2104}\inst{\ref{aff19}}
\and G.~Polenta\orcid{0000-0003-4067-9196}\inst{\ref{aff100}}
\and M.~Poncet\inst{\ref{aff24}}
\and F.~Raison\orcid{0000-0002-7819-6918}\inst{\ref{aff61}}
\and A.~Refregier\inst{\ref{aff101}}
\and A.~Renzi\orcid{0000-0001-9856-1970}\inst{\ref{aff14},\ref{aff59},\ref{aff3}}
\and J.~Rhodes\orcid{0000-0002-4485-8549}\inst{\ref{aff70}}
\and G.~Riccio\inst{\ref{aff28}}
\and I.~Risso\orcid{0000-0003-2525-7761}\inst{\ref{aff25},\ref{aff26},\ref{aff10}}
\and F.~Rizzo\orcid{0000-0002-9407-585X}\inst{\ref{aff21}}
\and E.~Romelli\orcid{0000-0003-3069-9222}\inst{\ref{aff21}}
\and B.~Rusholme\orcid{0000-0001-7648-4142}\inst{\ref{aff69}}
\and R.~Saglia\orcid{0000-0003-0378-7032}\inst{\ref{aff62},\ref{aff61}}
\and Z.~Sakr\orcid{0000-0002-4823-3757}\inst{\ref{aff102},\ref{aff13},\ref{aff103}}
\and A.~G.~S\'anchez\orcid{0000-0003-1198-831X}\inst{\ref{aff61}}
\and D.~Sapone\orcid{0000-0001-7089-4503}\inst{\ref{aff104}}
\and B.~Sartoris\orcid{0000-0003-1337-5269}\inst{\ref{aff105}}
\and M.~Schirmer\orcid{0000-0003-2568-9994}\inst{\ref{aff77}}
\and P.~Schneider\orcid{0000-0001-8561-2679}\inst{\ref{aff85}}
\and T.~Schrabback\orcid{0000-0002-6987-7834}\inst{\ref{aff106}}
\and M.~Scodeggio\inst{\ref{aff17}}
\and A.~Secroun\orcid{0000-0003-0505-3710}\inst{\ref{aff60}}
\and E.~Sefusatti\orcid{0000-0003-0473-1567}\inst{\ref{aff21},\ref{aff20},\ref{aff22}}
\and E.~Sihvola\orcid{0000-0003-1804-7715}\inst{\ref{aff80},\ref{aff81}}
\and C.~Sirignano\orcid{0000-0002-0995-7146}\inst{\ref{aff14},\ref{aff59}}
\and G.~Sirri\orcid{0000-0003-2626-2853}\inst{\ref{aff5}}
\and L.~Stanco\orcid{0000-0002-9706-5104}\inst{\ref{aff59}}
\and C.~Surace\orcid{0000-0003-2592-0113}\inst{\ref{aff1}}
\and P.~Tallada-Cresp\'{i}\orcid{0000-0002-1336-8328}\inst{\ref{aff36},\ref{aff37}}
\and A.~N.~Taylor\inst{\ref{aff45}}
\and H.~I.~Teplitz\orcid{0000-0002-7064-5424}\inst{\ref{aff107}}
\and I.~Tereno\orcid{0000-0002-4537-6218}\inst{\ref{aff54},\ref{aff55}}
\and N.~Tessore\orcid{0000-0002-9696-7931}\inst{\ref{aff52}}
\and S.~Toft\orcid{0000-0003-3631-7176}\inst{\ref{aff108},\ref{aff109}}
\and R.~Toledo-Moreo\orcid{0000-0002-2997-4859}\inst{\ref{aff110},\ref{aff111}}
\and F.~Torradeflot\orcid{0000-0003-1160-1517}\inst{\ref{aff37},\ref{aff36}}
\and A.~Tsyganov\inst{\ref{aff112}}
\and I.~Tutusaus\orcid{0000-0002-3199-0399}\inst{\ref{aff38},\ref{aff39},\ref{aff13}}
\and J.~Valiviita\orcid{0000-0001-6225-3693}\inst{\ref{aff80},\ref{aff81}}
\and T.~Vassallo\orcid{0000-0001-6512-6358}\inst{\ref{aff21},\ref{aff67}}
\and A.~Veropalumbo\orcid{0000-0003-2387-1194}\inst{\ref{aff10},\ref{aff26},\ref{aff25}}
\and Y.~Wang\orcid{0000-0002-4749-2984}\inst{\ref{aff69}}
\and J.~Weller\orcid{0000-0002-8282-2010}\inst{\ref{aff62},\ref{aff61}}
\and A.~Zacchei\orcid{0000-0003-0396-1192}\inst{\ref{aff21},\ref{aff20}}
\and F.~M.~Zerbi\orcid{0000-0002-9996-973X}\inst{\ref{aff10}}
\and I.~A.~Zinchenko\orcid{0000-0002-2944-2449}\inst{\ref{aff113}}
\and E.~Zucca\orcid{0000-0002-5845-8132}\inst{\ref{aff3}}
\and V.~Allevato\orcid{0000-0001-7232-5152}\inst{\ref{aff28}}
\and I.~T.~Andika\orcid{0000-0001-6102-9526}\inst{\ref{aff62}}
\and M.~Ballardini\orcid{0000-0003-4481-3559}\inst{\ref{aff114},\ref{aff115},\ref{aff3}}
\and P.~Bergamini\orcid{0000-0003-1383-9414}\inst{\ref{aff3}}
\and S.~Borgani\orcid{0000-0001-6151-6439}\inst{\ref{aff92},\ref{aff20},\ref{aff21},\ref{aff22},\ref{aff116}}
\and E.~Bozzo\orcid{0000-0002-8201-1525}\inst{\ref{aff56}}
\and C.~Burigana\orcid{0000-0002-3005-5796}\inst{\ref{aff117},\ref{aff64}}
\and R.~Cabanac\orcid{0000-0001-6679-2600}\inst{\ref{aff13}}
\and M.~Calabrese\orcid{0000-0002-2637-2422}\inst{\ref{aff118},\ref{aff17}}
\and A.~Cappi\inst{\ref{aff105},\ref{aff3}}
\and F.~Caro\orcid{0009-0003-1053-0507}\inst{\ref{aff40}}
\and T.~Castro\orcid{0000-0002-6292-3228}\inst{\ref{aff119},\ref{aff21},\ref{aff20}}
\and S.~Conseil\orcid{0000-0002-3657-4191}\inst{\ref{aff120},\ref{aff8}}
\and D.~Di~Ferdinando\inst{\ref{aff5}}
\and J.~A.~Escartin~Vigo\inst{\ref{aff61}}
\and G.~Fabbian\orcid{0000-0002-3255-4695}\inst{\ref{aff57}}
\and J.~Garc\'ia-Bellido\orcid{0000-0002-9370-8360}\inst{\ref{aff102}}
\and T.~Gasparetto\orcid{0000-0002-7913-4866}\inst{\ref{aff40}}
\and A.~Gregorio\orcid{0000-0003-4028-8785}\inst{\ref{aff92},\ref{aff21},\ref{aff22}}
\and C.~M.~Gutierrez\orcid{0000-0001-7854-783X}\inst{\ref{aff44},\ref{aff121}}
\and A.~Hall\orcid{0000-0002-3139-8651}\inst{\ref{aff45}}
\and Y.~Kang\orcid{0009-0000-8588-7250}\inst{\ref{aff106}}
\and A.~Loureiro\orcid{0000-0002-4371-0876}\inst{\ref{aff122},\ref{aff123}}
\and C.~Mancini\orcid{0000-0002-4297-0561}\inst{\ref{aff17}}
\and R.~Maoli\orcid{0000-0002-6065-3025}\inst{\ref{aff124},\ref{aff40}}
\and R.~B.~Metcalf\orcid{0000-0003-3167-2574}\inst{\ref{aff2},\ref{aff3}}
\and S.~Nadathur\orcid{0000-0001-9070-3102}\inst{\ref{aff125}}
\and M.~P\"ontinen\orcid{0000-0001-5442-2530}\inst{\ref{aff80}}
\and E.~Sarpa\orcid{0000-0002-1256-655X}\inst{\ref{aff21}}
\and V.~Scottez\orcid{0009-0008-3864-940X}\inst{\ref{aff126},\ref{aff127}}
\and M.~Sereno\orcid{0000-0003-0302-0325}\inst{\ref{aff3},\ref{aff5}}
\and K.~Tanidis\orcid{0000-0001-9843-5130}\inst{\ref{aff128}}
\and M.~Tenti\orcid{0000-0002-4254-5901}\inst{\ref{aff5}}
\and A.~Troja\orcid{0000-0003-0239-4595}\inst{\ref{aff21}}
\and M.~Tucci\inst{\ref{aff56}}
\and M.~Viel\orcid{0000-0002-2642-5707}\inst{\ref{aff20},\ref{aff21},\ref{aff23},\ref{aff22},\ref{aff116}}
\and M.~Wiesmann\orcid{0009-0000-8199-5860}\inst{\ref{aff68}}
\and Abdurro'uf\orcid{0000-0002-5258-8761}\inst{\ref{aff129}}
\and J.~A.~Acevedo~Barroso\orcid{0000-0002-9654-1711}\inst{\ref{aff70}}
\and Y.~Akrami\orcid{0000-0002-2407-7956}\inst{\ref{aff102},\ref{aff130}}
\and G.~Alguero\inst{\ref{aff131}}
\and G.~Angora\orcid{0000-0002-0316-6562}\inst{\ref{aff28},\ref{aff114}}
\and S.~Anselmi\orcid{0000-0002-3579-9583}\inst{\ref{aff59},\ref{aff14},\ref{aff132}}
\and M.~Archidiacono\orcid{0000-0003-4952-9012}\inst{\ref{aff9},\ref{aff11}}
\and G.~Aric\`o\orcid{0000-0002-2802-2928}\inst{\ref{aff5}}
\and F.~Atrio-Barandela\orcid{0000-0002-2130-2513}\inst{\ref{aff133}}
\and M.~Aubert\orcid{0009-0002-7667-8814}\inst{\ref{aff8}}
\and M.~Baes\orcid{0000-0002-3930-2757}\inst{\ref{aff134}}
\and N.~Bartolo\inst{\ref{aff14},\ref{aff59},\ref{aff15}}
\and L.~Bazzanini\orcid{0000-0003-0727-0137}\inst{\ref{aff114},\ref{aff3}}
\and J.~Bel\orcid{0009-0006-7837-1866}\inst{\ref{aff135}}
\and J.~R.~Bermejo-Climent\inst{\ref{aff136},\ref{aff137}}
\and D.~Bertacca\orcid{0000-0002-2490-7139}\inst{\ref{aff14},\ref{aff15},\ref{aff59}}
\and F.~Beutler\orcid{0000-0003-0467-5438}\inst{\ref{aff45}}
\and A.~Blanchard\orcid{0000-0001-8555-9003}\inst{\ref{aff13}}
\and M.~Bonici\orcid{0000-0002-8430-126X}\inst{\ref{aff97},\ref{aff17}}
\and F.~Bouch\`e\orcid{0000-0002-4663-1786}\inst{\ref{aff138},\ref{aff41}}
\and M.~L.~Brown\orcid{0000-0002-0370-8077}\inst{\ref{aff46}}
\and S.~Bruton\orcid{0000-0002-6503-5218}\inst{\ref{aff139}}
\and A.~Calabro\orcid{0000-0003-2536-1614}\inst{\ref{aff40}}
\and B.~Camacho~Quevedo\orcid{0000-0002-8789-4232}\inst{\ref{aff20},\ref{aff23},\ref{aff21}}
\and C.~S.~Carvalho\inst{\ref{aff140}}
\and F.~Cogato\orcid{0000-0003-4632-6113}\inst{\ref{aff3}}
\and T.~E.~Collett\orcid{0000-0001-5564-3140}\inst{\ref{aff125}}
\and A.~R.~Cooray\orcid{0000-0002-3892-0190}\inst{\ref{aff141}}
\and P.~Corcho-Caballero\orcid{0000-0001-6327-7080}\inst{\ref{aff142}}
\and M.~Correnti\orcid{0000-0001-6464-3257}\inst{\ref{aff40},\ref{aff100}}
\and B.~Csizi\orcid{0000-0003-3227-6581}\inst{\ref{aff106}}
\and O.~Cucciati\orcid{0000-0002-9336-7551}\inst{\ref{aff3}}
\and H.~Dannerbauer\orcid{0000-0001-7147-3575}\inst{\ref{aff44},\ref{aff121}}
\and T.~de~Boer\orcid{0000-0001-5486-2747}\inst{\ref{aff42}}
\and G.~Desprez\orcid{0000-0001-8325-1742}\inst{\ref{aff142}}
\and A.~D\'iaz-S\'anchez\orcid{0000-0003-0748-4768}\inst{\ref{aff143}}
\and S.~Di~Domizio\orcid{0000-0003-2863-5895}\inst{\ref{aff25},\ref{aff26}}
\and J.~M.~Diego\orcid{0000-0001-9065-3926}\inst{\ref{aff144}}
\and P.-A.~Duc\orcid{0000-0003-3343-6284}\inst{\ref{aff4}}
\and V.~Duret\orcid{0009-0009-0383-4960}\inst{\ref{aff145}}
\and A.~Eggemeier\orcid{0000-0002-1841-8910}\inst{\ref{aff85}}
\and A.~Enia\orcid{0000-0002-0200-2857}\inst{\ref{aff146},\ref{aff3}}
\and Y.~Fang\orcid{0000-0002-0334-6950}\inst{\ref{aff62}}
\and A.~Farina\orcid{0009-0000-3420-929X}\inst{\ref{aff10},\ref{aff26}}
\and A.~M.~N.~Ferguson\orcid{0000-0001-7934-1278}\inst{\ref{aff45}}
\and A.~Finoguenov\orcid{0000-0002-4606-5403}\inst{\ref{aff80}}
\and F.~Fontanot\orcid{0000-0003-4744-0188}\inst{\ref{aff21},\ref{aff20}}
\and A.~Franco\orcid{0000-0002-4761-366X}\inst{\ref{aff147},\ref{aff148},\ref{aff149}}
\and N.~Frusciante\orcid{0000-0002-7375-1230}\inst{\ref{aff150},\ref{aff151}}
\and Y.~Fu\orcid{0000-0002-0759-0504}\inst{\ref{aff35},\ref{aff142}}
\and K.~Ganga\orcid{0000-0001-8159-8208}\inst{\ref{aff88}}
\and E.~Gaztanaga\orcid{0000-0001-9632-0815}\inst{\ref{aff38},\ref{aff39},\ref{aff125}}
\and Z.~Ghaffari\orcid{0000-0002-6467-8078}\inst{\ref{aff21},\ref{aff20}}
\and F.~Giacomini\orcid{0000-0002-3129-2814}\inst{\ref{aff5}}
\and F.~Gianotti\orcid{0000-0003-4666-119X}\inst{\ref{aff3}}
\and G.~Gozaliasl\orcid{0000-0002-0236-919X}\inst{\ref{aff152},\ref{aff80}}
\and A.~Gruppuso\orcid{0000-0001-9272-5292}\inst{\ref{aff3},\ref{aff5}}
\and M.~Guidi\orcid{0000-0001-9408-1101}\inst{\ref{aff7},\ref{aff3}}
\and N.~A.~Hatch\orcid{0000-0001-5600-0534}\inst{\ref{aff153}}
\and C.~Hern\'andez-Monteagudo\orcid{0000-0001-5471-9166}\inst{\ref{aff121},\ref{aff44}}
\and J.~Hjorth\orcid{0000-0002-4571-2306}\inst{\ref{aff154}}
\and M.~Jauzac\orcid{0000-0003-1974-8732}\inst{\ref{aff155},\ref{aff87},\ref{aff156},\ref{aff157},\ref{aff13}}
\and J.~J.~E.~Kajava\orcid{0000-0002-3010-8333}\inst{\ref{aff158},\ref{aff159},\ref{aff160}}
\and D.~Karagiannis\orcid{0000-0002-4927-0816}\inst{\ref{aff161},\ref{aff114},\ref{aff162}}
\and J.~Kim\orcid{0000-0003-2776-2761}\inst{\ref{aff66}}
\and C.~C.~Kirkpatrick\inst{\ref{aff163}}
\and A.~Kov\'acs\orcid{0000-0002-5825-579X}\inst{\ref{aff137},\ref{aff136}}
\and I.~Kova{\v{c}}i{\'{c}}\orcid{0000-0001-6751-3263}\inst{\ref{aff134}}
\and K.~Koyama\orcid{0000-0001-6727-6915}\inst{\ref{aff125}}
\and S.~Kruk\orcid{0000-0001-8010-8879}\inst{\ref{aff18}}
\and M.~C.~Lam\orcid{0000-0002-9347-2298}\inst{\ref{aff45}}
\and M.~Lattanzi\orcid{0000-0003-1059-2532}\inst{\ref{aff115}}
\and F.~Leclercq\orcid{0000-0002-9339-1404}\inst{\ref{aff164}}
\and S.~Lee\orcid{0000-0002-8289-740X}\inst{\ref{aff165},\ref{aff70}}
\and M.~Lembo\orcid{0000-0002-5271-5070}\inst{\ref{aff164}}
\and M.~D.~Lepinzan\orcid{0000-0003-1287-9801}\inst{\ref{aff21}}
\and G.~F.~Lesci\orcid{0000-0002-4607-2830}\inst{\ref{aff117}}
\and J.~Lesgourgues\orcid{0000-0001-7627-353X}\inst{\ref{aff166}}
\and L.~Linke\orcid{0000-0002-2622-8113}\inst{\ref{aff106}}
\and S.~J.~Liu\orcid{0000-0001-7680-2139}\inst{\ref{aff63}}
\and F.~Mannucci\orcid{0000-0002-4803-2381}\inst{\ref{aff167}}
\and C.~J.~A.~P.~Martins\orcid{0000-0002-4886-9261}\inst{\ref{aff168},\ref{aff30}}
\and M.~Migliaccio\inst{\ref{aff169},\ref{aff170}}
\and M.~Miluzio\inst{\ref{aff18},\ref{aff171}}
\and G.~Morgante\inst{\ref{aff3}}
\and K.~Naidoo\orcid{0000-0002-9182-1802}\inst{\ref{aff125},\ref{aff77}}
\and A.~Navarro-Alsina\orcid{0000-0002-3173-2592}\inst{\ref{aff85}}
\and S.~Nesseris\orcid{0000-0002-0567-0324}\inst{\ref{aff102}}
\and L.~Nicastro\orcid{0000-0001-8534-6788}\inst{\ref{aff3}}
\and M.~Oguri\orcid{0000-0003-3484-399X}\inst{\ref{aff172},\ref{aff173}}
\and F.~Oppizzi\orcid{0000-0003-3904-8370}\inst{\ref{aff26},\ref{aff59},\ref{aff14}}
\and F.~Pace\orcid{0000-0001-8039-0480}\inst{\ref{aff32},\ref{aff33},\ref{aff34},\ref{aff55}}
\and M.~J.~Page\orcid{0000-0002-6689-6271}\inst{\ref{aff52}}
\and E.~Palazzi\orcid{0000-0002-8691-7666}\inst{\ref{aff3}}
\and D.~Paoletti\orcid{0000-0003-4761-6147}\inst{\ref{aff3},\ref{aff64}}
\and G.~Papini\orcid{0000-0002-5810-318X}\inst{\ref{aff3},\ref{aff2}}
\and F.~Passalacqua\orcid{0000-0002-8606-4093}\inst{\ref{aff59}}
\and K.~Paterson\orcid{0000-0001-8340-3486}\inst{\ref{aff77}}
\and C.~Pattison\orcid{0000-0003-3272-2617}\inst{\ref{aff125}}
\and R.~Paviot\orcid{0009-0002-8108-3460}\inst{\ref{aff19},\ref{aff24}}
\and A.~Pisani\orcid{0000-0002-6146-4437}\inst{\ref{aff60}}
\and L.~Pizzuti\orcid{0000-0001-5654-7580}\inst{\ref{aff174}}
\and J.~J.~Popp\orcid{0000-0002-3724-1727}\inst{\ref{aff175}}
\and D.~Potter\orcid{0000-0002-0757-5195}\inst{\ref{aff176}}
\and M.~Poulain\orcid{0000-0002-7664-4510}\inst{\ref{aff177}}
\and G.~W.~Pratt\inst{\ref{aff19}}
\and M.~Radovich\orcid{0000-0002-3585-866X}\inst{\ref{aff15}}
\and F.~Ricci\orcid{0000-0001-5742-5980}\inst{\ref{aff178},\ref{aff40}}
\and G.~Rodighiero\orcid{0000-0002-9415-2296}\inst{\ref{aff14},\ref{aff15}}
\and W.~Roster\orcid{0000-0002-9149-6528}\inst{\ref{aff61}}
\and M.~Sahl\'en\orcid{0000-0003-0973-4804}\inst{\ref{aff179}}
\and D.~B.~Sanders\orcid{0000-0002-1233-9998}\inst{\ref{aff42}}
\and A.~Schneider\orcid{0000-0001-7055-8104}\inst{\ref{aff176}}
\and M.~Schultheis\inst{\ref{aff105}}
\and D.~Sciotti\orcid{0009-0008-4519-2620}\inst{\ref{aff40},\ref{aff86}}
\and E.~Sellentin\orcid{0009-0002-2655-3458}\inst{\ref{aff180},\ref{aff35}}
\and S.~Serjeant\orcid{0000-0002-0517-7943}\inst{\ref{aff175}}
\and F.~Shankar\orcid{0000-0001-8973-5051}\inst{\ref{aff181}}
\and L.~C.~Smith\orcid{0000-0002-3259-2771}\inst{\ref{aff182}}
\and I.~Szapudi\orcid{0000-0003-2274-0301}\inst{\ref{aff42}}
\and F.~Tarsitano\orcid{0000-0002-5919-0238}\inst{\ref{aff183},\ref{aff101}}
\and G.~Testera\orcid{0000-0003-2970-766X}\inst{\ref{aff26}}
\and R.~Teyssier\orcid{0000-0001-7689-0933}\inst{\ref{aff184}}
\and S.~Tosi\orcid{0000-0002-7275-9193}\inst{\ref{aff25},\ref{aff10},\ref{aff26}}
\and C.~Uhlemann\orcid{0000-0001-7831-1579}\inst{\ref{aff185}}
\and C.~Valieri\inst{\ref{aff5}}
\and A.~Venhola\orcid{0000-0001-6071-4564}\inst{\ref{aff177}}
\and P.~Vielzeuf\orcid{0000-0003-2035-9339}\inst{\ref{aff60}}
\and A.~Viitanen\orcid{0000-0001-9383-786X}\inst{\ref{aff56},\ref{aff163},\ref{aff40}}
\and E.~Vilenius\orcid{0000-0002-6184-7681}\inst{\ref{aff52}}
\and S.~Vinciguerra\orcid{0009-0005-4018-3184}\inst{\ref{aff1}}
\and M.~von~Wietersheim-Kramsta\orcid{0000-0003-4986-5091}\inst{\ref{aff87},\ref{aff155}}
\and N.~A.~Walton\orcid{0000-0003-3983-8778}\inst{\ref{aff182}}
\and L.~Wang\orcid{0000-0002-6736-9158}\inst{\ref{aff186},\ref{aff142}}
\and A.~H.~Wright\orcid{0000-0001-7363-7932}\inst{\ref{aff187}}
\and D.~Scott\orcid{0000-0002-6878-9840}\inst{\ref{aff188}}}
										   
\institute{Aix-Marseille Universit\'e, CNRS, CNES, LAM, Marseille, France\label{aff1}
\and
Dipartimento di Fisica e Astronomia "Augusto Righi" - Alma Mater Studiorum Universit\`a di Bologna, via Piero Gobetti 93/2, 40129 Bologna, Italy\label{aff2}
\and
INAF-Osservatorio di Astrofisica e Scienza dello Spazio di Bologna, Via Piero Gobetti 93/3, 40129 Bologna, Italy\label{aff3}
\and
Universit\'e de Strasbourg, CNRS, Observatoire astronomique de Strasbourg, UMR 7550, 67000 Strasbourg, France\label{aff4}
\and
INFN-Sezione di Bologna, Viale Berti Pichat 6/2, 40127 Bologna, Italy\label{aff5}
\and
Astronomical Institute, Tohoku University, Sendai 980-8578, Japan\label{aff6}
\and
Dipartimento di Fisica e Astronomia, Universit\`a di Bologna, Via Gobetti 93/2, 40129 Bologna, Italy\label{aff7}
\and
Universit\'e Lyon 1, CNRS/IN2P3, IP2I Lyon, Villeurbanne, France\label{aff8}
\and
Dipartimento di Fisica "Aldo Pontremoli", Universit\`a degli Studi di Milano, Via Celoria 16, 20133 Milano, Italy\label{aff9}
\and
INAF-Osservatorio Astronomico di Brera, Via Brera 28, 20122 Milano, Italy\label{aff10}
\and
INFN-Sezione di Milano, Via Celoria 16, 20133 Milano, Italy\label{aff11}
\and
Minnesota Institute for Astrophysics, University of Minnesota, 116 Church St SE, Minneapolis, MN 55455, USA\label{aff12}
\and
Institut de Recherche en Astrophysique et Plan\'etologie (IRAP), Universit\'e de Toulouse, CNRS, UPS, CNES, 14 Av. Edouard Belin, 31400 Toulouse, France\label{aff13}
\and
Dipartimento di Fisica e Astronomia "G. Galilei", Universit\`a di Padova, Via Marzolo 8, 35131 Padova, Italy\label{aff14}
\and
INAF-Osservatorio Astronomico di Padova, Via dell'Osservatorio 5, 35122 Padova, Italy\label{aff15}
\and
STAR Institute, University of Li{\`e}ge, Quartier Agora, All\'ee du six Ao\^ut 19c, 4000 Li\`ege, Belgium\label{aff16}
\and
INAF-IASF Milano, Via Alfonso Corti 12, 20133 Milano, Italy\label{aff17}
\and
ESAC/ESA, Camino Bajo del Castillo, s/n., Urb. Villafranca del Castillo, 28692 Villanueva de la Ca\~nada, Madrid, Spain\label{aff18}
\and
Universit\'e Paris-Saclay, Universit\'e Paris Cit\'e, CEA, CNRS, AIM, 91191, Gif-sur-Yvette, France\label{aff19}
\and
IFPU, Institute for Fundamental Physics of the Universe, via Beirut 2, 34151 Trieste, Italy\label{aff20}
\and
INAF-Osservatorio Astronomico di Trieste, Via G. B. Tiepolo 11, 34143 Trieste, Italy\label{aff21}
\and
INFN, Sezione di Trieste, Via Valerio 2, 34127 Trieste TS, Italy\label{aff22}
\and
SISSA, International School for Advanced Studies, Via Bonomea 265, 34136 Trieste TS, Italy\label{aff23}
\and
Centre National d'Etudes Spatiales -- Centre spatial de Toulouse, 18 avenue Edouard Belin, 31401 Toulouse Cedex 9, France\label{aff24}
\and
Dipartimento di Fisica, Universit\`a di Genova, Via Dodecaneso 33, 16146, Genova, Italy\label{aff25}
\and
INFN-Sezione di Genova, Via Dodecaneso 33, 16146, Genova, Italy\label{aff26}
\and
Department of Physics "E. Pancini", University Federico II, Via Cinthia 6, 80126, Napoli, Italy\label{aff27}
\and
INAF-Osservatorio Astronomico di Capodimonte, Via Moiariello 16, 80131 Napoli, Italy\label{aff28}
\and
European Southern Observatory, Karl-Schwarzschild-Str.~2, 85748 Garching, Germany\label{aff29}
\and
Instituto de Astrof\'isica e Ci\^encias do Espa\c{c}o, Universidade do Porto, CAUP, Rua das Estrelas, PT4150-762 Porto, Portugal\label{aff30}
\and
Faculdade de Ci\^encias da Universidade do Porto, Rua do Campo de Alegre, 4150-007 Porto, Portugal\label{aff31}
\and
Dipartimento di Fisica, Universit\`a degli Studi di Torino, Via P. Giuria 1, 10125 Torino, Italy\label{aff32}
\and
INFN-Sezione di Torino, Via P. Giuria 1, 10125 Torino, Italy\label{aff33}
\and
INAF-Osservatorio Astrofisico di Torino, Via Osservatorio 20, 10025 Pino Torinese (TO), Italy\label{aff34}
\and
Leiden Observatory, Leiden University, Einsteinweg 55, 2333 CC Leiden, The Netherlands\label{aff35}
\and
Centro de Investigaciones Energ\'eticas, Medioambientales y Tecnol\'ogicas (CIEMAT), Avenida Complutense 40, 28040 Madrid, Spain\label{aff36}
\and
Port d'Informaci\'{o} Cient\'{i}fica, Campus UAB, C. Albareda s/n, 08193 Bellaterra (Barcelona), Spain\label{aff37}
\and
Institute of Space Sciences (ICE, CSIC), Campus UAB, Carrer de Can Magrans, s/n, 08193 Barcelona, Spain\label{aff38}
\and
Institut d'Estudis Espacials de Catalunya (IEEC),  Edifici RDIT, Campus UPC, 08860 Castelldefels, Barcelona, Spain\label{aff39}
\and
INAF-Osservatorio Astronomico di Roma, Via Frascati 33, 00078 Monteporzio Catone, Italy\label{aff40}
\and
INFN -- Sezione di Napoli, Via Cinthia 6, 80126, Napoli, Italy\label{aff41}
\and
Institute for Astronomy, University of Hawaii, 2680 Woodlawn Drive, Honolulu, HI 96822, USA\label{aff42}
\and
Dipartimento di Fisica e Astronomia "Augusto Righi" - Alma Mater Studiorum Universit\`a di Bologna, Viale Berti Pichat 6/2, 40127 Bologna, Italy\label{aff43}
\and
Instituto de Astrof\'{\i}sica de Canarias, E-38205 La Laguna, Tenerife, Spain\label{aff44}
\and
Institute for Astronomy, University of Edinburgh, Royal Observatory, Blackford Hill, Edinburgh EH9 3HJ, UK\label{aff45}
\and
Jodrell Bank Centre for Astrophysics, Department of Physics and Astronomy, University of Manchester, Oxford Road, Manchester M13 9PL, UK\label{aff46}
\and
European Space Agency/ESRIN, Largo Galileo Galilei 1, 00044 Frascati, Roma, Italy\label{aff47}
\and
Institut de Ci\`{e}ncies del Cosmos (ICCUB), Universitat de Barcelona (IEEC-UB), Mart\'{i} i Franqu\`{e}s 1, 08028 Barcelona, Spain\label{aff48}
\and
Instituci\'o Catalana de Recerca i Estudis Avan\c{c}ats (ICREA), Passeig de Llu\'{\i}s Companys 23, 08010 Barcelona, Spain\label{aff49}
\and
Institut de Ciencies de l'Espai (IEEC-CSIC), Campus UAB, Carrer de Can Magrans, s/n Cerdanyola del Vall\'es, 08193 Barcelona, Spain\label{aff50}
\and
Universit\'e Lyon 1, CNRS/IN2P3, IUF, IP2I Lyon, Villeurbanne, France \label{aff51}
\and
Mullard Space Science Laboratory, University College London, Holmbury St Mary, Dorking, Surrey RH5 6NT, UK\label{aff52}
\and
Canada-France-Hawaii Telescope, 65-1238 Mamalahoa Hwy, Kamuela, HI 96743, USA\label{aff53}
\and
Departamento de F\'isica, Faculdade de Ci\^encias, Universidade de Lisboa, Edif\'icio C8, Campo Grande, PT1749-016 Lisboa, Portugal\label{aff54}
\and
Instituto de Astrof\'isica e Ci\^encias do Espa\c{c}o, Faculdade de Ci\^encias, Universidade de Lisboa, Campo Grande, 1749-016 Lisboa, Portugal\label{aff55}
\and
Department of Astronomy, University of Geneva, ch. d'Ecogia 16, 1290 Versoix, Switzerland\label{aff56}
\and
Universit\'e Paris-Saclay, CNRS, Institut d'astrophysique spatiale, 91405, Orsay, France\label{aff57}
\and
Institut universitaire de France (IUF), 1 rue Descartes, 75231 PARIS CEDEX 05, France\label{aff58}
\and
INFN-Padova, Via Marzolo 8, 35131 Padova, Italy\label{aff59}
\and
Aix-Marseille Universit\'e, CNRS/IN2P3, CPPM, Marseille, France\label{aff60}
\and
Max Planck Institute for Extraterrestrial Physics, Giessenbachstr. 1, 85748 Garching, Germany\label{aff61}
\and
Universit\"ats-Sternwarte M\"unchen, Fakult\"at f\"ur Physik, Ludwig-Maximilians-Universit\"at M\"unchen, Scheinerstr.~1, 81679 M\"unchen, Germany\label{aff62}
\and
INAF-Istituto di Astrofisica e Planetologia Spaziali, via del Fosso del Cavaliere, 100, 00100 Roma, Italy\label{aff63}
\and
INFN-Bologna, Via Irnerio 46, 40126 Bologna, Italy\label{aff64}
\and
School of Physics, HH Wills Physics Laboratory, University of Bristol, Tyndall Avenue, Bristol, BS8 1TL, UK\label{aff65}
\and
Department of Physics, University of Oxford, Keble Road, Oxford OX1 3RH, UK\label{aff66}
\and
University Observatory, LMU Faculty of Physics, Scheinerstr.~1, 81679 Munich, Germany\label{aff67}
\and
Institute of Theoretical Astrophysics, University of Oslo, P.O. Box 1029 Blindern, 0315 Oslo, Norway\label{aff68}
\and
Caltech/IPAC, 1200 E. California Blvd., Pasadena, CA 91125, USA\label{aff69}
\and
Jet Propulsion Laboratory, California Institute of Technology, 4800 Oak Grove Drive, Pasadena, CA, 91109, USA\label{aff70}
\and
School of Physics and Astronomy, Lancaster University, Lancaster, LA1 4YB, UK\label{aff71}
\and
Felix Hormuth Engineering, Goethestr. 17, 69181 Leimen, Germany\label{aff72}
\and
Technical University of Denmark, Elektrovej 327, 2800 Kgs. Lyngby, Denmark\label{aff73}
\and
Cosmic Dawn Center (DAWN), Denmark\label{aff74}
\and
Universit\'e PSL, Observatoire de Paris, Sorbonne Universit\'e, CNRS, LERMA, 75014, Paris, France\label{aff75}
\and
Universit\'e Paris-Cit\'e, 5 Rue Thomas Mann, 75013, Paris, France\label{aff76}
\and
Max-Planck-Institut f\"ur Astronomie, K\"onigstuhl 17, 69117 Heidelberg, Germany\label{aff77}
\and
NASA Goddard Space Flight Center, Greenbelt, MD 20771, USA\label{aff78}
\and
Universit\'e de Gen\`eve, D\'epartement de Physique Th\'eorique and Centre for Astroparticle Physics, 24 quai Ernest-Ansermet, CH-1211 Gen\`eve 4, Switzerland\label{aff79}
\and
Department of Physics, P.O. Box 64, University of Helsinki, 00014 Helsinki, Finland\label{aff80}
\and
Helsinki Institute of Physics, Gustaf H{\"a}llstr{\"o}min katu 2, University of Helsinki, 00014 Helsinki, Finland\label{aff81}
\and
Laboratoire d'etude de l'Univers et des phenomenes eXtremes, Observatoire de Paris, Universit\'e PSL, Sorbonne Universit\'e, CNRS, 92190 Meudon, France\label{aff82}
\and
SKAO, Jodrell Bank, Lower Withington, Macclesfield SK11 9FT, UK\label{aff83}
\and
Centre de Calcul de l'IN2P3/CNRS, 21 avenue Pierre de Coubertin 69627 Villeurbanne Cedex, France\label{aff84}
\and
Universit\"at Bonn, Argelander-Institut f\"ur Astronomie, Auf dem H\"ugel 71, 53121 Bonn, Germany\label{aff85}
\and
INFN-Sezione di Roma, Piazzale Aldo Moro, 2 - c/o Dipartimento di Fisica, Edificio G. Marconi, 00185 Roma, Italy\label{aff86}
\and
Department of Physics, Institute for Computational Cosmology, Durham University, South Road, Durham, DH1 3LE, UK\label{aff87}
\and
Universit\'e Paris Cit\'e, CNRS, Astroparticule et Cosmologie, 75013 Paris, France\label{aff88}
\and
CNRS-UCB International Research Laboratory, Centre Pierre Bin\'etruy, IRL2007, CPB-IN2P3, Berkeley, USA\label{aff89}
\and
University of Applied Sciences and Arts of Northwestern Switzerland, School of Engineering, 5210 Windisch, Switzerland\label{aff90}
\and
Institute of Physics, Laboratory of Astrophysics, Ecole Polytechnique F\'ed\'erale de Lausanne (EPFL), Observatoire de Sauverny, 1290 Versoix, Switzerland\label{aff91}
\and
Dipartimento di Fisica - Sezione di Astronomia, Universit\`a di Trieste, Via Tiepolo 11, 34131 Trieste, Italy\label{aff92}
\and
Telespazio UK S.L. for European Space Agency (ESA), Camino bajo del Castillo, s/n, Urbanizacion Villafranca del Castillo, Villanueva de la Ca\~nada, 28692 Madrid, Spain\label{aff93}
\and
Institut de F\'{i}sica d'Altes Energies (IFAE), The Barcelona Institute of Science and Technology, Campus UAB, 08193 Bellaterra (Barcelona), Spain\label{aff94}
\and
School of Mathematics and Physics, University of Surrey, Guildford, Surrey, GU2 7XH, UK\label{aff95}
\and
European Space Agency/ESTEC, Keplerlaan 1, 2201 AZ Noordwijk, The Netherlands\label{aff96}
\and
Waterloo Centre for Astrophysics, University of Waterloo, Waterloo, Ontario N2L 3G1, Canada\label{aff97}
\and
Department of Physics and Astronomy, University of Waterloo, Waterloo, Ontario N2L 3G1, Canada\label{aff98}
\and
Perimeter Institute for Theoretical Physics, Waterloo, Ontario N2L 2Y5, Canada\label{aff99}
\and
Space Science Data Center, Italian Space Agency, via del Politecnico snc, 00133 Roma, Italy\label{aff100}
\and
Institute for Particle Physics and Astrophysics, Dept. of Physics, ETH Zurich, Wolfgang-Pauli-Strasse 27, 8093 Zurich, Switzerland\label{aff101}
\and
Instituto de F\'isica Te\'orica UAM-CSIC, Campus de Cantoblanco, 28049 Madrid, Spain\label{aff102}
\and
Universit\'e St Joseph; Faculty of Sciences, Beirut, Lebanon\label{aff103}
\and
Departamento de F\'isica, FCFM, Universidad de Chile, Blanco Encalada 2008, Santiago, Chile\label{aff104}
\and
Universit\'e C\^{o}te d'Azur, Observatoire de la C\^{o}te d'Azur, CNRS, Laboratoire Lagrange, Bd de l'Observatoire, CS 34229, 06304 Nice cedex 4, France\label{aff105}
\and
Universit\"at Innsbruck, Institut f\"ur Astro- und Teilchenphysik, Technikerstr. 25/8, 6020 Innsbruck, Austria\label{aff106}
\and
Infrared Processing and Analysis Center, California Institute of Technology, Pasadena, CA 91125, USA\label{aff107}
\and
Cosmic Dawn Center (DAWN)\label{aff108}
\and
Niels Bohr Institute, University of Copenhagen, Jagtvej 128, 2200 Copenhagen, Denmark\label{aff109}
\and
Universidad Polit\'ecnica de Cartagena, Departamento de Electr\'onica y Tecnolog\'ia de Computadoras,  Plaza del Hospital 1, 30202 Cartagena, Spain\label{aff110}
\and
European University of Technology EUt+, European Union\label{aff111}
\and
Centre for Information Technology, University of Groningen, P.O. Box 11044, 9700 CA Groningen, The Netherlands\label{aff112}
\and
Astronomisches Rechen-Institut, Zentrum f\"ur Astronomie der Universit\"at Heidelberg, M\"onchhofstr. 12-14, 69120 Heidelberg, Germany\label{aff113}
\and
Dipartimento di Fisica e Scienze della Terra, Universit\`a degli Studi di Ferrara, Via Giuseppe Saragat 1, 44122 Ferrara, Italy\label{aff114}
\and
Istituto Nazionale di Fisica Nucleare, Sezione di Ferrara, Via Giuseppe Saragat 1, 44122 Ferrara, Italy\label{aff115}
\and
ICSC - Centro Nazionale di Ricerca in High Performance Computing, Big Data e Quantum Computing, Via Magnanelli 2, Bologna, Italy\label{aff116}
\and
INAF, Istituto di Radioastronomia, Via Piero Gobetti 101, 40129 Bologna, Italy\label{aff117}
\and
Astronomical Observatory of the Autonomous Region of the Aosta Valley (OAVdA), Loc. Lignan 39, I-11020, Nus (Aosta Valley), Italy\label{aff118}
\and
Department of Mathematical Physics, Institute of Physics, University of S\~ao Paulo, R. do Mat\~ao 1371, 05508-090, S\~ao Paulo, SP, Brazil\label{aff119}
\and
Centre de Recherche Astrophysique de Lyon, UMR5574, CNRS, Universit\'e Claude Bernard Lyon 1, ENS de Lyon, 69230, Saint-Genis-Laval, France\label{aff120}
\and
Universidad de La Laguna, Dpto. Astrof\'\i sica, E-38206 La Laguna, Tenerife, Spain\label{aff121}
\and
Oskar Klein Centre for Cosmoparticle Physics, Department of Physics, Stockholm University, Stockholm, SE-106 91, Sweden\label{aff122}
\and
Astrophysics Group, Blackett Laboratory, Imperial College London, London SW7 2AZ, UK\label{aff123}
\and
Dipartimento di Fisica, Sapienza Universit\`a di Roma, Piazzale Aldo Moro 2, 00185 Roma, Italy\label{aff124}
\and
Institute of Cosmology and Gravitation, University of Portsmouth, Portsmouth PO1 3FX, UK\label{aff125}
\and
Institut d'Astrophysique de Paris, 98bis Boulevard Arago, 75014, Paris, France\label{aff126}
\and
ICL, Junia, Universit\'e Catholique de Lille, LITL, 59000 Lille, France\label{aff127}
\and
Center for Astrophysics and Cosmology, University of Nova Gorica, Nova Gorica, Slovenia\label{aff128}
\and
Department of Astronomy, Indiana University, 727 East Third Street, Bloomington, IN 47405, USA\label{aff129}
\and
CERCA/ISO, Department of Physics, Case Western Reserve University, 10900 Euclid Avenue, Cleveland, OH 44106, USA\label{aff130}
\and
Univ. Grenoble Alpes, CNRS, Grenoble INP, LPSC-IN2P3, 53, Avenue des Martyrs, 38000 Grenoble, France\label{aff131}
\and
Laboratoire Univers et Th\'eorie, Observatoire de Paris, Universit\'e PSL, Universit\'e Paris Cit\'e, CNRS, 92190 Meudon, France\label{aff132}
\and
Departamento de F{\'\i}sica Fundamental. Universidad de Salamanca. Plaza de la Merced s/n. 37008 Salamanca, Spain\label{aff133}
\and
Universiteit Gent, Department of Physics and Astronomy, Proeftuinstraat 86 N3, 9000 Ghent, Belgium
\label{aff134}
\and
Aix-Marseille Universit\'e, Universit\'e de Toulon, CNRS, CPT, Marseille, France\label{aff135}
\and
Konkoly Observatory, HUN-REN CSFK, MTA Centre of Excellence, Budapest, Konkoly Thege Mikl\'os {\'u}t 15-17. H-1121, Hungary\label{aff136}
\and
MTA-CSFK Lend\"ulet Large-Scale Structure Research Group, Konkoly-Thege Mikl\'os \'ut 15-17, H-1121 Budapest, Hungary\label{aff137}
\and
Scuola Superiore Meridionale, Via Mezzocannone 4, 80138, Napoli, Italy\label{aff138}
\and
California Institute of Technology, 1200 E California Blvd, Pasadena, CA 91125, USA\label{aff139}
\and
Instituto de Astrof\'isica e Ci\^encias do Espa\c{c}o, Faculdade de Ci\^encias, Universidade de Lisboa, Tapada da Ajuda, 1349-018 Lisboa, Portugal\label{aff140}
\and
Department of Physics \& Astronomy, University of California Irvine, Irvine CA 92697, USA\label{aff141}
\and
Kapteyn Astronomical Institute, University of Groningen, PO Box 800, 9700 AV Groningen, The Netherlands\label{aff142}
\and
Departamento F\'isica Aplicada, Universidad Polit\'ecnica de Cartagena, Campus Muralla del Mar, 30202 Cartagena, Murcia, Spain\label{aff143}
\and
Instituto de F\'isica de Cantabria, Edificio Juan Jord\'a, Avenida de los Castros, 39005 Santander, Spain\label{aff144}
\and
Univ. Grenoble Alpes, CNRS, IPAG, 38000 Grenoble, France\label{aff145}
\and
CEA Saclay, DFR/IRFU, Service d'Astrophysique, Bat. 709, 91191 Gif-sur-Yvette, France\label{aff146}
\and
Department of Mathematics and Physics E. De Giorgi, University of Salento, Via per Arnesano, CP-I93, 73100, Lecce, Italy\label{aff147}
\and
INFN, Sezione di Lecce, Via per Arnesano, CP-193, 73100, Lecce, Italy\label{aff148}
\and
INAF-Sezione di Lecce, c/o Dipartimento Matematica e Fisica, Via per Arnesano, 73100, Lecce, Italy\label{aff149}
\and
Dipartimento di Fisica "E. Pancini", Universita degli Studi di Napoli Federico II, Via Cinthia 6, 80126, Napoli, Italy\label{aff150}
\and
INFN Sezione di Napoli, Universita degli Studi di Napoli "Federico II", Compl. Univ. di Monte S. Angelo, Edificio G, Via Cinthia, I-80126, Napoli, Italy\label{aff151}
\and
Department of Computer Science, Aalto University, PO Box 15400, Espoo, FI-00 076, Finland\label{aff152}
\and
School of Physics and Astronomy, University of Nottingham, University Park, Nottingham NG7 2RD, UK\label{aff153}
\and
DARK, Niels Bohr Institute, University of Copenhagen, Jagtvej 155, 2200 Copenhagen, Denmark\label{aff154}
\and
Department of Physics, Centre for Extragalactic Astronomy, Durham University, South Road, Durham, DH1 3LE, UK\label{aff155}
\and
Astrophysics Research Centre, University of KwaZulu-Natal, Westville Campus, Durban 4041, South Africa\label{aff156}
\and
School of Mathematics, Statistics \& Computer Science, University of KwaZulu-Natal, Westville Campus, Durban 4041, South Africa\label{aff157}
\and
Department of Physics and Astronomy, Vesilinnantie 5, University of Turku, 20014 Turku, Finland\label{aff158}
\and
Finnish Centre for Astronomy with ESO (FINCA), Quantum, Vesilinnantie 5, University of Turku, 20014 Turku, Finland\label{aff159}
\and
Serco for European Space Agency (ESA), Camino bajo del Castillo, s/n, Urbanizacion Villafranca del Castillo, Villanueva de la Ca\~nada, 28692 Madrid, Spain\label{aff160}
\and
Van Swinderen Institute, University of Groningen, Nijenborgh 3, 9747 AG Groningen, The Netherlands\label{aff161}
\and
Department of Physics and Astronomy, University of the Western Cape, Bellville, Cape Town, 7535, South Africa\label{aff162}
\and
Department of Physics and Helsinki Institute of Physics, Gustaf H\"allstr\"omin katu 2, University of Helsinki, 00014 Helsinki, Finland\label{aff163}
\and
Institut d'Astrophysique de Paris, UMR 7095, CNRS, and Sorbonne Universit\'e, 98 bis boulevard Arago, 75014 Paris, France\label{aff164}
\and
Ohio University, Physics \& Astronomy Department,1 Ohio University, Athens, OH 45701, USA\label{aff165}
\and
Institute for Theoretical Particle Physics and Cosmology (TTK), RWTH Aachen University, 52056 Aachen, Germany\label{aff166}
\and
INAF-Osservatorio Astrofisico di Arcetri, Largo E. Fermi 5, 50125, Firenze, Italy\label{aff167}
\and
Centro de Astrof\'{\i}sica da Universidade do Porto, Rua das Estrelas, 4150-762 Porto, Portugal\label{aff168}
\and
Dipartimento di Fisica, Universit\`a di Roma Tor Vergata, Via della Ricerca Scientifica 1, Roma, Italy\label{aff169}
\and
INFN, Sezione di Roma 2, Via della Ricerca Scientifica 1, Roma, Italy\label{aff170}
\and
HE Space for European Space Agency (ESA), Camino bajo del Castillo, s/n, Urbanizacion Villafranca del Castillo, Villanueva de la Ca\~nada, 28692 Madrid, Spain\label{aff171}
\and
Center for Frontier Science, Chiba University, 1-33 Yayoi-cho, Inage-ku, Chiba 263-8522, Japan\label{aff172}
\and
Department of Physics, Graduate School of Science, Chiba University, 1-33 Yayoi-Cho, Inage-Ku, Chiba 263-8522, Japan\label{aff173}
\and
Dipartimento di Fisica ``G. Occhialini", Universit\`a degli Studi di Milano Bicocca, Piazza della Scienza 3, 20126 Milano, Italy\label{aff174}
\and
School of Physical Sciences, The Open University, Milton Keynes, MK7 6AA, UK\label{aff175}
\and
Department of Astrophysics, University of Zurich, Winterthurerstrasse 190, 8057 Zurich, Switzerland\label{aff176}
\and
Space physics and astronomy research unit, University of Oulu, Pentti Kaiteran katu 1, FI-90014 Oulu, Finland\label{aff177}
\and
Department of Mathematics and Physics, Roma Tre University, Via della Vasca Navale 84, 00146 Rome, Italy\label{aff178}
\and
Theoretical astrophysics, Department of Physics and Astronomy, Uppsala University, Box 524, 751 20 Uppsala, Sweden\label{aff179}
\and
Mathematical Institute, University of Leiden, Einsteinweg 55, 2333 CA Leiden, The Netherlands\label{aff180}
\and
School of Physics \& Astronomy, University of Southampton, Highfield Campus, Southampton SO17 1BJ, UK\label{aff181}
\and
Institute of Astronomy, University of Cambridge, Madingley Road, Cambridge CB3 0HA, UK\label{aff182}
\and
Kobayashi-Maskawa Institute for the Origin of Particles and the Universe, Nagoya University, Chikusa-ku, Nagoya, 464-8602, Japan\label{aff183}
\and
Department of Astrophysical Sciences, Peyton Hall, Princeton University, Princeton, NJ 08544, USA\label{aff184}
\and
Fakult\"at f\"ur Physik, Universit\"at Bielefeld, Postfach 100131, 33501 Bielefeld, Germany\label{aff185}
\and
SRON Netherlands Institute for Space Research, Landleven 12, 9747 AD, Groningen, The Netherlands\label{aff186}
\and
Ruhr University Bochum, Faculty of Physics and Astronomy, Astronomical Institute (AIRUB), German Centre for Cosmological Lensing (GCCL), 44780 Bochum, Germany\label{aff187}
\and
Department of Physics and Astronomy, University of British Columbia, Vancouver, BC V6T 1Z1, Canada\label{aff188}}


\abstract{
 The European Space Agency's \Euclid mission has issued its first major data release (Data Release 1, DR1), providing a transformative view of the extragalactic sky through deep photometry and near-infrared slitless spectroscopy. This paper presents the Euclid DR1 spectroscopic data set and details the evolution of the spectroscopic processing function (SPE PF) pipeline, which is responsible for the automated analysis of 1-dimensional spectra acquired by the Near-Infrared Spectro-Photometer (NISP). The SPE PF pipeline workflow comprises spectroscopic classification, redshift determination, and the systematic measurement of spectral features. Emission line fluxes are derived using two independent methodologies: direct integration (DI) and Gaussian fitting (GF). For sources exhibiting broad $\ha$ profiles (velocity dispersion larger than $1500$ km s$^{-1}$), a dedicated `QSO mode' is implemented to robustly deconvolve nuclear and host galaxy components. The data set was validated by cross-matching \Euclid results with external spectroscopic catalogues, specifically  Dark Energy Spectroscopic Instrument (DESI) DR1, and through an extensive visual inspection campaign of approximately 4000 sources within the Euclid Deep Field North (EDF-N). For high-quality samples selected in the EDF-N, we find a 98.5\% agreement between \Euclid and DESI spectroscopic redshifts. This success rate increases to 98.9\% when incorporating pipeline probability and reliability estimators. Visual inspection enabled the definition of distinct quality tiers (uncertain, marginal, reliable, and secure), which are further validated by median-stacked spectra demonstrating consistent spectral features and robust line profiles across all tiers.  The Euclid DR1 spectroscopic data set provides reliable measurements for millions of extragalactic sources, establishing a fundamental baseline for studies of large-scale structure and galaxy evolution. It also allows the selection of high~purity samples of stars and quasars thanks to dedicated template fitting. Despite the inherent challenges of slitless spectroscopy, the pipeline performance and validated data quality ensure a high-precision resource for the scientific community. }
%
%
   \titlerunning{The spectroscopic data set}
   \authorrunning{Euclid Collaboration: V. Le Brun et al.: DR1}
   \maketitle
%

%
%
%
   
\section{\label{sc:Intro}Introduction}
One of the two main objectives of the \Euclid mission \citep{EuclidSkyOverview} is to measure galaxy clustering in the redshift interval $0.84<z<1.88$ (hereafter, the `target range'), using spectroscopic redshifts measured from the spectra provided by the red grism (RGS) of the Near Infrared Spectro-Photometer (NISP) instrument \citep{EuclidSkyNISP} in the framework of the Euclid Wide Survey (EWS). The spectra have a resolution of $R\simeq 500$, cover the wavelength range 1206--1892\,\si{\nm}, and redshifts in the target range are measured by identifying the \ha\ emission line. The quality (purity, completeness) of spectroscopic samples from the EWS is estimated using the Euclid Deep Survey (EDS) spectra, obtained with both RGS and the Blue grism (BGS), through repeated observations (25 visits with BGS and 15 with RGS at the end of the survey). NISP is a slitless spectrometer, which means that the spectra of all objects on the sky potentially leave a trace on the spectroscopic exposures. The conditions for a spectrum to be extracted are that a corresponding object be identified in the photometric catalogue constructed from the \Euclid imaging, and the spectrum fall at least partly on the detectors. In the framework of the first \Euclid data release (DR1;  \citealp{DR1cite}), this paper presents the quality of the redshifts, the spectroscopic measurements, and a classification of sources based on their spectroscopic properties only. The spectra are extracted for objects brighter than $\HE=23.5$ and observed over the area of the three Euclid Deep Fields (EDFs), as well as over 1900 square degrees of the EWS. This paper focuses only on the quality of the measurements, and not on the criteria to be used to build statistical samples for cosmological studies, which will be presented in a separate paper.

The paper is organised as follows. Section~\ref{sc:Pipeline} describes the methods of redshift calculation, spectral feature measurements, and spectroscopic classification. Section~\ref{Sect:classif_results} presents the results of the spectroscopic classification, with special attention to the spectroscopic templates used in the quasar category, which also include other types to allow the identification of specific objects like quiescent galaxies. In Sect.\,\ref{Sect:DeepField}, we describe the results obtained in the EDF North (EDF-N) field, which has the most complete spectral coverage among the three EDFs \citep{EuclidSkyOverview}. They include the results of a visual inspection campaign and provide a quality criterion independent of any external redshift catalogue, which could suffer various kinds of biases. Section~\ref{Sect:wide} presents the results obtained for the EWS spectra, in comparison not only to external catalogues, but also to the EDF-N redshifts, which is the methodology planned for the EWS  data set. This includes  validation of the emission line flux measurements, in comparison with external data.

\section{\label{sc:Pipeline}Pipeline description}
\begin{figure*}
    \centering
    \includegraphics[width=0.9\textwidth]{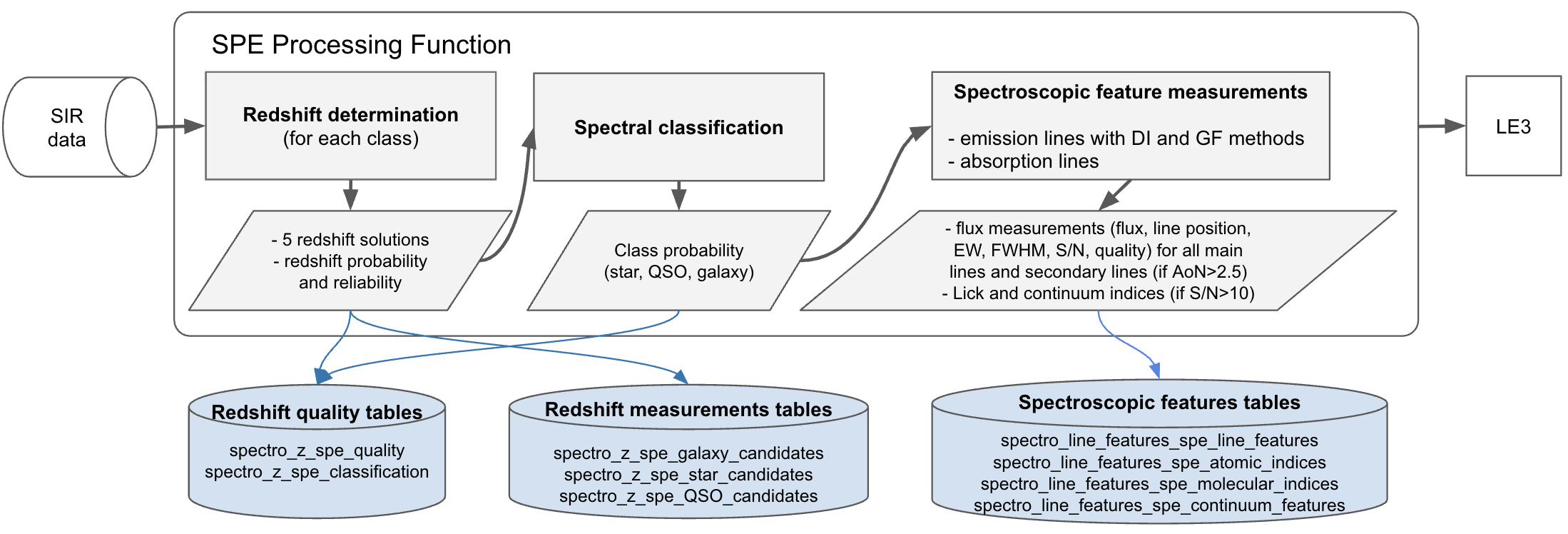}
    \caption{SPE Processing Function workflow. The grey blocks represent the different pipeline components, and the blue ones the tables produced.}
    \label{fig:SPE_workflow}
\end{figure*}

 The SPE processing function (SPE PF) comprises three main steps: (i)  redshift determination for different categories of objects,  (ii)  spectroscopic classification; and (iii)  spectral feature measurement. Here we present the main concepts and describe their updates with respect to Q1, especially the spectral feature measurements. The main workflow is shown in Fig.~\ref{fig:SPE_workflow}.

 The SPE PF ingests the combined spectra as produced by the SIR PF (in charge of the extraction, calibration and combination of individual spectra), which include the signal, its variance, and flags describing the quality of each pixel \citep{DR1-TP023}. SIR processing is run for all sources present in the catalogues produced by the MER PF \citep{DR1-TP009}, in batches that correspond to the sky-tiles defined there. The SPE PF then considers only data products that the SIR PF has declared valid. The SPE PF first excludes pixels flagged as {\sc `do not use'}, as well as the pixels with poor absolute flux calibration, mainly located at the extremities of the spectra, where the total transmission is close to zero. Isolated pixels are also discarded, since the analysis of Q1 results showed that they often induce errors in the solution, and at best do not improve the results. The calibrated flux for these pixels is the result of the division by two very small numbers, leading to large uncertainty. After this pixel selection, the redshift measurement is performed only on spectra containing more than 100 valid pixels. After processing, the SPE PF products are validated, again at the scale of MER tiles. We have invalidated the tiles in which no reliable measurement for galaxies was present, using the criterion ${\tt gal.spe\_z\_prob}>0.99$ in the `galaxy' category for reliability (see Sect.\,\ref{sc:redshifts} for a description of the parameters used here).

\subsection{\label{sc:redshifts}Redshift determination}

In order to properly identify the different populations of objects through their spectral properties, the pipeline can process each object in different configurations, each corresponding to a category of objects; there are currently the `galaxy', `qso' and `stars' categories. The redshift determination is based on fitting a model  over a discrete grid of redshifts. The limits and spacing of the grid can be different for each category, and the various elements of the models can also differ (single template, combination of continuum template and emission lines or only emission lines). For each value of the redshift grid, the parameters of the best spectroscopic model are derived by $\chi^2$ minimisation, and ultimately a redshift probability distribution function (PDF)  is built over the redshift grid, calculating a joint likelihood over all models \citep{Amazed_2026}. The peaks of the PDF are then detected by a simple local maximum detection, then fitted by a Gaussian function whose width gives the redshift precision, and ranked following their integral over a $\pm 3\,\sigma$ range. The first five redshift solutions are then provided. The value of the integral of the PDF under each peak (over a $\pm 3\,\sigma$ interval) is used to compute the {\tt RedshiftProba} parameter, which gives an indication of the relative significance of the different solutions. However, it does not give an absolute indication of the quality of the measurement.  A redshift prior can be applied to the PDF, in the form of a multiplicative factor over dedicated redshift intervals,  in order to penalise solutions that are unlikely or not targeted. This will then favour solutions that may be of lower quality, but that could be filtered out afterwards based on the {\tt RedshiftProba} or any other relevant parameter. The possible priors are the so-called \ha\ prior, which favours solutions for which the \ha\  line is observable, the `strong line prior', which acts the same way for the main strong lines that are likely to be observable (\oii, \oiii, \ha), or a prior on the redshift distribution based on the models described in \cite{Pozzetti2016}. All the tests made for  galaxy clustering showed that the \ha\ prior provided the best results, it is therefore applied to the `galaxy' category.

The models used for each spectral category are the following.

\begin{itemize}
    \item Galaxy: The redshift grid spans from 0.01 (to avoid overlap with the star model) to 5, with steps of $10^{-4}$ on a logarithmic scale of $(1+z)$. The values of the redshift are thus discretised over this grid. This model has been optimised for cosmology, which targets strong \ha emitters at redshifts between 0.84 and 1.88. Tests have been done with models including a physical continuum and emission lines, as well as models forcing the continuum to be zero where it is not significantly detected (below a given threshold that is a parameter of the pipeline); the best results are obtained when the continuum is filtered out from the data using an iterative kernel median filtering \citep{Garilli_2010}, and only emission lines are modelled. The detection of emission is based on templates of the relative strength of all emission lines from different models from the VVDS survey \citep{LeFevre2013}. As a consequence, no information from the continuum is used to derive the redshift, and the corresponding model only contains information on the emission lines. In addition, to improve the purity and completeness of the sample used for the galaxy clustering measurements, the `\ha\ prior' is used to build the list of redshift solutions. This allows building an initial sample as complete as possible, making sure that all galaxies showing strong \ha\ emission have the correct redshift, and reaching the highest purity by discarding low-quality measurements. However, a consequence of this procedure is that an emission line can be wrongly identified as \ha, leading to `line interlopers', examples of which can be seen in Fig.~\ref{fig:z_vs_z_purity_deep}. This model is not particularly  suitable for low-redshift ($z<0.84$) objects or for quiescent galaxies, whose spectra exhibit no (or faint) emission lines in the observed wavelength range. Given the constraints on the data model that was frozen at the time these tests were conducted, we have mitigated this weakness by extending the quasar model discussed below. This was not done at the time of Q1, which focused on the galaxy clustering results.
    \item Star: The redshift grid spans from $-0.01$ to $+0.01$ with linear steps of $10^{-3}$. The model is a pure template fit, with a stellar template library built from the ESO Infrared Spectra collection \citep{Pickles_1998}, and complemented by ultra-cool dwarfs (M, L, and T types) collected by the Primeval Universe Science Working Group \citep{Banados}, since these objects have photometric properties that mimic high-redshift quasars. Since there is no emission feature in the templates, the cross-correlation is not effective when the continuum is not significantly detected, which happens for objects fainter than $\HE=19$ in the EWS and in the Fornax Deep Field (EDF-F) and $\HE \simeq 19.5$ in EDF-N and Euclid Deep Field South (EDF-S). For objects fainter than those limits, there is no solution for the star category except in cases where strong residuals from the decontamination or other spatially extended artefacts produce spurious results. 
    \item QSO: The redshift grid spans from 0.01 to 8 with a logarithmic step of $10^{-4}$, and the model is based on a library of templates representative of the full spectrum, including continuum and emission lines. The initial goal was to use only two templates of broad emission-line quasars: the SDSS mean quasar template \citep{VandenBerk_2001}, extended past 1~$\mu$m by a simple power law, as for the Q1 release; and the mean quasar spectrum built by \cite{Fu2025} using 3000 QSO spectra observed by \Euclid. However, given the limitations of the so-called galaxy model described above, with respect to Q1 results, we have also added five templates of various types of galaxies, described in Sect.~\ref{sc:templates}. It must then be noted that the category name might be misleading, since some of the objects for which this is the preferred classification are not actually quasars.
\end{itemize}

\subsection{\label{sc:Classif}Spectroscopic classification}
The classification is based on the Bayesian model selection which compares the probabilities of the three model categories.
The object is assigned to the category having the highest probability. The probability of a model category is deduced from the Bayesian evidence or marginal likelihood of the model. It is the likelihood of the spectrum data integrated over the whole parameter space of the considered model category.
However, if we used the continuum of the `galaxy' model computed by median filtering instead of a template fit, the fit would obviously be much better than for the other categories, whose model is built on templates that may not reproduce exactly the full spectrum. We therefore used a specific tuning for the `galaxy model': the continuum is first fit using a continuum library based on the BC03 models \citep{BC03}, without emission lines, and the $\chi^2$ value is computed with respect to this continuum, in the same way as for the QSO and star models. The classification is then based on this specific calculation of $\chi^2$. After that calculation, we use the continuum estimated by median filtering to detect the emission lines in the way described in Sect.~\ref{sc:redshifts}. For the `galaxy' model, the continuum is estimated only at locations where no emission line is expected at the considered redshift; hence the amplitude of the emission lines has no impact on the $\chi^2$ value.  Conversely, for the two other models the full spectrum is used, and the relative weight of the continuum and the emission or absorption lines is fixed. A residual bias therefore remains in favour of the `galaxy' model, which does not allow us to give a fair estimation of the completeness within each category, some real quasars or stars being wrongly classified as galaxies. This does not prevent the redshift solution of the corresponding category from being correct, and the quality of the classification can be assessed using only  the spectrum or after adding additional  morphological information (see Sect.\ref{Sect:classif_results}).  

\subsection {\label{sc:LineFlux}Spectral feature measurements}

Following the SPE PF procedures described in \cite{Q1-TP007}, line fluxes are measured using both direct integration (hereafter DI) and Gaussian fit (GF) methods for each of the five estimated redshift solutions. 
The DI method evaluates a global spectral continuum with multi-scale smoothing, first suppressing the most significant line peaks with median filtering and then estimating the continuum with a boxcar filter. To further reduce the impact of potentially spurious artefacts and of the strongest emission lines, before continuum estimation, the spectrum is smoothed with a wavelet transform and the 10 strongest peaks are removed; this procedure is then repeated on the resulting spectrum to remove any additional peaks that emerge after the first pass.\footnote{During the analysis of DR1 data, we found out that this procedure was too aggressive, and the corresponding continuum was lower than the expected one, resulting in fluxes about 15\% larger in DI compared to the GF ones. For DR2, this procedure will be removed, having verified that this way the measurements from the 2 methods are fully compatible.}
The flux of each line is then measured on the continuum-subtracted spectrum, integrating symmetrically in both directions from the line centre provided by the redshift solution until the continuum level is reached.
The GF method fits a Gaussian model to the lines, considering as free parameters the line position ($\lambda_0$), width ($\sigma$), and amplitude ($I_0$), together with a constant continuum value ($c$). The line position is allowed to vary within $\pm\sigma_z/(1+z)=10^{-3}$ (the nominal accuracy on the redshift), and the continuum is estimated over a 30~nm  window centred on the line (roughly 20 pixels). The values of the window and the choice of a constant continuum have been validated using spectroscopic simulations, which show that at the expected signal-to-noise ratio (S/N) of \Euclid data, the best results are obtained with these parameter choices.
The integrated flux of the line is evaluated analytically as $F=\sqrt{2\,\pi}\,I_0\,\sigma$, and its full width at half maximum as $\text{FWHM}=2\sqrt{2\ln2}\,\sigma$. 

For line complexes expected to be blended at the \Euclid resolution (e.g., H\,$\alpha$ and the \nii~doublet), or lines whose flux ratios are fixed by atomic physics (e.g., the \oiii~or \sii\ doublets), the models are obtained as a combination of multiple Gaussians, and the lines are fitted together. In this case, we assume the widths of the lines of the complex to be the same, and the relative line positions to be fixed by their rest-frame wavelengths, with only the global redshift left free. In the case of doublets, the measured flux is the total flux of the doublet; to obtain the measurement for each component, it has to be multiplied by the corresponding ratios fixed by atomic physics (1/3:2/3 for \oiii~and \nii, and 1/2:1/2 for \sii). We note here that in this case, the GF method will provide the total flux of the doublets: $F([\ion{O}{iii}])_{\rm{GF}}=F([\ion{O}{iii}]\lambda4959)_{\rm{GF}}+F([\ion{O}{iii}]\lambda5007)_{\rm{GF}}$.
The main difference between DI and GF measurements is seen for the blended \ha+\nii\ complex: the DI will provide the total flux of the complex, while GF will provide the deconvolved flux of \ha and the \nii\ doublet separately.

In the EWS, only the RGS spectra are provided, while in the EDS both the RGS and BGS spectra can be available. In the second case, spectral feature measurements are performed independently on each grism data set, and a best combined measurement is also provided, where fluxes are averaged for the lines present in the overlapping spectral region when both grisms yield valid measurements.

For each method, we provide a measurement of the integrated flux, line position, equivalent width, and value of the continuum and average number of dithers under the line, each with its associated uncertainty. For GF, we also provide an estimate of the width (FWHM) of the line. In addition to these quantities, we provide, for each method, several estimators to quantify the quality of the line measurement.  The S/N is defined as the ratio between the integrated line flux within $\pm2\,\sigma$ of the line centre and the mean integrated noise estimated over the same area from the error spectrum around the line. In addition, a quality flag is associated with each measurement, as described in Table~\ref{tab:qual_feats}.

Line measurements are provided for all spectra for lines considered as `main' (Ly\,$\alpha$, \oii, H\,$\beta$, \oiii, H\,$\alpha$, \nii$\lambda6585$, \sii$\lambda6718$, \siii$\lambda9068$, \siii$\lambda9530$, Pa\,$\gamma$, and Pa\,$\beta$), while for all other secondary lines a measurement is stored only if the amplitude-over-noise (AoN, defined as the peak amplitude of the line divided by the local RMS of the spectrum) is larger than 2.5. 

In the case of very broad H\,$\alpha$ lines, identified with an internal velocity dispersion larger than $1500$ km\,s$^{-1}$ (computed by convolving the instrumental width with a Gaussian profile corresponding to the velocity dispersion during the fit), the pipeline also has a \texttt{QSO mode}.
In this case, we adopt a simplified modelling approach for the H\,$\alpha$+\nii$\lambda\lambda 6548, 6584$ complex, excluding a separate \nii~doublet from the model to avoid degeneracies between the narrow \nii~lines and the wings of a broad H\,$\alpha$ component. Instead, we use a dual-Gaussian configuration for \ha\ consisting of a narrow component (associated with the host galaxy or the narrow line region) and a broad component (tracing the broad line region, BLR), and to reach beyond the line wings, we extended the width of the fitted wavelength range to 80 nm. This choice ensures a more robust estimation of the spectral features of the broad-line component. 
Extensive tests were performed to optimise this part of the code, verifying that with the current settings a broad-line component is correctly recovered.

Finally, the pipeline is also designed to measure absorption lines, namely Lick indices and continuum breaks (like the $D4000$), following the same procedure also applied in \cite{Q1-Mao}. Standard Lick indices and continuum feature definitions have been updated to vacuum wavelengths. These features are measured only in spectra where the S/N of the continuum estimated in the window [1350--1750]\,nm\ is larger than~10.

\subsection{\label{sc:templates}Templates used in the QSO category}

As stated before, the `galaxy' category is tuned for the detection of the \ha\ emission line, in order to optimise the cosmological constraints, which are the primary goals of \Euclid. It is therefore not suitable for all types of galaxy, like quiescent ones, or galaxies at $z<0.85$, where no prominent emission lines are present in the NISP observing range except for star-forming galaxies. In order to improve the value of the spectroscopy-based catalogs for galaxy-evolution studies without changing the data model (which was frozen early in the operational timeline for DR1), we have added to the templates of the `QSO' category several templates for otherwise lacking galaxy types.

We utilised the  \texttt{SpectraPyle} code introduced in \citet{EP-Quai} for the spectral stacking procedure, using \Euclid spectra from Q1 with available external spectroscopic redshifts, for different classes of galaxy. 
We used the quiescent galaxies (QGs) and star-forming galaxies (SFGs) at $z_{\rm{spec,ext}}<1.5$, as classified by Mao et al. (2026, submitted) based on rest-frame $UVJ$ colours estimated at $z_{\rm{spec,ext}}$. The SFG population is further divided into blue and red SFGs, with $U-V<1.27$ and $U-V>1.27$, respectively. The samples thus obtained contain \num{12190} QGs, \num{21641} blue SFGs, and \num{7810} red SFGs.
In addition, we used the sample of type 2 AGN classified using DESI spectra (Lopez Lopez et al. 2026, submitted). 

For each class, spectra were normalised in the common rest-frame wavelength range before stacking, removing strong line features where necessary, and were resampled to $0.7$\,nm per pixel. Sigma clipping was applied at $4\,\sigma$, with removal of bad pixels. These samples, covering a rather wide redshift range ($0<z<1.5$), allowed us to reconstruct global templates from rest-frame optical to near-IR. We further extended the templates to shorter wavelengths using population synthesis models \citep{Vazdekis2016}. 

\begin{figure*}
    \centering
    \includegraphics[width=0.95\textwidth]{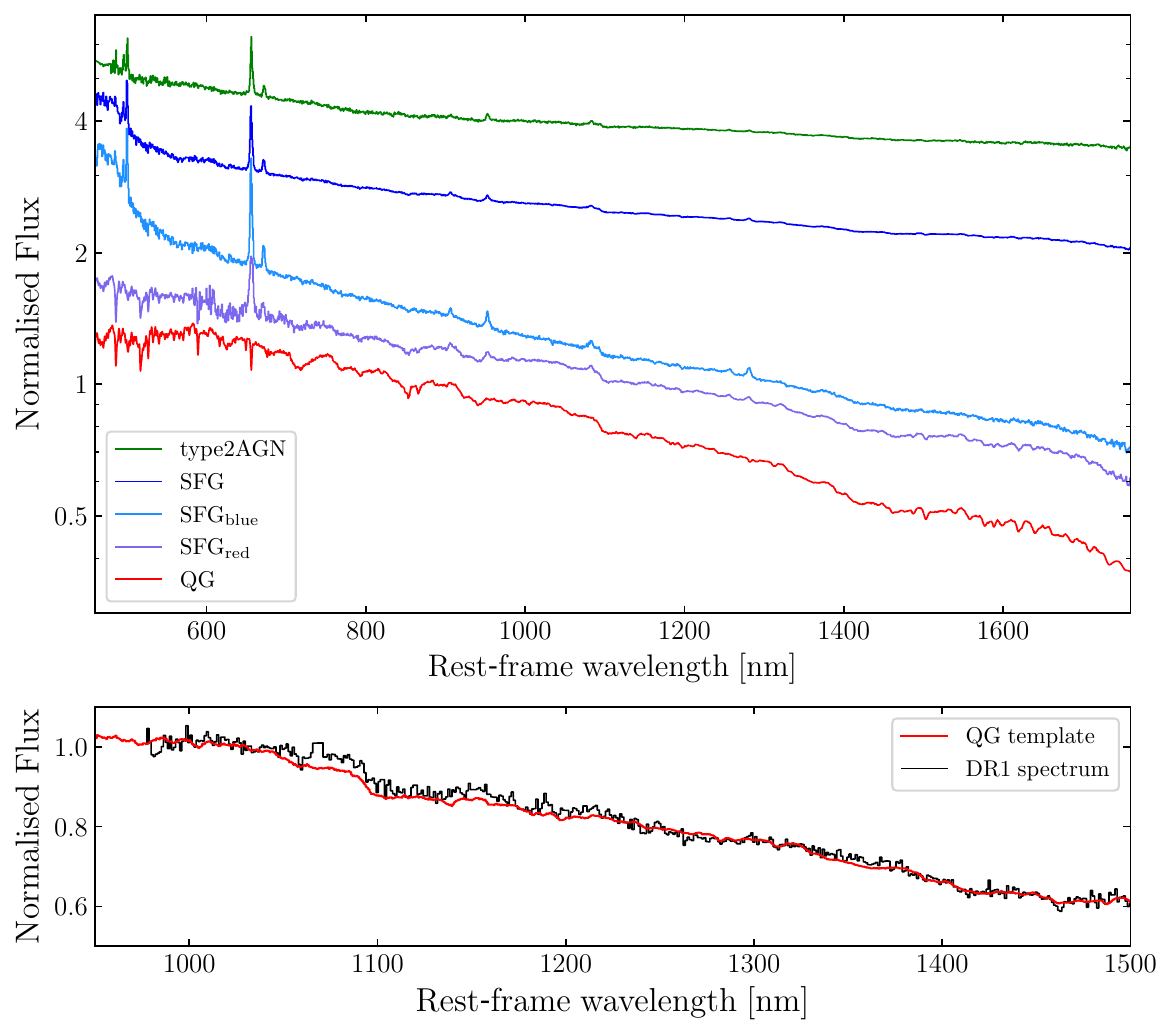}
    \caption{The top panel shows templates of five populations. Type 2 AGNs are classified using DESI spectra, QGs and SFGs are classified using the rest-frame $UVJ$ diagram, and red and blue SFGs are separated by $U-V=1.27$. The bottom panel shows a \Euclid DR1 spectrum with $z=0.232$ obtained from QG template fitting. The object has a magnitude $\HE=15.5$.}
    \label{fig:templates}
\end{figure*}

Figure\,\ref{fig:templates} shows the final templates that we obtained. These are used in the SPE PF for spectral continuum and feature identification. In Fig.\,\ref{fig:templates} we also show an example of a quiescent galaxy identified by the pipeline at $z=0.232$.

\subsection{Description of the data products}
\label{sec:datadescription}
In the Euclid Science Archive,\footnote{\url{ https://www.cosmos.esa.int/web/euclid/data-release-dr1}} the SPE PF products are stored in 12 tables for each of the four survey configurations. The table names include a {\tt <SURVEY>} suffix to identify these configurations. The {\tt deep\_survey} string refers to the EDS, with measurements performed simultaneously on the RGS and BGS stacked spectra; the {\tt wide\_survey} extension applies solely to the EWS data. The catalogues with the {\tt deep\_visits} suffix contain the results on the individual visits made to the EDFs with RGS only, namely eight observations in EDF-N (seven being complete), one in the EDF-F, and ten in EDF-S. In these catalogues a given object can appear as many times as the number of individual visits, with slightly different identifiers, since those are built by the MER PF from the measured position on each individual image. The {\tt patch\_id\_list} keyword relative to each visit enables the retrieval of data for each of them. The last type of catalogue is {\tt wide} and contains all the single-visit observations, hence including EWS, individual visits to EDFs, and some calibration fields.  Among the 12 tables, two contain information relative to individual spectra: 
\begin{itemize}
    \item {\tt spectro\_z\_spe\_quality} gives information about the quality of the redshift measurement process, such as the number of pixels used, the wavelength range, the number of individual dithers, and any errors that occurred during the processing;
    \item  {\tt spectro\_z\_spe\_classification} contains the results of the classification process, namely the probabilities associated with each class and the resulting output. 
\end{itemize}

Three other tables contain the results of the redshift measurements for each of the models, named {\tt spectro\_z\_spe\_<CLASS>\_candidates\_<SURVEY>}, where {\tt <CLASS>} has one of the three possible values, {\tt galaxy, star} or {\tt qso}. There can be up to five solutions per object in each table, identified by the {\tt spe\_rank} parameter, with the 0 value being the most likely solution. These tables also contain the computations from the model fit, as described in the SPE data model.\footnote{\url{http://st-dm.pages.euclid-sgs.uk/data-product-doc/dmdr1/spedpd/dpcards/spe_spepfoutputcatalog.html}} The next three tables ({\tt spectro\_model\_z\_spe\_<CLASS>\_models\_<SURVEY>}) have been filled in the same way as the previous ones, and contain, for each of the three classes, the model parameters that were used to derive each redshift solution for each object. 

Four additional tables contain the spectral feature measurements performed on the spectra, namely emission-line fluxes and corresponding quantities: continuum features, and atomic and molecular indices (see Sect.\ref{sc:LineFlux} for an extended discussion). For each object, there are as many lines in each table as the number of measured lines for each of the redshift solutions provided. This number can be variable since for secondary lines, measurements are computed only for the most reliable ones (AoN $>2.5$). In each table, the measured feature is identified by {\tt SPE\_<MEAS TYPE>\_NAME} (where {\tt <MEAS TYPE>} is either {\tt LINE},  {\tt AT\_IDX},  {\tt MOL\_IDX}, or {\tt CNT\_FEAT}) and the corresponding redshift rank ({SPE\_RANK=0} for the best redshift solution). 

The large number of tables present in the archive comes from the direct copy of the files produced by the pipeline into the Science Archive. This will be improved in the next releases; in the meantime, a sample query that allows one to retrieve the most common results is given in Appendix~\ref{Sec:ADQL}.

\section{\label{Sect:classif_results}Classification results}

The SPE PF pipeline for DR1 classifies \Euclid spectra into three categories, namely   `galaxies', `stars', or `quasars'. In addition to giving the most probable class, a probability is attributed to each one.

The main caveats of the classification, already described in Sects.\ref{sc:Classif} and \ref{sc:templates}, can be summarised as follows. The `galaxy' category is meant to gather galaxies with emission-line spectra, with a cap on line width, which is optimal for building the cosmological sample (the complete description of this sample is described in  Euclid Collaboration: Markovic et al. in prep.). For technical reasons, the 'QSO' category of DR1 collects not only quasars, but also other types of galaxies whose spectra display a strong continuum component, in particular QGs and type II AGNs. Because of details of the $\chi^2$ calculations used for the classification procedure, there may be biases in its results. In the rest of this section, including Table~\ref{tab:classification}, we call galaxies only the objects fit using the `galaxy' class after continuum subtraction, and ignore the galaxies fitted by galaxy templates stored in the QSO class. We call QSOs only those objects that are best-fitted by one of the two quasar templates: \texttt{optical\_nir\_qso\_template\_v1.txt}; and \texttt{ComposantesPCA4\_meanqso.txt}. These names are the values of the {\tt spe\_cnt\_tpl\_id} column in the {\tt catalogue.spectro\_model\_spe\_qso\_models\_<SURVEY>} table that allow users to actually identify quasars. 

To validate our classification, we use the DESI survey \citep{Adame2024} as a reference. We match the DESI catalogue with the SPE PF EWS and EDS catalogues using a $0 \arcsecf 2$  radius, a cut based on the distance histogram to avoid random associations. As the DESI survey mainly focused on emission line galaxies and quasars, it is expected that the classification from the two surveys match. In Table\,\ref{tab:classification}, we present the fraction of objects whose classification agrees with DESI. We compute it both using the most probable class for all objects (third column) and restricting our analysis to objects with one class having a probability higher than 0.99 (fourth column). Since most objects have a high probability for a single class, this has a small impact on the results. To ensure that only reliable spectra are used in our analysis, we set minimum numbers of valid pixels and of  median number of dithers: 300 and 3 for the EWS, respectively, and 800 and 12 for the EDS.

In the EWS, the SPE PF galaxy sample is highly contaminated by stars, with only 64\,\% of objects also classified by DESI as galaxies. Since galaxies are expected to be extended objects compared to stars and QSOs, which are point-like, we matched our sample with the MER point-like probability based on the \Euclid visible imaging. If we consider only objects with a point-like probability lower than 0.5, the fraction of objects in agreement with DESI jumps to 98\,\%. In the EDS, the fraction of agreement with DESI is 79\,\% using only spectroscopy, and 98\,\% when adding the point-like criterion.

Objects classified as stars by the SPE PF are, with an excellent rate of agreement (99\% in both EDS and EWS), also labelled stars in DESI. If we assume that the DESI sample is reasonably representative, we can thus conclude that the star sample has a high purity. However, a large fraction of DESI stars are classified as galaxies (35\,082 compared to 21\,997 stars classified as stars), and the completeness is thus limited. These spurious galaxies could be caused by random lines produced by the noise in continuum-subtracted stellar spectra. Good performance in terms of purity is also obtained for QSOs, with 88\,\% agreement in the EWS, and 95\,\% in the EDS. However, no more than 5\,\% of DESI quasars are similarly classified by SPE, and most of the QSOs leak into the galaxy class.

\begin{table}
\caption{\label{tab:classification} Performance of the SPE PF spectral classification, estimated using DESI as a reference. See text of Sect.\ref{Sect:classif_results} for the definitions of the object types used in this table}
\begin{tabular}{llcc}
\hline
\hline
\noalign{\vskip 2pt}
Main SPE PF & Additional & \multicolumn{2}{c}{Fraction agreeing with DESI} \\
type & criterion &  $P_{\rm type}^{\rm SPE}$ & $P_{\rm type}^{\rm SPE}>0.99$\\  
\hline
\noalign{\vskip 2pt}
\multicolumn{4}{c}{EWS} \\
\hline
\noalign{\vskip 2pt}
Galaxies & None & 0.646 & 0.647\\
Galaxies & Extended & 0.982 & 0.982\\
Stars & None & 0.996 & 0.996\\
Stars & Point-like & 0.997 & 0.997\\
QSOs & None & 0.882 & 0.880\\
QSOs & Point-like & 0.881 & 0.878\\
\hline
\noalign{\vskip 2pt}
\multicolumn{4}{c}{EDS} \\
\hline
\noalign{\vskip 2pt}
Galaxies & None & 0.786 & 0.787\\
Galaxies & Extended & 0.984 & 0.984\\
Stars & None & 0.986 & 0.986\\
Stars & Point-like & 0.997 & 0.997\\
QSOs & None & 0.955 & 0.954\\
QSOs & Point-like & 0.951 & 0.951\\
\hline
\end{tabular}

\end{table}

\section{\label{Sect:DeepField}Redshift measurements in Euclid Deep Field North}

For the Q1 data release, only a single RGS visit of the EDFs was available \citep{Q1-TP001}. For DR1, the EDF-N benefits from seven complete RGS and three BGS visits. In this section, we characterise the performance on redshift measurement of the EDS data by combining all the DR1 data in the EDF-N using a comparison with DESI (Sect.\,\ref{sc:deep_comp_desi}). We also performed a campaign of visual inspection focused only on the redshift and flux range used for cosmology in the EWS ($0.85<z<1.85$, \ha\,flux larger than $1.5\times  10^{-16}$~\ecgs, Sect.\,\ref{sc:visual_deep}).

\subsection{\label{sc:deep_comp_desi} Comparison with DESI}

The combination of depth and wavelength coverage in the EDF-N corresponds to an uncharted parameter space. Thus, it is hard to fully validate the performance using ancillary data. However, the DESI survey \citep{Adame2024} provides  us with a reliable reference catalogue up to $z=1.6$ \citep{Q1-TP007}.

For this comparison, we focus on a sample of high-quality spectra classified as `galaxy', with a high number of valid pixels {\tt spe\_npix}$>$800 and a median of at least 12 dithers (${\tt spe\_n\_dith\_med}\ge 12$). Similarly to \cite{Q1-TP007}, we also apply a criterion on the velocity dispersion to reduce the number of false systems ($<1500$\,km\,s$^{-1}$), and restrict our analysis to spatially extended objects (probability of being point-like from MER lower than 0.05) to avoid contamination by stars. We then match our catalogue with DESI using a maximal search radius of $0 \arcsecf 2$.

\begin{figure*}
\centering
\begin{tabular}{cc}
\includegraphics{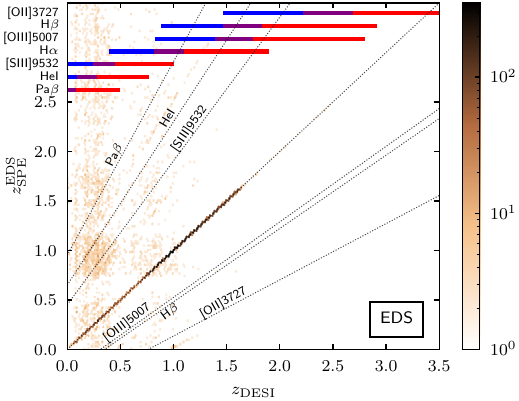} & \includegraphics{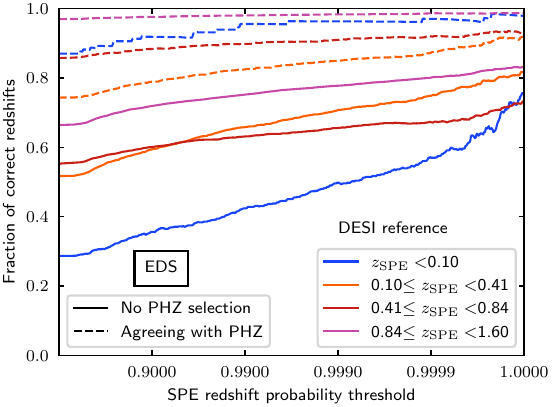}\\
\end{tabular}
\caption{\label{fig:z_vs_z_purity_deep} \textit{Left}: Comparison between the SPE PF high-quality EDS galaxy redshifts in the EDF-N and their DESI counterparts. Only redshifts selected for their high quality (see Sect.\,\ref{sc:deep_comp_desi}) are displayed in this figure. The colour coding indicates the density on a logarithmic scale. The dotted lines show the 1:1 relation and expected position of the main lines that could be misidentified as \ha. The top of the panel contains the range covered by the BGS and RGS for each line (overlapping region in purple). \textit{Right}: Fraction of correct EDS redshifts in the EDF-N as a function of the SPE PF redshift probability cut in various redshift slices and estimated using DESI redshifts as a reference: $z<0.1$ (blue); $0.1 \le z<0.41$ (orange); $0.41 \le z<0.84$ (red); and $0.84 \le z<1.6$ (purple). The solid lines correspond to all the high-quality galaxy spectra. The dashed lines correspond to a high-purity sample used as a reference for the EWS for which the redshift from SPE PF and PHZ agree (see Sect.\,\ref{sc:specphot}).}
\end{figure*}

In Fig.\,\ref{fig:z_vs_z_purity_deep}, we compare the EDS redshifts with DESI. To avoid crowding the figures with incorrect redshifts, we restricted our analysis to galaxies with a good SPE PF probability (${\tt gal.spe\_z\_prob}>0.99$). A large fraction of the objects are on the one-to-one relation, where DESI and SPE PF redshifts agree. The precision obtained for these specific objects is better than that obtained in Q1 ($\sigma_{\Delta z / (1+z)} = 0.0003$ in DR1 EDFs versus 0.0008 in Q1), which is discussed in Appendix\,\ref{sc:deep_precision}.

We observe catastrophic outlier redshifts distributed all over the plane. As explained in \cite{Q1-TP007}, these correspond to noise interlopers \citep{Granett26,EP-Quai}, which are wrong redshifts caused by a noise peak interpreted mainly as \ha. We would also have expected the presence of line interlopers (also discussed in the previous references), corresponding to one line misinterpreted as another. These interlopers are expected to form diagonal features, along the dotted lines of Fig.\ref{fig:z_vs_z_purity_deep}. The Q1 data contained such features, but we do not identify any of them in the DR1 EDS data. This shows how the addition of the BGS and the deeper data avoid redshifts based on a single line, and reduce the presence of line interlopers dramatically.

The fraction of correct redshifts depends on the measured redshift. We chose to use the following limits for our redshift bins: 0.1 (arbitrary value to select very low redshifts), 0.41 (\ha entering the BGS), 0.84 (\ha\ entering the RGS), and 1.6 (approximate upper limit of the DESI galaxy sample). We then varied the SPE PF probability threshold, and, similarly to \cite{Q1-TP007}, we considered a redshift to be correct if $| z_{\rm SPE} - z_{\rm DESI} | < 0.003 $. The results are shown in the right-hand panel of Fig.\,\ref{fig:z_vs_z_purity_deep} (solid lines). 

As expected, the fraction of correct redshifts increases with {\tt spe\_z\_prob}. However, even for high probability values, the fraction of correct redshifts does not reach values above 80\%;  while the requirement is above 99\% in the redshift range used for galaxy clustering measurements. This result is most likely due to the incomplete observing time for the EDF-N. The performance is best in the $0.84 \le z<1.6$ range (75\,\% of correct redshifts for ${\tt gal.spe\_z\_prob} \ge 0.99$). This is expected, since the \ha\ line is then detected in the RGS, where the sensitivity is better. There is no significant change of performance in the two redshift bins above and below 0.41, indicating that the presence of \ha\ in the BGS is not essential, since other lines (such as \siii), can be used to derive the redshift. In contrast, the fraction of correct redshifts remains relatively low at $z<0.1$ (42\,\% of correct redshifts for an SPE PF probability higher than 0.99). This could be caused by either the fact that low-$z$ galaxies tend to have large angular sizes, thus diluting the signal, or the lack of bright lines in this redshift range, or both.

\subsection{\label{sc:specphot}Combination of spectroscopic and photometric redshifts}

Finally, we combined the SPE PF information in the EDF-N with the photometric redshifts from PF-PHZ \citep{DR1-TP011} to build a reliable reference spectroscopic sample over a wide range of redshifts. We select an object only if its SPE PF redshift is within the 95\,\% confidence interval of PHZ. With this selection, the fraction of correct redshifts is always higher than 80\,\% for an SPE PF probability threshold of 0.99, and reaches 98\,\% for the $0.84<z<1.6$ range, where \ha\ is detected in the RGS. It highlights how powerful the combination of spectroscopic and photometric information is. In Sect.\,\ref{Sect:wide}, we use this reference sample to measure the performance of the redshift measurement in single EWS-like visits in the EDF-N to estimate EWS performance, and in Appendix\,\ref{sc:redshift_distibutions}, we discuss the redshift distributions of the high-quality spectra with and without using PHZ for the selection.

\subsection{\label{sc:visual_deep}Redshift measurement quality visual inspection}

The only external reference sample that can be used on EDS is DESI, which suffers some biases regarding the redshift range and the galaxy population, as the spectroscopic sample is built after selection on dedicated photometric quantities, while EDS is  limited only by magnitude and geometrical constraints. It was therefore necessary to evaluate the performance of the redshift measurement in an independent way, and the only possible one is a human inspection and validation of the results. 
The completeness and quality of DR1 observations are very dependent on the field: the EDF-F has been covered by only one visit in each grism and EDF-S by 10 RGS visits, while the EDF-N has received seven complete RGS and three BGS visits. The latter is then the only field that could currently  be used as reference to evaluate the purity and completeness of the EWS. We therefore have decided to conduct a visual inspection and redshift validation campaign on this field. This exercise explicitly focuses on the cosmology target sample ($0.85\le z \le 1.85$ and $F(\ha) > 2\times 10^{-16}\ecgs$), for which the EDF sample is intended to be used as a reference, but can be extended to other redshifts by using similar criteria for emission lines other than \ha.  

The first step was to establish criteria on spectrum quality, object properties, and spectroscopic measurements identical to the criteria used by the galaxy clustering SWG and check they produce the highest agreement with DESI DR1.    
The criteria used are ${\tt spe\_n\_dith\_med} \ge 15$ and  ${\tt spe\_npix} \ge 800$ to ensure complete spectra built from a sufficient number of individual visits. We also restricted the study to objects satisfying ${\tt mer.point\_like\_prob} < 0.05$, ${\tt mer.spurious\_prob} < 0.5$, and ${\tt mer.kron\_radius} < 50$. These criteria are those used by the Galaxy Clustering Science Working Group to exclude point-like objects (mainly stars), large objects (for which the redshift is unlikely to be actually greater than 0.85), and also exclude objects whose reality is uncertain.
We selected objects with a measured redshift between 0.85 and 1.85 to ensure that the \ha~line is detected in the RGS spectrum, and with an \ha~flux greater than $1.5\times  10^{-16}$~\ecgs to ensure completeness at the level of $2\times  10^{-16}$~\ecgs when taking the uncertainties in the measurement into account. We discarded low-quality flux measurements using the ${\tt spe\_line\_snr\_gf}>0$ criterion and selected measurements made on either the RGS or combined between the two grisms. 
With this selection, based on the spectral quality and the target redshift range, there is 98.5\% agreement between the \Euclid and DESI redshifts over 3985 objects.  To further improve the results, the $\tt{gal.spe\_z\_prob}$ and $\tt{gal.spe\_z\_rel}$ parameters can be used. The latter is a probability calculated by an ML algorithm trained on the PDFs of objects from which the redshift is known. This latter quantity is intended to be applied to EWS objects using the EDS redshifts as references for the training, and the current values have been calculated using simulated data from the EL-COSMOS \citep{Saito2020} field. The results on DR1 data are therefore not optimal, since the training data need to be representative of the data to which the model is applied; however, excluding small values for $\tt{gal.spe\_z\_rel}$ allows the exclusion of a small fraction of wrong redshifts without losing the correct ones. Combining the criteria $\tt{gal.spe\_z\_prob} > 0.94 $ and 
$\tt{gal.spe\_z\_rel} > 0.05 $ gives a sample of 3863 objects with a 98.9\% success rate compared to DESI. The difference with the numbers obtained in Sect.\,\ref{sc:deep_comp_desi} comes from the criterion added on the \ha\ flux, which leads to better measurements of redshift. 

We then applied these criteria to the full EDF-N catalogue, resulting in a sample of 76\,002 objects. In order to allow visual inspection and evaluation of the redshift measurements in a reasonable amount of time, we first randomly selected a subsample of 4000 objects. This list was made available for visual inspection to a set of volunteers, regardless of past experience in redshift measurement. The tool developed within the Galaxy Evolution Science Working Group \citep{EP-Quai} randomly assigns objects to each participant, and allows for multiple assignments for each object to take human biases into account. The process was intended to last for 6 weeks, and was extended to 8. At the end of the 6th week, the random assignment algorithm was modified in order to make sure that all objects would be inspected at least once and to increase the fraction of objects inspected at least twice. At the end of the process, all objects had been inspected, and 2044 (51\%) had been inspected three times or more. The participants were asked to visualise the spectra together with the redshift solutions provided by the PF SPE, and assign a quality index between 1 and 4 to the redshift solution, with the following meaning.
\begin{itemize}
    \item 1 indicates that the assigned redshift could not be confirmed, and that no reliable redshift could actually be assigned given the quality of the spectrum;
    \item 2 indicates evidence that an emission line is present in the spectrum, but no other line could be used to get a reliable value of the redshift. We noticed that, depending on the expertise and confidence of the participant,  spectra showing an asymmetric profile, indicating the presence of the \ha-\nii{} complex, could enter into  this category
    \item 3 indicates that more than one emission line is detected, leading to a reliable redshift value, but that either poor S/N for one of them, or other features in the spectrum can induce some doubt. An asymmetric profile, indicating the \ha-\nii{} complex, often falls into this category
    \item 4 indicates an unambiguous value of the redshift. 
\end{itemize}

\begin{table}
\centering
\caption{\label{tab:deep_quality} Visually assigned quality of redshifts in EDF-N.}
\begin{tabular}{lrr}
\hline 
\hline
\noalign{\vskip 2pt}
Quality   & Total sample & $\ge 3$ inspections\\
   & (4000 objects) & (2044 objects)\\
\hline
\noalign{\vskip 2pt}
Uncertain & 51 (1.3\%)& 37 (2\%) \\
Marginal  & 695 (17.4\%) & 376  (18\%) \\
Reliable & 2036 (50.9\%) & 1053 (52\%) \\
Secure & 1218 (30\%) & 578 (28\%) \\
\hline
\end{tabular}
\label{tab:deep_flags}
\end{table}

\begin{figure*}
    \includegraphics[width=\textwidth]{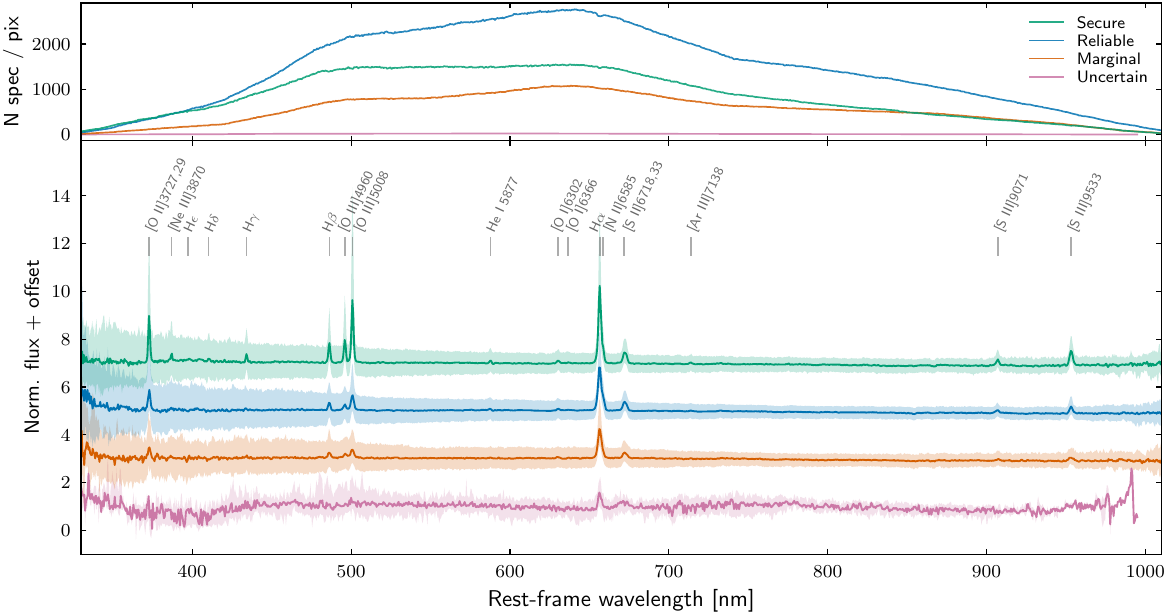}
    \caption{Median-composite spectra of the 4000 galaxies in the visual-inspection sample, grouped by the median redshift-quality flag assigned during the visual-inspection campaign (Sect.~\ref{sc:visual_deep}). The shaded regions represent the uncertainties on the median-composite spectra, estimated through bootstrap resampling with replacement. The upper panel shows the number of spectral pixels contributing to the composite in each wavelength bin, including the BGS  and RGS.\label{fig:stacked_spectra_final}}
\end{figure*}

To formalise these results, the individual grades are combined into a mean grade $\langle g \rangle$, on which we define four quality categories:
\begin{itemize}
\item \emph{uncertain}  ($\langle g \rangle < 2.5$); a majority of inspectors considered the redshift as uncertain (even with the presence of a significant emission line);
    \item \emph{marginal} ($2.5 \leq \langle g \rangle < 3$); at least
    one inspector considered the redshift likely correct;
    \item \emph{reliable} ($3 \leq \langle g \rangle < 3.5$); inspectors
    agreed on average that the redshift is likely correct, with at least one
    clear spectral feature identified;
    \item \emph{secure} ($\langle g \rangle \geq 3.5$); strong consensus
    that the redshift is correct, with multiple unambiguous emission lines
    detected.
\end{itemize}

In order to take into account the individual biases of inspectors, and to mitigate extreme optimism or conservatism, we used the median value of the grade for the objects that were rated three times or more. For some specific objects, 
that showed a unique significant emission line but whose profile asymmetry strongly favours the identification as \ha\ distorted by the neighbouring \nii{} doublet, they may have been assigned a grade 2 by reviewers who tended to give a conservative validation.
The partition in raw numbers and fractions of the four categories is shown in Table~\ref {tab:deep_quality}. As can be seen, the distribution is consistent whether objects have been graded at least three times or not, and one can ultimately consider that about 80\% of the redshifts are correct in the EDF-N cosmological sample, as long as an emission line is detected with a flux larger than $1.5\times 10^{-16}$\ecgs\ in the RGS spectrum. Indeed, for other strong emission lines, the detection of \oiii\ would imply the presence of the second line of the doublet and possibly \hb , rendering the confusion with \ha\ or \oii\ quite rare given the quality of the spectra. The main confusion that could arise is an \oii\ line being identified as \ha\ at redshift greater than 2.6, when \oiii\ is not detected in the RGS spectrum. However, our validation shows that the asymmetry of the \ha-\nii{} complex is most of the time correctly identified, and the presence of \ha\ in the RGS at wavelengths above 1300\,nm would most likely be linked to \oiii\ in the BGS spectrum, reducing the risk of mistake. The fact that the performance is lower than the initial value obtained in comparison with DESI is likely due to the bias of the selection in DESI, targeting brighter and more star-forming galaxies than \Euclid.

To illustrate the visual-inspection classification, we used \texttt{SpectraPyle} \citep{EP-Quai} to construct median-stacked spectra for the four quality categories defined above. The BGS and RGS spectra were treated as separate spectral entries throughout the stacking procedure, rather than being combined on a source-by-source basis. Thus, when both grisms are available for a given galaxy, the corresponding BGS and RGS spectra can both contribute to the stack, including in the wavelength region where their rest-frame coverage overlaps. Each spectrum was individually shifted to the rest frame and normalised by its median flux. The normalised spectra were then resampled onto a common linear wavelength grid with a fixed pixel size of $0.5$\,nm. Within each quality category, the final stacked spectrum was obtained by taking the median flux of all contributing spectra at each wavelength pixel. The resulting stacks are shown in Fig.~\ref{fig:stacked_spectra_final}.
A clear qualitative sequence emerges across the four visual-inspection classes. The `Secure', `Reliable', and `Marginal' categories all yield coherent rest-frame stacked spectra, with a progressive decrease in the continuum S/N and in the detectability of the weakest spectral features toward lower-confidence classes. The `Secure' and `Reliable' stacks display the main optical nebular emission lines, including \oii{$\lambda$3727}, \hb{}, \oiii{$\lambda\lambda$4960,5008}, the \ha+\nii{$\lambda\lambda$6549,6585}\,complex, and \sii{$\lambda\lambda$6716,6731}, together with weaker transitions such as \neiii{$\lambda$3869}, \hei{$\lambda$5877}, \oi{$\lambda\lambda$6300,6364}, \ariii{$\lambda$7138}, and \siii{$\lambda\lambda$9069,9532}. Several of these weaker features are difficult to identify robustly in individual \Euclid spectra, but become clearly visible in the stacked data. This shows both the quality of the \Euclid slitless spectra and the strong potential of stacking techniques to extend their exploitation beyond the primary cosmological goals of the mission, enabling the study of intrinsically faint spectral diagnostics.

The `Marginal' stack remains fully consistent with this picture: the main emission-line pattern is still clearly recovered, while the continuum is noisier and the weakest transitions become progressively less evident. The ordered evolution of the stacked spectra from the `Secure' to the `Marginal' category therefore provides a qualitative validation of the visual-inspection classification, with decreasing confidence corresponding to a progressive loss of spectral coherence and information content.
The `Uncertain' stack behaves differently; although some emission-like features are still visible, their presence cannot by itself be interpreted as evidence that the corresponding redshift estimates are correct. As discussed in section ~5.2 of \citet{EP-Quai}, low-quality redshift solutions can arise when random noise fluctuations are interpreted as emission lines by the template-fitting procedure. Since the redshift determination relies on templates containing multiple emission features, spectra assigned an incorrect redshift can nevertheless produce apparently coherent features when shifted and stacked: noise fluctuations that contributed to the template match become aligned at the expected rest-frame wavelengths, allowing features such as H$\beta$, [O\,III], H\,$\alpha$, and [S\,II] to emerge even in a population of redshift contaminants.

\section{\label{Sect:wide} Redshift quality in Wide Survey}

\begin{figure*}
\centering
\begin{tabular}{cc}
\includegraphics[width=0.485\textwidth]{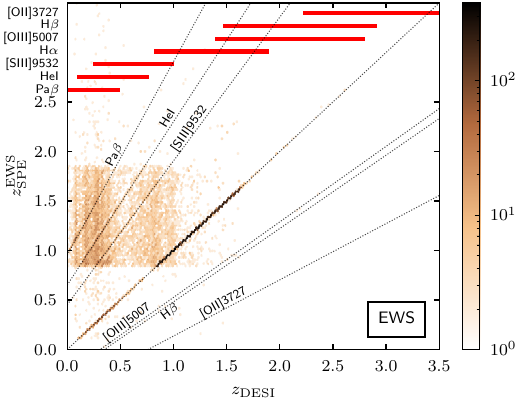} &  \includegraphics[width=0.485\textwidth]{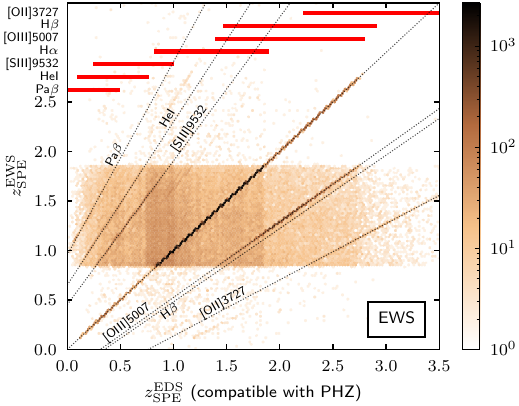}\\
\includegraphics[width=0.485\textwidth]{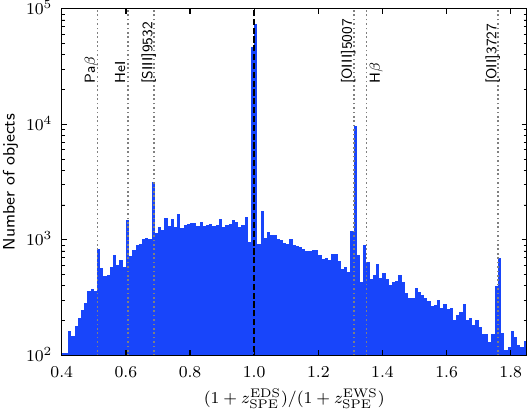} & \includegraphics[width=0.485\textwidth]{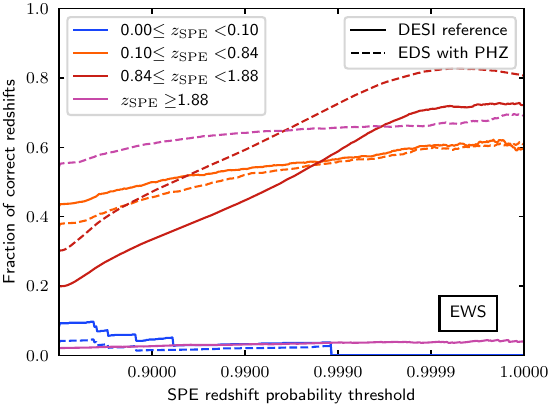} \\
\end{tabular}
\caption{\label{fig:z_vs_z_purity_wide} \textit{Top}: Same figure as Fig.\,\ref{fig:z_vs_z_purity_deep} (left panel), but for the single visits of the EDF-N corresponding to the EWS observations.  Only the RGS is shown, since the BGS is not used in EWS. In the left panel, we use DESI as a reference, and the high-purity sample from the EDF-N redshifts combined with PHZ in the right panel. \textit{Bottom left}: Distribution of $(1+z_{\rm SPE}^{\rm EDS}) / ( 1+ z_{\rm SPE}^{\rm EWS})$ for our high-purity reference sample with a similar EDF-N SPE PF and PHZ redshift (see Sect.\,\ref{sc:deep_comp_desi}). A line misidentification between two given lines will always be located at the same position on the $x$-axis (ratio of the wavelengths of the lines). The position corresponding to the misidentification of a given line with \ha\ is shown as a dotted vertical line, and the correct redshift 1:1 relation is shown as a dashed vertical line. \textit{Bottom right}: Analog of Fig.\,\ref{fig:z_vs_z_purity_deep} (right panel) but for the EWS. We use different limits between the redshift bins: 0.1 (arbitrary); 0.84 (\ha\ entering the RGS); and 1.88 (\ha\ exiting the RGS). We use as references both DESI (solid line) and the high-purity sample built using EDF-N SPE PF and PHZ data (dashed line).}

\end{figure*}

\subsection{Comparison with DESI redshifts}

To estimate the SPE PF performance on redshift measurement in the EWS, we matched the redshifts obtained from the RGS data of individual visits to the EDF-N fields with the DESI DR1 results. Indeed, those individual visits have the same exposure set-ups as those of the EWS. We have considered the seven complete visits, which were observed at different dates with different orientations of the spacecraft, as independent observations and matched them with the DESI values, after applying the quality criteria recommended by the DESI collaboration to ensure a failure rate of a few percent at most. The success rate we derive is thus a lower bound if one takes into account those few wrong DESI values, but likely to be very close to reality. Each DESI object could then be matched with up to seven \Euclid measurements using a procedure similar to that described in Sect.\,\ref{sc:deep_comp_desi}. We used the same criteria on all parameters but {\tt spe\_z\_prob} as in the EDF-N to build a high-quality sample, but we require only 300 valid pixels and a median of more than three dithers. The redshift distribution is discussed in Appendix\,\ref{sc:redshift_distibutions}.

In Fig.\,\ref{fig:z_vs_z_purity_wide} (top left), we compare the SPE PF redshifts with DESI. For $0.84<z<1.88$, where \ha\ is detected in the RGS, a large fraction of them are on the 1:1 relation with a correct redshift. Contrary to the EDF-N (Fig.\,\ref{fig:z_vs_z_purity_deep}), most of the noise interlopers are located in the $0.84<z<1.88$ range. This is caused by the \ha\ prior used in the EWS to maximise the completeness of the cosmological sample, pushing them into this redshift range. We can also observe line interlopers corresponding to \Pb, \hei, and \siii{$\lambda$9532} misidentified as \ha.

\subsection{Comparison with the Euclid Deep Survey}

The EDS provides a more complete sample than DESI to characterise the EWS. In the EDF-N, the EWS high-quality sample contains 974\,487 sources (about 140\,000 per visit), but there are only 29\,394 sources with a DESI counterpart. The EDS high-quality reference with compatible photometric redshifts (Sect.\,\ref{sc:specphot}) contains 236\,740 objects and extends to higher redshift (up to $z=4.07$ with \oii). This sample is only a fraction of the full sample, but these objects have a much higher median SPE PF redshift probability than what is obtained without using PHZ (0.99995 versus 0.996). Using this quality criterion, we are thus mainly excluding objects with lower reliability.

In Fig.\,\ref{fig:z_vs_z_purity_wide} (top right), we present the comparison between the EWS and the EDS high-quality reference sample (Sect.\,\ref{sc:deep_comp_desi}) in the EDF-N. At $z \lesssim 1$, the results are similar to the comparison with DESI. At higher redshift, we observe that the 1:1 feature corresponding to correct redshifts extends up to $z=2.78$, where \oiii\ exits the RGS. Using the EDS as a reference, we can see that the reference redshift of random interlopers extends beyond $z\simeq  3$. The most important result is the presence of a sizable population of \oiii\ interlopers, which were missing in the Q1 results but expected from SPE PF pre-flight simulations \citep{Granett26}. We also have \hb\ and \oii\ interlopers, but they are rarer and barely visible in the figure. This confirms that the redshift range used for cosmology can be contaminated by objects at higher redshift, but we also know that the value adopted for the strength of the \ha\ prior (Sect.\,\ref{sc:redshifts}) is probably not optimal. It was tuned on the Q1 data, which were of lower quality, and it may be too strong for the DR1 data. A dedicated tuning using up-to-date data is planned for DR2.

To better visualise the interlopers, we computed the distribution of the following quantity
\begin{equation}
\frac{1+z_{\rm SPE}^{\rm EDS}}{1+z_{\rm SPE}^{\rm EWS}} = \frac{\lambda_{\rm EWS}}{\lambda_{\rm EDS}},
\end{equation}
where $z_{\rm SPE}^{\rm EDS}$ are the EDS redshifts assuming the main line has a wavelength $\lambda_{\rm EDS}$. Similarly, $z_{\rm SPE}^{\rm EWS}$ is the redshift from EWS, while $\lambda_{\rm EWS}$ is the wavelength of the assumed main line, which is usually \ha\ considering the prior applied by the algorithm. This distribution is shown in Fig.\,\ref{fig:z_vs_z_purity_wide} (bottom left). In this histogram, the correct redshifts are located at unity, and noise interlopers are smoothly distributed around this value. The vertical dotted lines correspond to the expected positions of the various line interlopers. The two most prominent peaks correspond to \oiii\, and \siii{$\lambda$9532}\, misidentified as \ha, and  contain, respectively, 12 and 44 times fewer objects than the peak of correct redshifts. There are other secondary peaks that can be clearly identified:  \hei, \Pb, \hb, and \oii. Other weaker features cannot be unambiguously disentangled from the noise.

\subsection{Estimated fraction of correct redshifts}

Both DESI and the EDS can be used as a reference to estimate the fraction of correct redshifts in the EWS. In Fig.\,\ref{fig:z_vs_z_purity_wide} (bottom right), we show our results in various redshift bins. Since the EDS can be used as a reference at higher redshift than DESI, and $z=0.41$ corresponding to \ha\ entering the BGS is irrelevant for EWS data, we now use different borders between bins than in Sect.\,\ref{sc:deep_comp_desi}: 0.1, 0.84, and 1.88 (\ha\ exiting the RGS).

Below $z=0.84$, where both DESI and the EDS provide solid reference samples, the estimated fractions of correct redshifts agree with each other. At $z<0.1$ (blue lines in the bottom-right panel of Fig.\ref{fig:z_vs_z_purity_wide}), similarly to what was found in \cite{Q1-TP007}, the fraction of correct galaxy redshifts for galaxies is very low ($<10$\,\%), according to both reference samples. Between $z=0.1$ and $z=0.84$ (orange lines), this fraction increases from about 40\,\% to 60\,\% when we increase the SPE PF probability threshold. This is a large improvement compared to the $\lesssim$10\,\% obtained in Q1. Nevertheless, our success rate remains modest, highlighting the challenge of deriving reliable redshifts when \ha\ is not in the RGS. The two \siii\ lines, \hi\, and  \Pb\ are useful, but usually rather faint.

In the redshift range where \ha\ is in the RGS ($0.84<z<1.88$, red lines), the estimate of the fraction of correct redshifts is approximately 10 percentage points larger using the EDS as a reference, whatever the SPE PF probability threshold. This might seem surprising considering the EDS reference contains fainter and higher redshift objects (up to $z=1.88$). However, as shown in Appendix A of \cite{Q1-TP007}, the correct fraction tends to be paradoxically higher for objects with a fainter \HE\ magnitude. In this study, we do not consider other criteria on the \ha\ flux or S/N to keep our analysis homogeneous over the various redshift ranges, but we previously showed that a tailored selection can lead to even higher performance \citep{Q1-TP007}. 

Finally, we examined the redshifts above $z=1.88$ where \ha\ is no longer available, but \oiii, \hb, and \oii\ are. While the very small fraction of objects with a DESI counterpart suggests poor performance ($\lesssim$3\,\% of correct redshifts, solid purple line), the comparison with the EDS spectra suggests that we can efficiently identify high-$z$ candidates in EWS (64\,\% of correct redshifts for an SPE PF probability threshold of 0.99, dashed purple line). At this redshift, the EDS reference is much more suitable as a reference sample. These results are thus very encouraging, but need to be confirmed when the full depth of the EDS will be reached.

\subsection{Validation of the flux measurements in the EWS}

\begin{figure}
\centering
\includegraphics[width=0.485\textwidth]{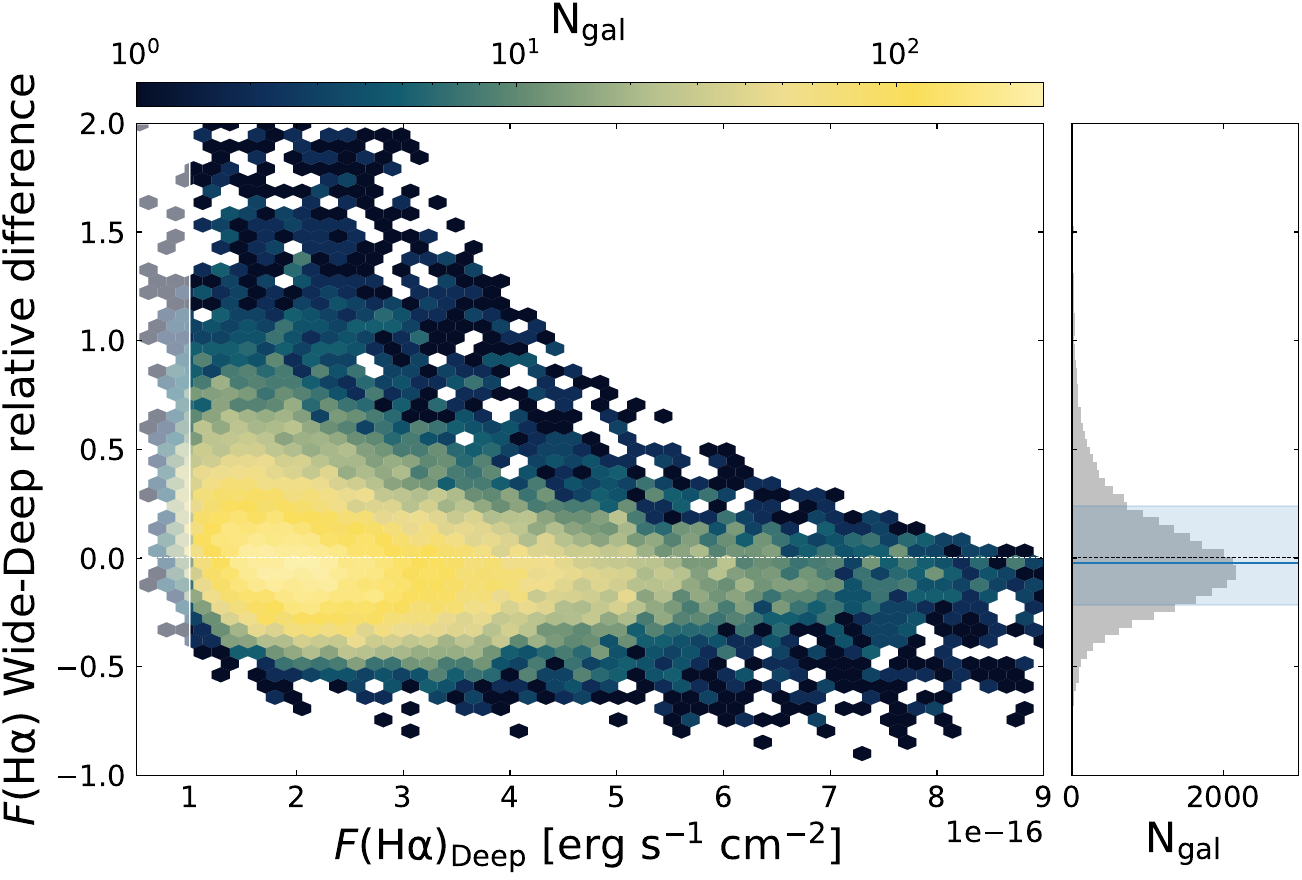}
\includegraphics[width=0.485\textwidth]{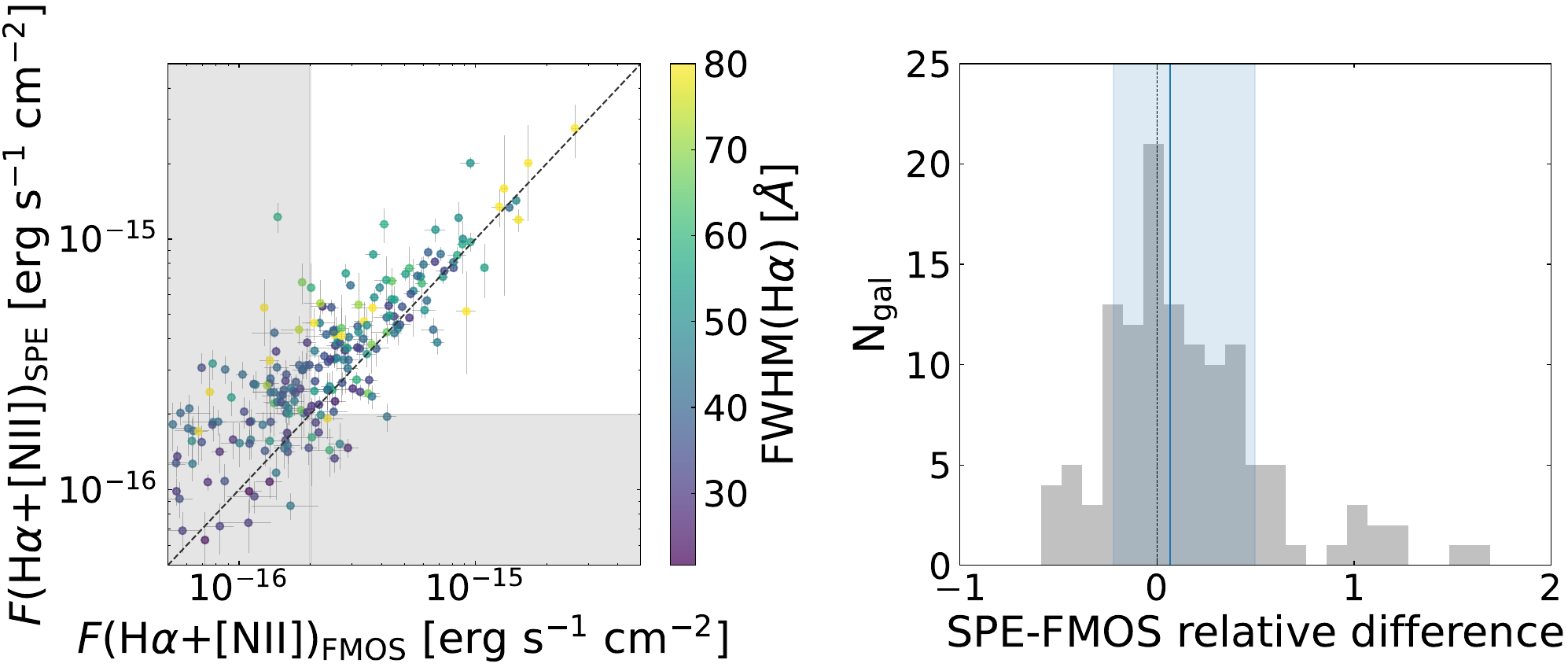}
\caption{Validation of the \ha~flux measurements. The top-left plot shows the relative difference between the flux measurements performed on EWS and EDS data, defined as the difference between the EWS and EDS measurements divided by the EDS one, binned into hexbins; points are coloured depending on the number of objects inside the pixel. 
In the top-right plot, the histogram of the distribution is presented.
The bottom-left plot shows a comparison of \ha+\nii~flux measurements from EWS and FMOS data, colour-coded by the measured line FWHM; the white area corresponds to the requirements on the flux and S/N for the EWS. The bottom-right plot presents the histogram of the relative difference, defined as the difference between the EWS and FMOS flux measurements divided by the FMOS one. The blue lines and shaded area show the median and 16th and 84th percentiles of the distribution.
In all these plots, we considered measurements performed with the GF method.}
\label{fig:flux_acc}
\end{figure}

\begin{figure*}
\centering
\includegraphics[width=\textwidth]{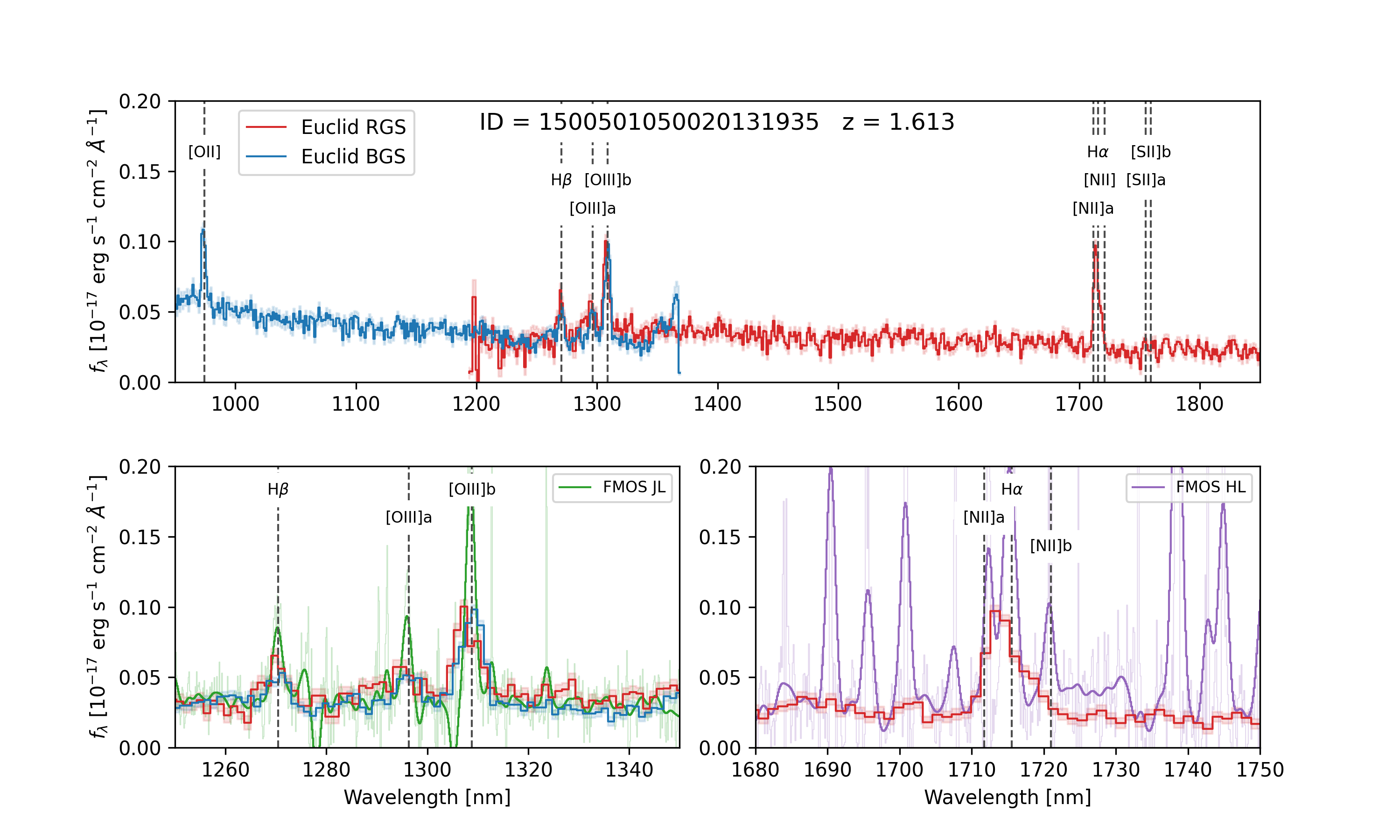}
\caption{Comparison with Subaru FMOS spectra. The upper panel shows the spectrum of a galaxy at $z=1.613$ with $F$(\ha)=$3.4\times 10^{-16}$~\ecgs~observed by \Euclid in the EDS (red and blue lines show the RGS and BGS spectra, respectively). The bottom panels present a comparison of the \Euclid\ spectrum (blue and red lines) with the FMOS spectrum. The bottom left plot shows a zoom in the region of the FMOS J-long grating (JL, optimised to observe the \hb{} and \oiii{} lines), where the green thinner line represents the original spectrum and the thicker one the spectrum smoothed with a Gaussian filter of 5 pixels. The bottom right panel shows the same, but for the FMOS H-long grating (HL, optimised to observe the \ha{} line), in violet.}
\label{fig:spec_EuclidFMOS}
\end{figure*}

We validated the flux measurements of our pipeline with two independent approaches.
Firstly, we used \Euclid data from the EDS as reference values, since in these fields we have flux measurements performed on the same objects as in the EWS, but at greater depths. In particular, we used the EDF-N data described in Sect. \ref{sec:datadescription}. To remove potentially spurious results, we selected a high-quality sample, imposing a cut ${\tt spe\_z\_prob} \ge 0.999$ for both redshift measurements, $\text{S/N}(\ha) > 5$ (10) in the EWS (EDS), and with a well-measured flux ({\tt spe\_line\_qual\_gf}=[212,252], including also 222 to be conservative, since, despite the lower S/N, the measurement can be good, see Tables~\ref{tab:qualflag} and ~\ref{tab:seconc_digit}). 
Secondly, we cross-correlated the \Euclid data in the COSMOS field with the public catalogue from the FMOS-COSMOS Survey \citep{Silverman2015,Kashino2019}. That survey is based on the near-infrared Fibre Multi-Object Spectrograph mounted on the Subaru Telescope, aiming to measure \ha~emission lines at medium resolution ($R\simeq3000$) to characterise the population of star-forming galaxies at $z\simeq1.6$. The results are shown in Fig.~\ref{fig:flux_acc}. We find that the \ha fluxes measured with the GF method show remarkable agreement between measurements performed in the EWS and EDS. Above the nominal flux limit of the EWS, $F(\ha) > 2\times 10^{-16}$~\ecgs, the median difference in flux between EWS and EDS measurements is $-2^{+19}_{-26}$\% (16th--84th percentile of the distribution). The errors on the flux scale perfectly with the number of observations, the EWS ones being a factor $\sqrt{17/4}$ larger than the EDS ones (17 and 4 being the number of observations in the EDS and EWS). The comparison with FMOS is intrinsically noisier, but shows the same result, with a median difference of $7^{+43}_{-28}$\%.
Figure~\ref{fig:spec_EuclidFMOS} shows a comparison of the spectrum of a galaxy observed by \Euclid in the EDS in the RGS and BGS, compared to the corresponding FMOS spectrum of the same object.

In addition, we also validated line-flux measurements in the rest-frame NIR, by comparing the results obtained with the software \texttt{slinefit}\footnote{The \texttt{slinefit} code is publicly available at \url{https://github.com/ cschreib/slinefit}} \citep{Schreiber_2018, Talia_2023} in a subsample selected with high S/N at low redshift. We find good agreement for \siii$\lambda9530$, Pa\,$\gamma$, and Pa\,$\beta$ on spectra like the one on Fig.~\ref{fig:AGN2}.  

\section{\label{Sect:concl} Conclusions}

We have described the basic functionalities of the SPE PF pipeline, in charge of the automatic redshift measurements of the \Euclid mission, and presented the properties of the spectroscopic sample of the Euclid DR1. This includes more than 160 million redshifts in the EWS and 7.0 million in the EDFs, among which are 2.7 million in EDF-N, whose spectroscopic depth is the most advanced. In both surveys, the accuracy of the redshift measurement for emission line galaxies is excellent, either compared to the DESI catalogue or by comparing EWS and EDS data, with a median value of the relative redshift error about $10^{-5}$ and a $1\,\sigma$ dispersion of $0.4\times 10^{-3}$.  We have shown that the results in EDF-N are very promising, since it is possible, for the galaxy clustering measurement, to extract a sample for which 75\% of the objects have a redshift in agreement  with DESI based on spectroscopy only, and up to 98\% when combining spectroscopic and photometric redshifts. Even if lower, the agreement reaches 90\% outside the cosmological sample when combining with the photometric redshifts. 
Using this high-purity EDS sample as a reference, we show that, after a tailored selection,  the purity in the EWS reaches the requested 80\% value in the cosmological target redshift interval, based on spectroscopy alone. 

Regarding other categories of objects, the use of templates built from \Euclid spectra for quasars and some specific categories of galaxies allows us to build high-quality samples based on spectroscopy alone. For stars, including ultra-cool dwarfs, whose templates are still coming from external sources, results are not as good; the integration of \Euclid -based templates is planned for the future DR2 release.

The line-flux measurements calculate the emission line properties with two methods: a direct integration that provides the flux as directly measured on the spectrum itself; and a Gaussian fit that provides additional properties like position, full width at half maximum, and equivalent width. Measurements are in excellent agreement with external data as well as between EWS and EDS. The future evolution of the pipeline will include a specific model for cosmology, based on the current `galaxy' category, together with a complete model dedicated to galaxy evolution studies, to allow the optimal exploitation of the spectroscopy of \Euclid. The components of this model, like continuum templates and emission line flux distributions, will be built in collaboration with the relevant science working groups. We will also upgrade the stellar templates to improve the classification results and provide more accurate radial velocity measurements. 

The evolution for the future DR2 release will include redshift calculation for four different models; the dedicated emission line detection for cosmology, stars, quasar with only specific templates, and a galaxy model including physical continuum templates and emission lines, and without the \ha\ prior. The emission line galaxy will not be considered in the classification, in order to minimise the bias described in Sect.\ref{Sect:classif_results}
\begin{acknowledgements}
\AckDRone
\AckEC 

MM acknowledges the financial contribution from the grant PRIN-MUR 2022 2022NY2ZRS 001 “Optimizing the extraction of cosmological information from Large Scale Structure analysis in view of the next large spectroscopic surveys” supported by Next Generation EU and from the grant ASI n. 2024-10-HH.0 “Attività scientifiche per la missione Euclid – fase E”. MT, LP, SQ, PC, MZ, DV acknowledge support from the ELSA project. ELSA: Euclid Legacy Science Advanced analysis tools (Grant Agreement no. 101135203) is funded by the European Union. Views and opinions expressed are however those of the author(s) only and do not necessarily reflect those of the European Union or Innovate UK. Neither the European Union nor the granting authority can be held responsible for them.
This research used data obtained with the Dark Energy Spectroscopic Instrument (DESI). DESI construction and operations is managed by the Lawrence Berkeley National Laboratory. This material is based upon work supported by the U.S. Department of Energy, Office of Science, Office of High-Energy Physics, under Contract No. DE–AC02–05CH11231, and by the National Energy Research Scientific Computing Center, a DOE Office of Science User Facility under the same contract. Additional support for DESI was provided by the U.S. National Science Foundation (NSF), Division of Astronomical Sciences under Contract No. AST-0950945 to the NSF’s National Optical-Infrared Astronomy Research Laboratory; the Science and Technology Facilities Council of the United Kingdom; the Gordon and Betty Moore Foundation; the Heising-Simons Foundation; the French Alternative Energies and Atomic Energy Commission (CEA); the National Council of Science and Technology of Mexico (CONACYT); the Ministry of Science and Innovation of Spain (MICINN), and by the DESI Member Institutions: www.desi.lbl.gov/collaborating-institutions. The DESI collaboration is honoured to be permitted to conduct scientific research on Iolkam Du’ag (Kitt Peak), a mountain with particular significance to the Tohono O’odham Nation. Any opinions, findings, and conclusions or recommendations expressed in this material are those of the author(s) and do not necessarily reflect the views of the U.S. National Science Foundation, the U.S. Department of Energy, or any of the listed funding agencies.

\end{acknowledgements}

\bibliography{Euclid,Others}

\begin{thebibliography}{27}
\expandafter\ifx\csname natexlab\endcsname\relax\def\natexlab#1{#1}\fi

\bibitem[{{Ba{\~n}ados} {et~al.}(2025){Ba{\~n}ados}, {Le Brun}, {Belladitta}, {Momcheva}, {Stern}, {Wolf}, {Ezziati}, {Mortlock}, {Humphrey}, {Smart}, {Casewell}, {P{\'e}rez-Garrido}, {Goldman}, {Mart{\'\i}n}, {Mohandasan}, {Reyl{\'e}}, {Dominguez-Tagle}, {Copin}, {Lusso}, {Matsuoka}, {McCarthy}, {Ricci}, {Rix}, {Rottgering}, {Schindler}, {Weaver}, {Allaoui}, {Bedrine}, {Castellano}, {Chabaud}, {Daste}, {Dufresne}, {Gracia-Carpio}, {K{\"u}mmel}, {Moresco}, {Scodeggio}, {Surace}, {Vibert}, {Balestra}, {Bonnefoi}, {Caillat}, {Cogato}, {Costille}, {Dusini}, {Ferriol}, {Franceschi}, {Gillard}, {Jahnke}, {Le Mignant}, {Ligori}, {Medinaceli}, {Morgante}, {Passalacqua}, {Paterson}, {Pires}, {Sirignano}, {Andika}, {Atek}, {Barrado}, {Bisogni}, {Conselice}, {Dannerbauer}, {Decarli}, {Dole}, {Dupuy}, {Feltre}, {Fotopoulou}, {Gillis}, {Lopez}, {Onoue}, {Rodighiero}, {Sedighi}, {Shankar}, {Siudek}, {Spinoglio}, {Vergani}, {Vietri}, {Walter}, {Zamorani}, {Zapatero Osorio}, {Zhang}, {Bethermin}, {Aghanim}, {Altieri},
  {Amara}, {Andreon}, {Baccigalupi}, {Baldi}, {Bardelli}, {Basset}, {Battaglia}, {Biviano}, {Bonchi}, {Bonino}, {Branchini}, {Brescia}, {Brinchmann}, {Camera}, {Capobianco}, {Carbone}, {Carretero}, {Casas}, {Castignani}, {Cavuoti}, {Cimatti}, {Colodro-Conde}, {Congedo}, {Conversi}, {Courbin}, {Courtois}, {Cropper}, {Cuby}, {Da Silva}, {Degaudenzi}, {De Lucia}, {Giorgio}, {Dolding}, {Dubath}, {Duncan}, {Dupac}, {Ealet}, {Farina}, {Faustini}, {Fourmanoit}, {Frailis}, {Galeotta}, {George}, {Giocoli}, {Granett}, {Grazian}, {Grupp}, {Guzzo}, {Haugan}, {Hoar}, {Hoekstra}, {Holmes}, {Hook}, {Hormuth}, {Hornstrup}, {Hudelot}, {Jhabvala}, {Joachimi}, {Keih{\"a}nen}, {Kermiche}, {Kubik}, {Kuijken}, {Kunz}, {Kurki-Suonio}, {Lilje}, {Lindholm}, {Lloro}, {Mainetti}, {Maino}, {Maiorano}, {Mansutti}, {Marggraf}, {Markovic}, {Martinelli}, {Martinet}, {Marulli}, {Massey}, {Mei}, {Mellier}, {Meneghetti}, {Merlin}, {Meylan}, {Mora}, {Moscardini}, {Neissner}, {Niemi}, {Nightingale}, {Padilla}, {Paltani}, {Pasian}, {Pedersen},
  {Percival}, {Pettorino}, {Polenta}, {Poncet}, {Popa}, {Pozzetti}, {Raison}, {Rebolo}, {Renzi}, {Rhodes}, {Riccio}, {Romelli}, {Roncarelli}, {Rossetti}, {Saglia}, {Sakr}, {Sapone}, {Sartoris}, {Schewtschenko}, {Schirmer}, {Schneider}, {Schrabback}, {Secroun}, {Sefusatti}, \& {Seidel}}]{Banados}
{Ba{\~n}ados}, E., {Le Brun}, V., {Belladitta}, S., {et~al.} 2025, \mnras, 542, 1088

\bibitem[{Bruzual \& Charlot(2003)}]{BC03}
Bruzual, G. \& Charlot, S. 2003, \mnras, 344, 1000

\bibitem[{{DESI Collaboration: Adame} {et~al.}(2024){DESI Collaboration: Adame}, Aguilar, Ahlen, Alam, Aldering, Alexander, Alfarsy, Allende~Prieto, Alvarez, Alves, Anand, Andrade-Oliveira, Armengaud, Asorey, Avila, Aviles, Bailey, Balaguera-Antolínez, Ballester, Baltay, Bault, Bautista, Behera, Beltran, BenZvi, Beraldo~e Silva, Bermejo-Climent, Berti, Besuner, Beutler, Bianchi, Blake, Blum, Bolton, Brieden, Brodzeller, Brooks, Brown, Buckley-Geer, Burtin, Cabayol-Garcia, Cai, Canning, Cardiel-Sas, Carnero~Rosell, Castander, Cervantes-Cota, Chabanier, Chaussidon, Chaves-Montero, Chen, Chen, Chuang, Claybaugh, Cole, Cooper, Cuceu, Davis, Dawson, de~Belsunce, de~la Cruz, de~la Macorra, Della~Costa, de~Mattia, Demina, Demirbozan, DeRose, Dey, Dey, Dhungana, Ding, Ding, Doel, Doshi, Douglass, Edge, Eftekharzadeh, Eisenstein, Elliott, Ereza, Escoffier, Fagrelius, Fan, Fanning, Fawcett, Ferraro, Flaugher, Font-Ribera, Forero-Romero, Forero-Sánchez, Frenk, Gänsicke, García, García-Bellido, Garcia-Quintero,
  Garrison, Gil-Marín, Golden-Marx, Gontcho A~Gontcho, Gonzalez-Morales, Gonzalez-Perez, Gordon, Graur, Green, Gruen, Guy, Hadzhiyska, Hahn, Han, Hanif, Herrera-Alcantar, Honscheid, Hou, Howlett, Huterer, Iršič, Ishak, Jacques, Jana, Jiang, Jimenez, Jing, Joudaki, Joyce, Jullo, Juneau, Karaçaylı, Karim, Kehoe, Kent, Khederlarian, Kim, Kirkby, Kisner, Kitaura, Kizhuprakkat, Kneib, Koposov, Kovács, Kremin, Krolewski, L’Huillier, Lahav, Lambert, Lamman, Lan, Landriau, Lang, Lange, Lasker, Leauthaud, Le~Guillou, Levi, Li, Linder, Lyons, Magneville, Manera, Manser, Margala, Martini, McDonald, Medina, Medina-Varela, Meisner, Mena-Fernández, Meneses-Rizo, Mezcua, Miquel, Montero-Camacho, Moon, Moore, Moustakas, Mueller, Mundet, Muñoz-Gutiérrez, Myers, Nadathur, Napolitano, Neveux, Newman, Nie, Nikutta, Niz, Norberg, Noriega, Paillas, Palanque-Delabrouille, Palmese, Pan, Parkinson, Penmetsa, Percival, Pérez-Fernández, Pérez-Ràfols, Pieri, Poppett, Porredon, Pothier, Prada, Pucha, Raichoor,
  Ramírez-Pérez, Ramirez-Solano, Rashkovetskyi, Ravoux, Rocher, Rockosi, Ross, Rossi, Ruggeri, Ruhlmann-Kleider, Sabiu, Said, Saintonge, Samushia, Sanchez, Saulder, Schaan, Schlafly, Schlegel, Scholte, Schubnell, Seo, Shafieloo, Sharples, Sheu, Silber, Sinigaglia, Siudek, Slepian, Smith, Soumagnac, Sprayberry, Stephey, Suárez-Pérez, Sun, Tan, Tarlé, Tojeiro, Ureña-López, Vaisakh, Valcin, Valdes, Valluri, Vargas-Magaña, Variu, Verde, Walther, Wang, Wang, Weaver, Weaverdyck, Wechsler, White, Xie, Yang, Yèche, Yu, Yuan, Zhang, Zhang, Zhao, Zheng, Zhou, Zhou, Zou, Zou, \& Zu}]{Adame2024}
{DESI Collaboration: Adame}, A.~G., Aguilar, J., Ahlen, S., {et~al.} 2024, \aj, 168, 58

\bibitem[{{Euclid Collaboration: Aussel} {et~al.}(2026){Euclid Collaboration: Aussel}, {Tereno}, {Schirmer}, {et~al.}}]{Q1-TP001}
{Euclid Collaboration: Aussel}, H., {Tereno}, I., {Schirmer}, M., {et~al.} 2026, \aap, 711, A1

\bibitem[{{Euclid Collaboration: Fu} {et~al.}(2025){Euclid Collaboration: Fu}, {Bouwens}, {Caputi}, {Vergani}, {Scialpi}, {Margalef-Bentabol}, {Wang}, {Bolzonella}, {Banerji}, {Ba{\~n}ados}, {Feltre}, {Toba}, {Calhau}, {Tarsitano}, {Cunha}, {Humphrey}, {Vietri}, {Mannucci}, {Bisogni}, {Ricci}, {Landt}, {Spinoglio}, {Matamoro Zatarain}, {Stern}, {Page}, {Alexander}, {Zamorani}, {Roster}, {Salvato}, {Copin}, {Sorce}, {Scott}, {Zhang}, {Lusso}, {Wolf}, {Yang}, {Rottgering}, {Laloux}, {Siudek}, {Belladitta}, {Liu}, {Allevato}, {Kuijken}, {Andreon}, {Auricchio}, {Baccigalupi}, {Baldi}, {Balestra}, {Bardelli}, {Battaglia}, {Biviano}, {Branchini}, {Brescia}, {Brinchmann}, {Camera}, {Ca{\~n}as-Herrera}, {Capobianco}, {Carbone}, {Carretero}, {Casas}, {Castellano}, {Castignani}, {Cavuoti}, {Chambers}, {Cimatti}, {Colodro-Conde}, {Congedo}, {Conselice}, {Conversi}, {Costille}, {Courbin}, {Courtois}, {Cropper}, {Da Silva}, {Degaudenzi}, {De Lucia}, {Dolding}, {Dole}, {Dubath}, {Duncan}, {Dupac}, {Dusini}, {Escoffier},
  {Fabricius}, {Farina}, {Farinelli}, {Ferriol}, {Finelli}, {Fosalba}, {Fourmanoit}, {Frailis}, {Franceschi}, {Franzetti}, {Fumana}, {Galeotta}, {George}, {Gillard}, {Gillis}, {Giocoli}, {Gracia-Carpio}, {Grazian}, {Grupp}, {Guzzo}, {Haugan}, {Hoekstra}, {Holmes}, {Hook}, {Hormuth}, {Hornstrup}, {Jahnke}, {Jhabvala}, {Joachimi}, {Keih{\"a}nen}, {Kermiche}, {Kiessling}, {Kubik}, {K{\"u}mmel}, {Kunz}, {Kurki-Suonio}, {Laureijs}, {Le Brun}, {Ligori}, {Lilje}, {Lindholm}, {Lloro}, {Mainetti}, {Maino}, {Maiorano}, {Mansutti}, {Marcin}, {Marggraf}, {Markovic}, {Martinelli}, {Martinet}, {Marulli}, {Massey}, {Medinaceli}, {Mei}, {Melchior}, {Mellier}, {Meneghetti}, {Merlin}, {Meylan}, {Mora}, {Moresco}, {Moscardini}, {Nakajima}, {Neissner}, {Nichol}, {Niemi}, {Padilla}, {Paltani}, {Pasian}, {Pedersen}, {Percival}, {Pettorino}, {Pires}, {Polenta}, {Poncet}, {Popa}, {Pozzetti}, {Raison}, {Rebolo}, {Renzi}, {Rhodes}, {Riccio}, {Romelli}, {Roncarelli}, {Rossetti}, {Saglia}, {Sakr}, {Sapone}, {Sartoris}, {Schirmer},
  {Schneider}, {Schrabback}, {Scodeggio}, {Secroun}, {Sefusatti}, {Seidel}, {Serrano}, {Simon}, {Sirignano}, {Sirri}, {Stanco}, {Starck}, {Steinwagner}, {Surace}, {Tallada-Cresp{\'\i}}, {Tavagnacco}, {Taylor}, {Teplitz}, {Tereno}, {Tessore}, {Toft}, {Toledo-Moreo}, {Torradeflot}, {Tutusaus}, \& {Valenziano}}]{Fu2025}
{Euclid Collaboration: Fu}, Y., {Bouwens}, R., {Caputi}, K.~I., {et~al.} 2025, arXiv e-prints, arXiv:2512.08803

\bibitem[{Euclid Collaboration:~Granett {et~al.}(2026)Euclid Collaboration:~Granett, de~la Torre, Moresco, Farina, Bethermin, {et~al.}}]{Granett26}
Euclid Collaboration:~Granett, B.~R., de~la Torre, S., Moresco, M., {et~al.} 2026, \aap, submitted [\eprint[arXiv]{2610.26990}]

\bibitem[{{Euclid Collaboration: Hartley} {et~al.}(2026){Euclid Collaboration: Hartley}, {Paltani}, {Dubath}, {et~al.}}]{DR1-TP011}
{Euclid Collaboration: Hartley}, W.~G., {Paltani}, S., {Dubath}, F., {et~al.} 2026, A\&A, submitted, arXiv:2611.15987

\bibitem[{{Euclid Collaboration: Jahnke} {et~al.}(2025){Euclid Collaboration: Jahnke}, {Gillard}, {Schirmer}, {et~al.}}]{EuclidSkyNISP}
{Euclid Collaboration: Jahnke}, K., {Gillard}, W., {Schirmer}, M., {et~al.} 2025, A\&A, 697, A3

\bibitem[{{Euclid Collaboration: K{\"u}mmel} {et~al.}(2026){Euclid Collaboration: K{\"u}mmel}, {Romelli}, {Gracia-Carpio}, {et~al.}}]{DR1-TP009}
{Euclid Collaboration: K{\"u}mmel}, M., {Romelli}, E., {Gracia-Carpio}, J., {et~al.} 2026, A\&A, submitted, arXiv:2611.17832

\bibitem[{{Euclid Collaboration: Le Brun} {et~al.}(2026){Euclid Collaboration: Le Brun}, {Bethermin}, {Moresco}, {et~al.}}]{Q1-TP007}
{Euclid Collaboration: Le Brun}, V., {Bethermin}, M., {Moresco}, M., {et~al.} 2026, \aap, 711, A7

\bibitem[{{Euclid Collaboration: Mao} {et~al.}(2026){Euclid Collaboration: Mao}, {Pozzetti}, {Enia}, {et~al.}}]{Q1-Mao}
{Euclid Collaboration: Mao}, Z., {Pozzetti}, L., {Enia}, A., {et~al.} 2026, A\&A, submitted, arXiv:2610.08999

\bibitem[{{Euclid Collaboration: Mellier} {et~al.}(2025){Euclid Collaboration: Mellier}, {Abdurro'uf}, {Acevedo~Barroso}, {et~al.}}]{EuclidSkyOverview}
{Euclid Collaboration: Mellier}, Y., {Abdurro'uf}, {Acevedo~Barroso}, J., {et~al.} 2025, A\&A, 697, A1

\bibitem[{{Euclid Collaboration: Quai} {et~al.}(2026){Euclid Collaboration: Quai}, {Pozzetti}, {Talia}, {et~al.}}]{EP-Quai}
{Euclid Collaboration: Quai}, S., {Pozzetti}, L., {Talia}, M., {et~al.} 2026, \aap, 707, A232

\bibitem[{{Euclid Collaboration: Scodeggio} {et~al.}(2026){Euclid Collaboration: Scodeggio}, {Copin}, {et~al.}}]{DR1-TP023}
{Euclid Collaboration: Scodeggio}, M., {Copin}, Y., {et~al.} 2026, in preparation

\bibitem[{{Euclid Data Release 1}(2026)}]{DR1cite}
{Euclid Data Release 1}. 2026, \url{https://doi.org/10.57780/esa-d079c51}

\bibitem[{Garilli {et~al.}(2010)Garilli, Fumana, Franzetti, Paioro, Scodeggio, Le~Fèvre, Paltani, \& Scaramella}]{Garilli_2010}
Garilli, B., Fumana, M., Franzetti, P., {et~al.} 2010, \pasp, 122, 827

\bibitem[{{Kashino} {et~al.}(2019){Kashino}, {Silverman}, {Sanders}, {Kartaltepe}, {Daddi}, {Renzini}, {Rodighiero}, {Puglisi}, {Valentino}, {Juneau}, {Arimoto}, {Nagao}, {Ilbert}, {Le F{\`e}vre}, \& {Koekemoer}}]{Kashino2019}
{Kashino}, D., {Silverman}, J.~D., {Sanders}, D., {et~al.} 2019, \apjs, 241, 10

\bibitem[{{Le Brun} {et~al.}(2026)}]{Amazed_2026}
{Le Brun}, V. {et~al.} 2026, \aap, in preparation

\bibitem[{{Le Fèvre} {et~al.}(2013){Le Fèvre}, {Cassata, P.}, {Cucciati, O.}, {Garilli, B.}, {Ilbert, O.}, {Le Brun, V.}, {Maccagni, D.}, {Moreau, C.}, {Scodeggio, M.}, {Tresse, L.}, {Zamorani, G.}, {Adami, C.}, {Arnouts, S.}, {Bardelli, S.}, {Bolzonella, M.}, {Bondi, M.}, {Bongiorno, A.}, {Bottini, D.}, {Cappi, A.}, {Charlot, S.}, {Ciliegi, P.}, {Contini, T.}, {de la Torre, S.}, {Foucaud, S.}, {Franzetti, P.}, {Gavignaud, I.}, {Guzzo, L.}, {Iovino, A.}, {Lemaux, B.}, {López-Sanjuan, C.}, {McCracken, H. J.}, {Marano, B.}, {Marinoni, C.}, {Mazure†, A.}, {Mellier, Y.}, {Merighi, R.}, {Merluzzi, P.}, {Paltani, S.}, {Pellò, R.}, {Pollo, A.}, {Pozzetti, L.}, {Scaramella, R.}, {Tasca, L.}, {Vergani, D.}, {Vettolani, G.}, {Zanichelli, A.}, \& {Zucca, E.}}]{LeFevre2013}
{Le Fèvre}, O., {Cassata, P.}, {Cucciati, O.}, {et~al.} 2013, \aap, 559, A14

\bibitem[{Pickles(1998)}]{Pickles_1998}
Pickles, A.~J. 1998, \pasp, 110, 863

\bibitem[{{Pozzetti} {et~al.}(2016){Pozzetti}, {Hirata}, {Geach}, {Cimatti}, {Baugh}, {Cucciati}, {Merson}, {Norberg}, \& {Shi}}]{Pozzetti2016}
{Pozzetti}, L., {Hirata}, C.~M., {Geach}, J.~E., {et~al.} 2016, \aap, 590, A3

\bibitem[{Saito {et~al.}(2020)Saito, de la Torre, Ilbert, Dubois, Yabe, \& Coupon}]{Saito2020}
Saito, S., de la Torre, S., Ilbert, O., {et~al.} 2020, \mnras, 494, 199

\bibitem[{{Schreiber} {et~al.}(2018){Schreiber}, {Glazebrook}, {Nanayakkara}, {Kacprzak}, {Labb{\'e}}, {Oesch}, {Yuan}, {Tran}, {Papovich}, {Spitler}, \& {Straatman}}]{Schreiber_2018}
{Schreiber}, C., {Glazebrook}, K., {Nanayakkara}, T., {et~al.} 2018, \aap, 618, A85

\bibitem[{{Silverman} {et~al.}(2015){Silverman}, {Kashino}, {Sanders}, {Kartaltepe}, {Arimoto}, {Renzini}, {Rodighiero}, {Daddi}, {Zahid}, {Nagao}, {Kewley}, {Lilly}, {Sugiyama}, {Baronchelli}, {Capak}, {Carollo}, {Chu}, {Hasinger}, {Ilbert}, {Juneau}, {Kajisawa}, {Koekemoer}, {Kovac}, {Le F{\`e}vre}, {Masters}, {McCracken}, {Onodera}, {Schulze}, {Scoville}, {Strazzullo}, \& {Taniguchi}}]{Silverman2015}
{Silverman}, J.~D., {Kashino}, D., {Sanders}, D., {et~al.} 2015, \apjs, 220, 12

\bibitem[{{Talia} {et~al.}(2023){Talia}, {Schreiber}, {Garilli}, {Pentericci}, {Pozzetti}, {Zamorani}, {Cullen}, {Moresco}, {Calabr{\`o}}, {Castellano}, {Fynbo}, {Guaita}, {Marchi}, {Mascia}, {McLure}, {Mignoli}, {Pompei}, {Vanzella}, {Bongiorno}, {Vietri}, {Amor{\'\i}n}, {Bolzonella}, {Carnall}, {Cimatti}, {Cresci}, {Cristiani}, {Cucciati}, {Dunlop}, {Fontanot}, {Franzetti}, {Gargiulo}, {Hamadouche}, {Hathi}, {Hibon}, {Iovino}, {Koekemoer}, {Mannucci}, {McLeod}, \& {Saldana-Lopez}}]{Talia_2023}
{Talia}, M., {Schreiber}, C., {Garilli}, B., {et~al.} 2023, \aap, 678, A25

\bibitem[{Vanden~Berk {et~al.}(2001)Vanden~Berk, Richards, Bauer, Strauss, Schneider, Heckman, York, Hall, Fan, Knapp, Anderson, Annis, Bahcall, Bernardi, Briggs, Brinkmann, Brunner, Burles, Carey, Castander, Connolly, Crocker, Csabai, Doi, Finkbeiner, Friedman, Frieman, Fukugita, Gunn, Hennessy, Ivezić, Kent, Kunszt, Lamb, Leger, Long, Loveday, Lupton, Meiksin, Merelli, Munn, Newberg, Newcomb, Nichol, Owen, Pier, Pope, Rockosi, Schlegel, Siegmund, Smee, Snir, Stoughton, Stubbs, SubbaRao, Szalay, Szokoly, Tremonti, Uomoto, Waddell, Yanny, \& Zheng}]{VandenBerk_2001}
Vanden~Berk, D.~E., Richards, G.~T., Bauer, A., {et~al.} 2001, \aj, 122, 549

\bibitem[{{Vazdekis} {et~al.}(2016){Vazdekis}, {Koleva}, {Ricciardelli}, {R{\"o}ck}, \& {Falc{\'o}n-Barroso}}]{Vazdekis2016}
{Vazdekis}, A., {Koleva}, M., {Ricciardelli}, E., {R{\"o}ck}, B., \& {Falc{\'o}n-Barroso}, J. 2016, \mnras, 463, 3409

\end{thebibliography}

\begin{appendix}

\section{Redshift precision and accuracy}

\label{sc:deep_precision}

\begin{figure}
\centering
\includegraphics[width=8.5cm]{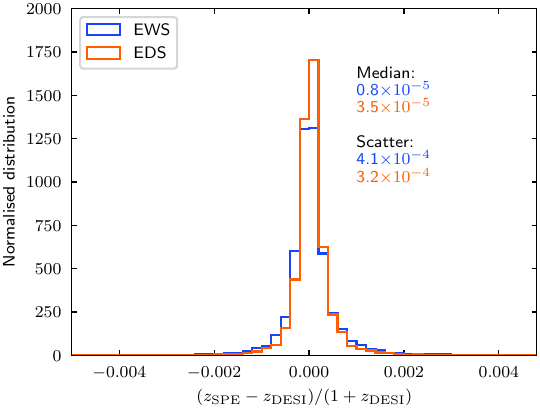}
\caption{\label{fig:z_precision} Normalised distribution of $(z_{\rm SPE} - z_{\rm DESI}) / (1+z_{\rm DESI})$ for the high-quality spectra classified as galaxies in the EWS (blue) and in the EDS (orange).}
\end{figure}

Redshift precision was discussed in \cite{Q1-TP007}. Results have not changed dramatically. However, we update the results and show the EDS results in this Appendix. As can be seen in Fig.~\ref{fig:z_precision}, the offset values are comparable to the Q1 results, and the scatter is improved by more than a factor of two. Performance is similar in the EWS and EDS, despite the much larger number of individual exposures that are stacked to produce the final spectra of the latter. Indeed, because the high-quality sample in the EDS is fainter than its counterpart in EWS, the resulting S/N in the two cases are similar. Both values are well within the requirements needed for the precision of the cosmological measurements of galaxy clustering. This demonstrates again the very satisfactory behaviour of the NISP instrument.

\begin{figure}
\centering
\includegraphics[width=8.5cm]{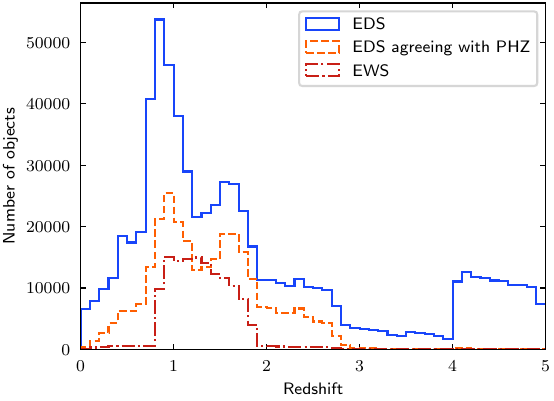}
\caption{\label{fig:z_distrib}  Redshift distribution of various spectroscopic samples of galaxies in the EDF-N. The blue solid line corresponds to the high-quality spectra from the EDS described in Sect.\,\ref{sc:deep_comp_desi}, and the orange dashed line is obtained after discarding objects with an incompatible PHZ redshift. The red dash-dot line represents the results for the EWS high-quality galaxy spectra (see Sect.\,\ref{Sect:wide}).}
\end{figure}

\section{\label{sc:redshift_distibutions} Redshift distributions}

In Fig.\,\ref{fig:z_distrib}, we present the redshift distribution of several samples of galaxies selected from the SPE outputs. We focused on high-quality spectra with ${\tt gal.spe\_z\_prob}>0.99$ in the EDS and EWS as defined in Sect.\,\ref{sc:deep_comp_desi} and Sect.\,\ref{Sect:wide} and shown in Fig.\,\ref{fig:z_vs_z_purity_deep} (left panel) and Fig.\,\ref{fig:z_vs_z_purity_wide} (top panels), respectively. These distributions are mainly illustrative, since any change in the exact selection criteria can significantly impact the results. For instance, the SPE redshift probability used for the selection is sensitive to the number of observable spectral lines at a specific redshift, and the redshift distribution thus reflects detection biases rather than physical properties. Consequently, it should be viewed as a demonstration of the typical redshifts recovered by SPE, without physical interpretation.

The EDS redshift distribution exhibits two peaks at $z=0.8$ and $z=1.6$. If we also consider objects for which SPE disagrees with PHZ, we observe complex patterns at higher redshift. The number of objects decreases with increasing redshift up to $z=4.1$, where \oii\ exits the RGS. We can also notice a steep drop at $z\simeq 2.8$ corresponding to \oiii\ and \hb\ exiting the RGS. At higher redshift, we can observe a large plateau between $z=4.1$ and $z=5$ (maximum redshift tested by SPE), where there is no strong line. This feature is likely artificial. If we consider only objects with a compatible spectroscopic and photometric redshift, this plateau totally disappears. The drop at $z=2.8$ is also much stronger, suggesting that PHZ is necessary to select reliable \oii\ emitters. The other features at lower redshift are unchanged, but normalisation decreases.

The distribution in the EWS is more simple with most of the objects between $z=0.84$ and $z=1.87$, corresponding to redshifts where \ha\ is in the RGS. It is not surprising, since a \ha\ prior is applied in the EWS, favouring these solutions. The distribution is mostly flat between $z=0.85$ and $z=1.4$, and decreases at higher redshifts.

\section{\label{Sec:ADQL}Examples of ADQL requests}
The description of all variables is given in the Data Product Description Document \url{https://euclid.esac.esa.int/dr/dr1/dpdd/spedpd/dpcards/spe_spepfoutputcatalog.html}. This technical documentation is much more exhaustive than the content of the Science Archive, but contains the column description. The following request allows one to retrieve information from the various tables stored in the \Euclid Science Archive, together with some coordinates and photometric data from the MER catalogue. The {\sc JOIN} command makes the link between data contained in different tables, and has to use both  the object identifier and the rank of the redshift solution to correctly link values for the candidate and the model table. This request returns values obtained for the best redshift solution (${\tt spe\_rank}  =0$)  for the three categories, but it could be extended to other solutions with a condition like

\begin{lstlisting}[]
WHERE
  (gal.spe_rank = 0 AND qso.spe_rank = 0 AND star.spe_rank = 0)
  OR (gal.spe_rank = 1 AND qso.spe_rank = 1 AND star.spe_rank = 1)
\end{lstlisting}

Each object with the required solutions would then appear twice in the catalogue, one line containing the results for the first solutions in the three categories and the other the second solutions. In order to produce a catalogue with a single line per object and several line property measurements, the {\tt spectro\_line\_features\_spe\_line\_features\_wide\_survey} has to be addressed for each desired line. The link is also made simultaneously on the object identifier and  the rank of the solution, with the condition to retrieve the correct line properties using the name of the line. 
\begin{lstlisting}[basicstyle=\footnotesize\ttfamily]
SELECT classif.object_id as object_id, mer.right_ascension as RA, mer.declination as dec, mer.flux_vis_sersic as flux_vis_sersic, mer.flux_y_sersic as flux_y_sersic, mer.flux_j_sersic as flux_j_sersic, mer.flux_h_sersic as flux_h_sersic, mer.kron_radius as kron_radius, mer.point_like_prob as point_like_proba, mer.spurious_prob as spurious_prob, mer.tile_index as Tile, 
classif.spe_class as Classification, classif.spe_gal_prob as gal_prob,classif.spe_qso_prob as qso_prob, classif.spe_star_prob as star_prob,
qual.spe_npix as N_pix, qual.spe_lsf_red as LSF, qual.spe_n_dith_max as N_dith_max, qual.spe_n_dith_med as N_dith_med,
gal.spe_z as galaxy_redshift, gal.spe_z_err as galaxy_z_err, gal.spe_z_prob as galaxy_z_proba, gal.spe_z_rel as galaxy_reliability, gal.spe_z_reduced_chi2 as galaxy_reduced_chi2, gal.spe_z_p_value as gal_p_value, gal.spe_z_residual_normality as gal_residual_norm,
gal_mod.spe_vel_disp_e as galaxy_vel_disp,
qso.spe_z as qso_redshift, qso.spe_z_err as qso_z_err, qso.spe_z_prob as qso_z_proba, qso.spe_z_reduced_chi2 as qso_Reduced_chi2, qso.spe_z_p_value as qso_p_value, qso.spe_z_residual_normality as qso_residual_norm, qso_mod.spe_cnt_tpl_id as qso_template,
star.spe_rvel/299792.458 as star_z, star.spe_rvel_err/299792.458 as star_z_err, star.spe_model_prob as star_z_proba, star.spe_rvel_reduced_chi2 as star_Reduced_chi2, star.spe_rvel_p_value as star_p_value, star.spe_rvel_residual_normality as star_residual_norm, star_mod.spe_cnt_tpl_id as star_template,
Halpha.spe_line_flux_gf as Halpha_gf_flux, Halpha.spe_line_flux_err_gf as Halpha_gf_flux_err, Halpha.spe_line_snr_gf as Halpha_gf_snr, Halpha.spe_line_n_dith as Halpha_n_dith, Halpha.spe_line_qual_gf as Halpha_gf_qual,
Hbeta.spe_line_flux_gf as Hbeta_gf_flux, Hbeta.spe_line_flux_err_gf as Hbeta_gf_flux_err, Hbeta.spe_line_snr_gf as Hbeta_gf_snr, Hbeta.spe_line_n_dith as Hbeta_n_dith,Hbeta.spe_line_qual_gf as Hbeta_gf_qual,
OIII.spe_line_flux_gf as OIII_gf_flux, OIII.spe_line_flux_err_gf as OIII_gf_flux_err, OIII.spe_line_snr_gf as OIII_gf_snr, OIII.spe_line_n_dith as OIII_n_dith,OIII.spe_line_qual_gf as OIII_gf_qual,
OII.spe_line_flux_gf as OII_gf_flux, OII.spe_line_flux_err_gf as OII_gf_flux_err, OII.spe_line_snr_gf as OII_gf_snr, OII.spe_line_n_dith as OII_n_dith, OII.spe_line_qual_gf as OII_gf_qual,
SIII.spe_line_flux_gf as SIII_gf_flux, SIII.spe_line_flux_err_gf as SIII_gf_flux_err, SIII.spe_line_snr_gf as SIII_gf_snr, SIII.spe_line_n_dith as SIII_n_dith,SIII.spe_line_qual_gf as SIII_gf_qual
FROM catalogue.spectro_z_spe_classification_wide_survey as classif
JOIN catalogue.spectro_z_spe_galaxy_candidates_wide_survey AS gal
  ON classif.object_id=gal.object_id
JOIN catalogue.spectro_model_spe_galaxy_models_wide_survey AS gal_mod
  ON classif.object_id=gal_mod.object_id and gal_mod.spe_rank=gal.spe_rank
JOIN catalogue.spectro_z_spe_qso_candidates_wide_survey AS qso
  ON classif.object_id=qso.object_id
JOIN catalogue.spectro_model_spe_qso_models_wide_survey AS qso_mod
  ON classif.object_id=qso_mod.object_id and qso_mod.spe_rank=qso.spe_rank
JOIN catalogue.spectro_z_spe_star_candidates_wide_survey AS star
  ON classif.object_id=star.object_id 
JOIN catalogue.spectro_model_spe_star_models_wide_survey AS star_mod
  ON classif.object_id=star_mod.object_id and star_mod.spe_rank=star.spe_rank
JOIN catalogue.
spectro_line_features_spe_line_features_wide_survey AS Halpha
  ON Halpha.object_id = gal.object_id and Halpha.spe_rank=gal.spe_rank
JOIN catalogue.
spectro_line_features_spe_line_features_wide_survey AS Hbeta
  ON Hbeta.object_id = gal.object_id and Hbeta.spe_rank=gal.spe_rank
JOIN catalogue.
spectro_line_features_spe_line_features_wide_survey AS OIII
  ON OIII.object_id = gal.object_id and OIII.spe_rank=gal.spe_rank
JOIN catalogue.
spectro_line_features_spe_line_features_wide_survey AS OII
  ON OII.object_id = gal.object_id and OII.spe_rank=gal.spe_rank
JOIN catalogue.
spectro_line_features_spe_line_features_wide_survey AS SIII
  ON SIII.object_id = gal.object_id and SIII.spe_rank=gal.spe_rank
JOIN catalogue.
spectro_z_spe_quality_wide_survey as qual
  ON gal.object_id = qual.object_id
JOIN catalogue.mer_catalogue_wide_survey AS mer
  ON mer.object_id = gal.object_id
WHERE
  (gal.spe_rank = 0 
  AND qso.spe_rank=0 
  and star.spe_rank=0)
  AND (Halpha.spe_line_name = 'Halpha' 
  and OIII.spe_line_name = 'OIII5008' 
  and OII.spe_line_name = 'OII3727' 
  and Hbeta.spe_line_name='Hbeta' 
  and SIII.spe_line_name = 'SIII9530')
  
\end{lstlisting}

\section{Line flux measurements quality flags}
As discussed in Sect. \ref{sc:LineFlux}, each flux measurement is provided with an associated quality flag that helps the user to assess the robustness of the measurement. This flag is composed of a 3-digit number that allows tracing the method with which a measurement has been performed, the goodness of the measurement (or the reason for failure), and the grism from which the line has been measured. Tables \ref{tab:qualflag} and \ref{tab:seconc_digit} provide the detailed information explaining the quality flag, with examples for possible values.

\begin{table*}[h]
\centering
\caption{Definition of the {\tt SPE\_LINE\_QUAL} three-digit quality flag.}
\label{tab:qualflag}
\begin{tabular}{p{4cm}|p{4cm}|p{4cm}}
\hline
\hline
1${\rm st}$ digit&
2${\rm nd}$ digit &
3${\rm rd}$ digit \\
& & \\
{1} (Direct Integration) \newline
{2} (Gaussian Fit) \newline
{3} (H\,$\alpha$\_GF for QSOs) \newline
``{$-$}'' (minus sign, the line is outside the observable wavelength range given the object's redshift) &  can assume one of the following values: \textbf{[0, 1, 2, 3, 4, 5, 7, 8, 9]} \newline\newline
(see table below) &
{1} (BGS) \newline
{2} (RGS) \newline
{3} (BGS+RGS) \\
\hline
\end{tabular}
\end{table*}

\begin{table*}
    \centering
\caption{Definition of the second digit of the {\tt SPE\_LINE\_QUAL}  quality flag.}
\label{tab:seconc_digit}
\begin{tabular}{p{0.8cm} p{2.5cm} p{9cm} p{2.5cm}}
\hline\hline
{2$^{\rm nd}$ digit} & {Definition} & {Description} & {Applicable to} \\
\hline
0 & Not measured &
The standard pipeline does not provide a measurement (for DI) or a fit (for GF or H\,$\alpha$\_QSO).
Possible values for the {\tt SPE\_LINE\_QUAL}: \textbf{101, 102, 103, 201, 202, 203, 301, 302, 303} &
all \\
\hline
1 & Good measure &
The measure is considered reliable.
Possible values for the {\tt SPE\_LINE\_QUAL}: \textbf{111, 112, 113, 211, 212, 213, 311, 312, 313} &
all \\
\hline
2 & Low S/N &
The measure might be good but the line has a low S/N ($<2.0$).
Possible values for the {\tt SPE\_LINE\_QUAL}: \textbf{121, 122, 123, 221, 222, 223, 321, 322, 323} &
all \\
\hline
3 & EW $>0$  &
The EW has a positive value (probably due to a negative continuum), while by convention EW should be negative for emission lines. The value is set to zero but the other measures (e.g.\ fluxes) are however reported.
Possible values for the {\tt SPE\_LINE\_QUAL}: \textbf{231, 232, 233, 331, 332, 333} &
GF or H\,$\alpha$\_QSO \\
\hline
4 & Issue for combination &
The 2 measures are not both good (e.g.\ one might be ``good'' and the other might have a low S/N). When possible, the weighted average is provided.
Possible values for the {\tt SPE\_LINE\_QUAL}: \textbf{143, 243, 343} &
all (only for BGS+RGS) \\
\hline
5 & Acceptable &
The measure is acceptable but up to 15\% of the pixels within the line are missing.
Possible values for the {\tt SPE\_LINE\_QUAL}: \textbf{251, 252, 351, 352} &
GF or H\,$\alpha$\_QSO \\
\hline
7 & Merging error &
Error in combination (i.e.\ blue+red output merging).
Possible values for the {\tt SPE\_LINE\_QUAL}: \textbf{173, 273} &
only for BGS+RGS \\
\hline
8 & Unreliable or wrong fluxes &
The measure is unreliable because more than 15\% of the pixels are missing (GF or QSO) or it's wrong because more than 1 pixel is missing (DI).
Possible values for the {\tt SPE\_LINE\_QUAL}: \textbf{181, 182, 281, 282, 381, 382} &
all \\
\hline
9 & Outside range &
The line is not measurable in the wavelength range given the redshift of the source.
Possible values for the {\tt SPE\_LINE\_QUAL}: \textbf{$-$91, $-$92, $-$93} &
all \\
\hline
\end{tabular}
\label{tab:qual_feats}
\end{table*}

\vfill\eject
\section{Example spectra}
We show here some examples of spectra randomly selected from the EWS, for different categories of objects. In order to display the highest quality results, selection criteria on number of pixels, probabilities or S/N were quite conservative.
\begin{figure*}
\centering
\includegraphics[width=18cm]{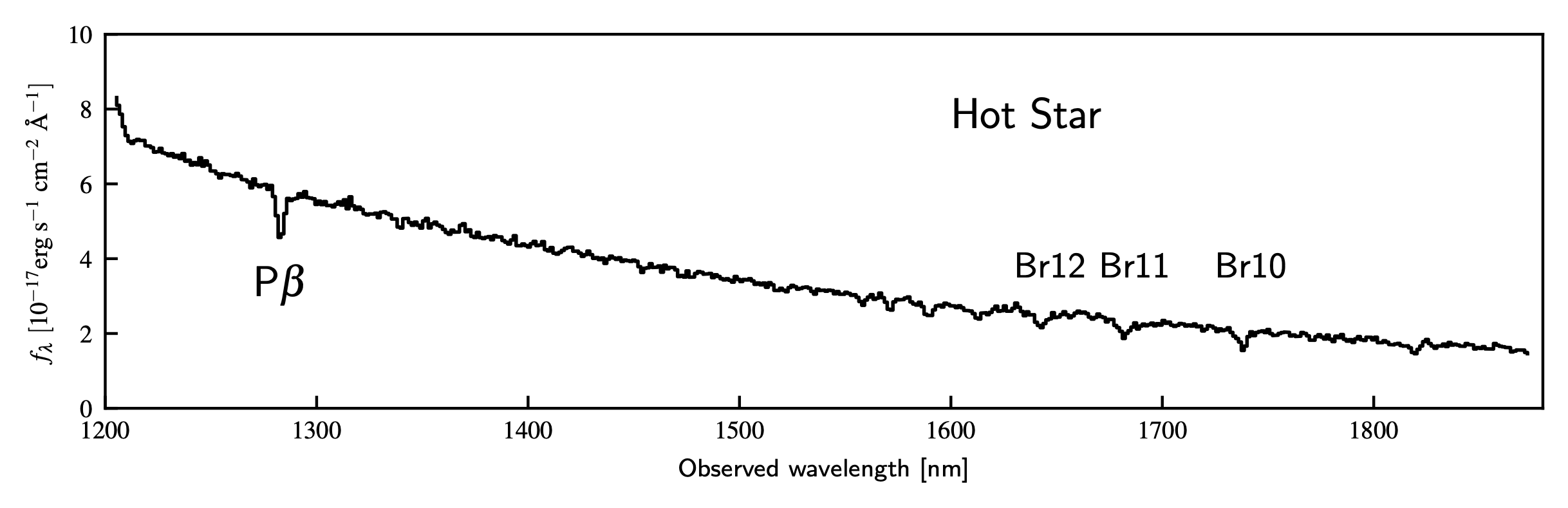}
\includegraphics[width=18cm]{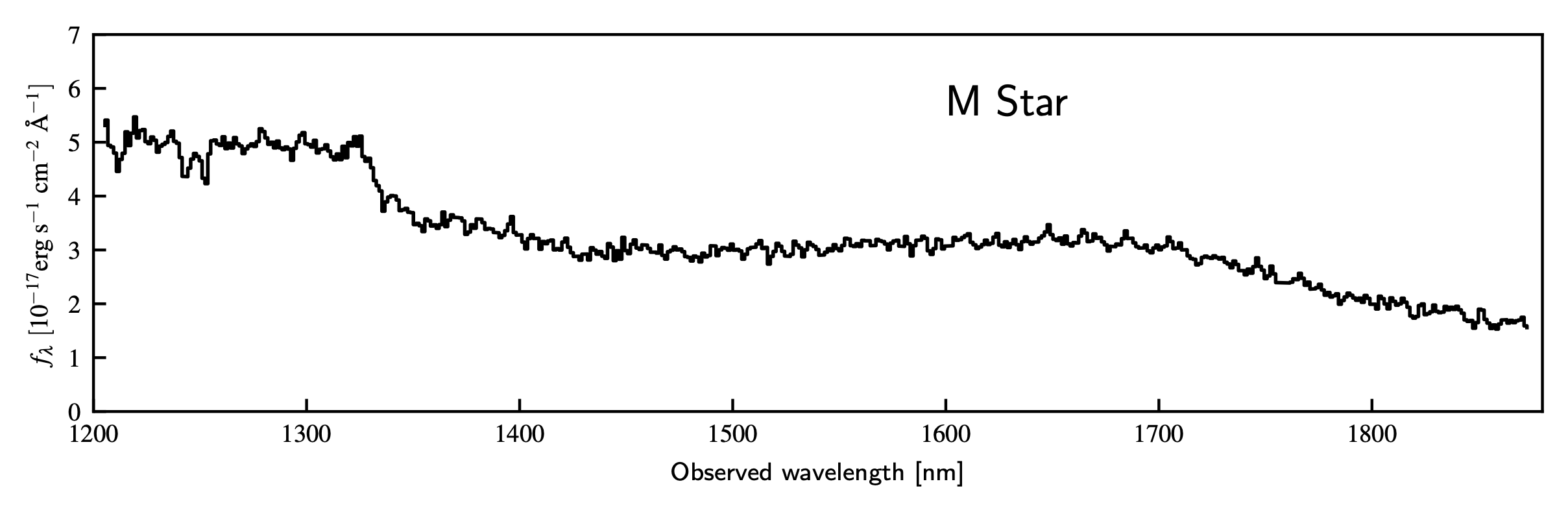}
\includegraphics[width=18cm]{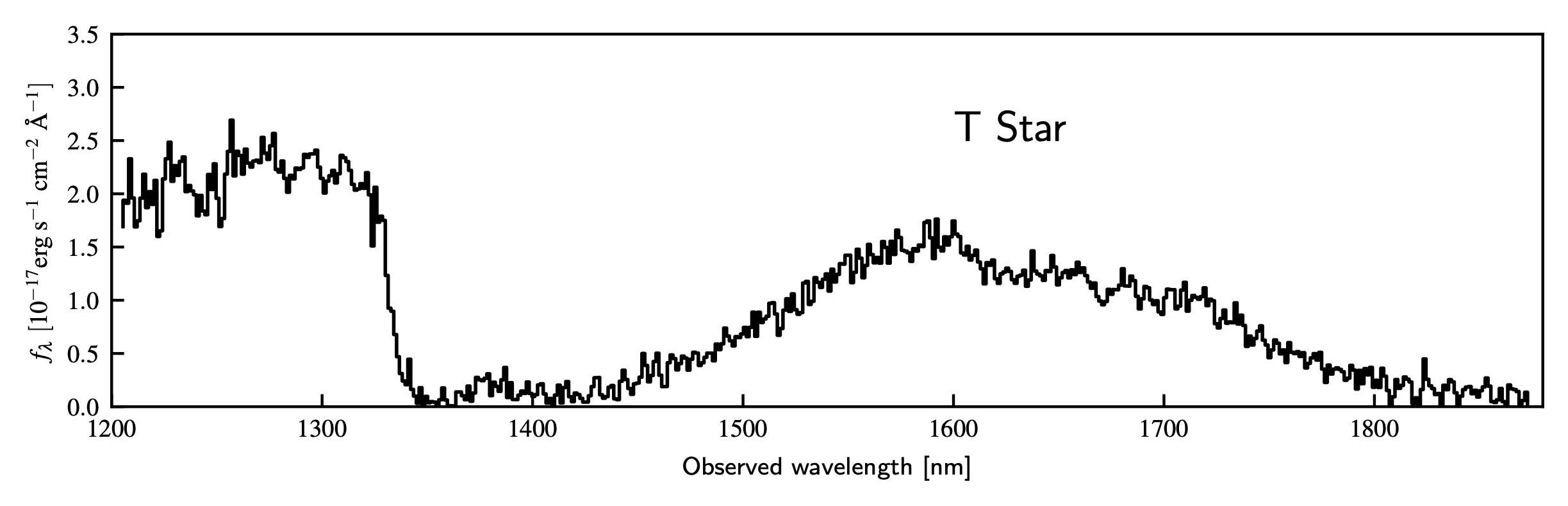}
\caption{\label{fig:Star_examples} Examples of stellar spectra from the EWS. From top to bottom : Hot ($\HE = 18.2$), M-type ($\HE = 18.0$)  and ultracool T type ($\HE = 19.1$). }
\end{figure*}
\begin{figure*}
\centering
\includegraphics[width=18cm]{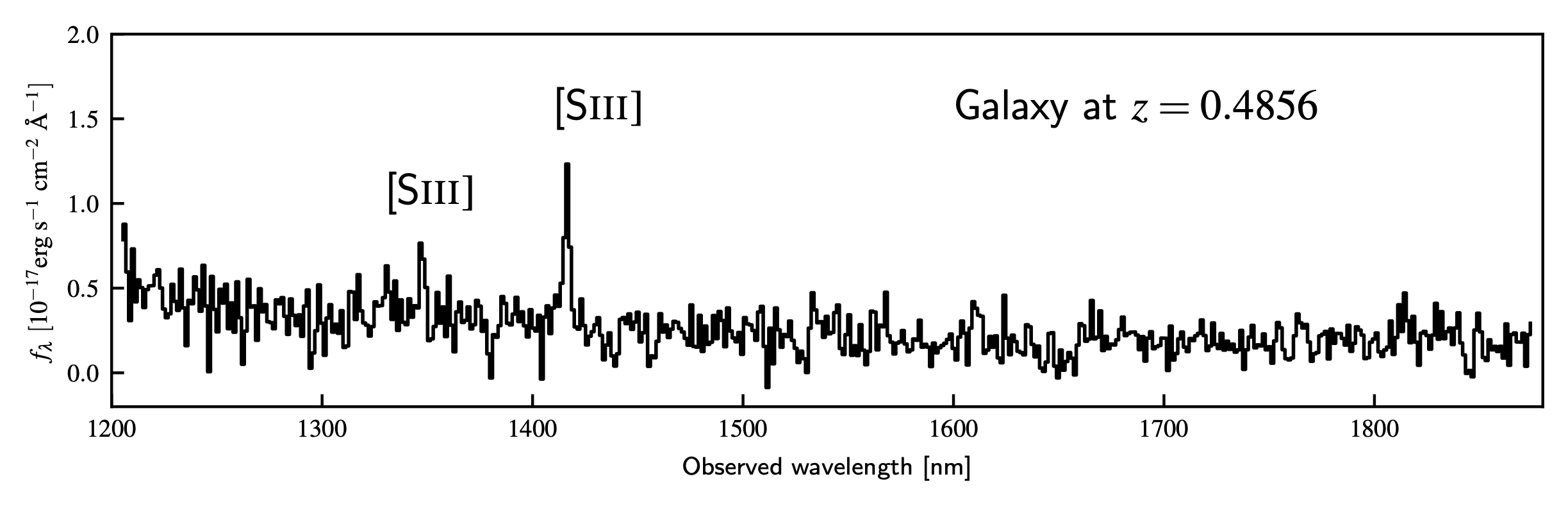}
\includegraphics[width=18cm]{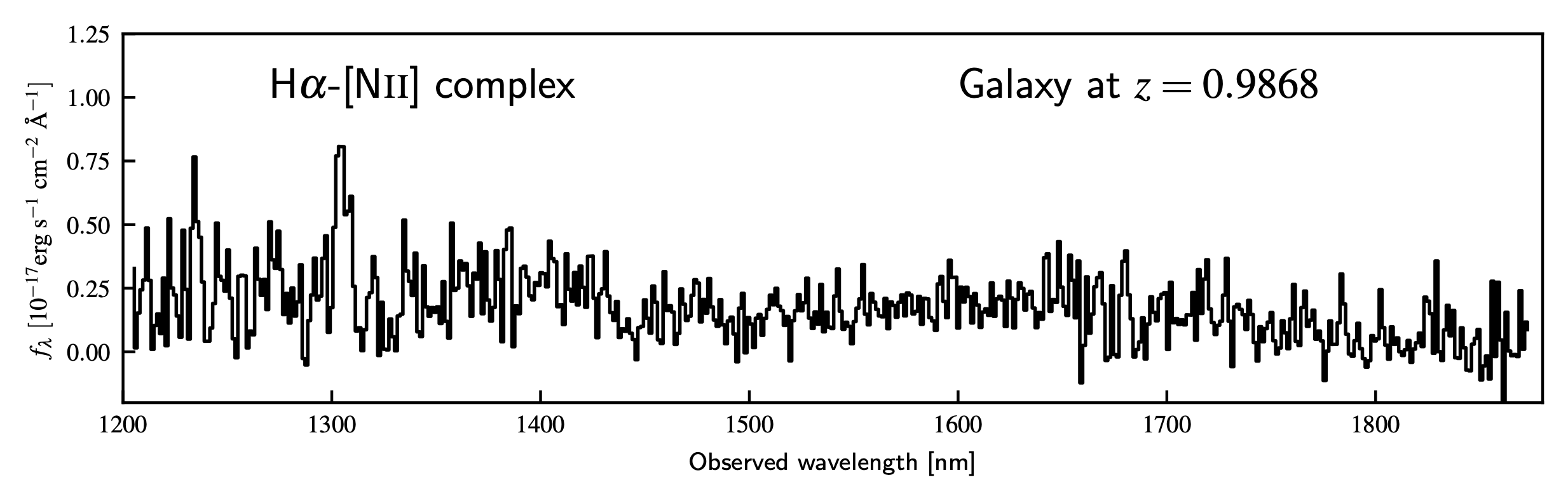}
\includegraphics[width=18cm]{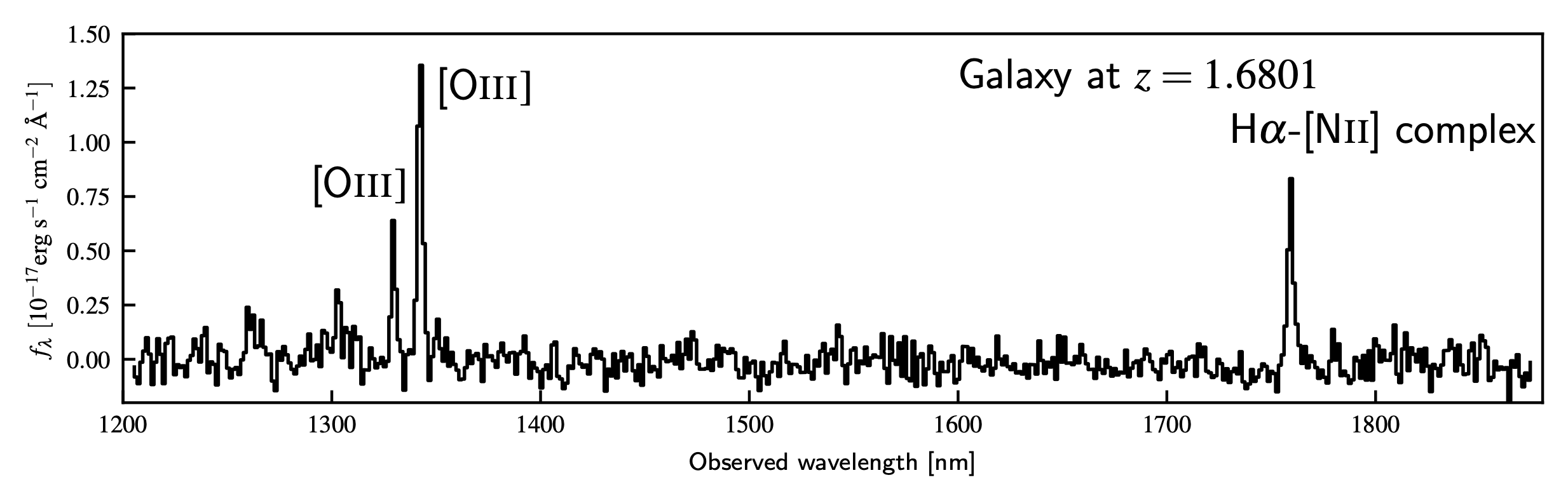}
\includegraphics[width=18cm]{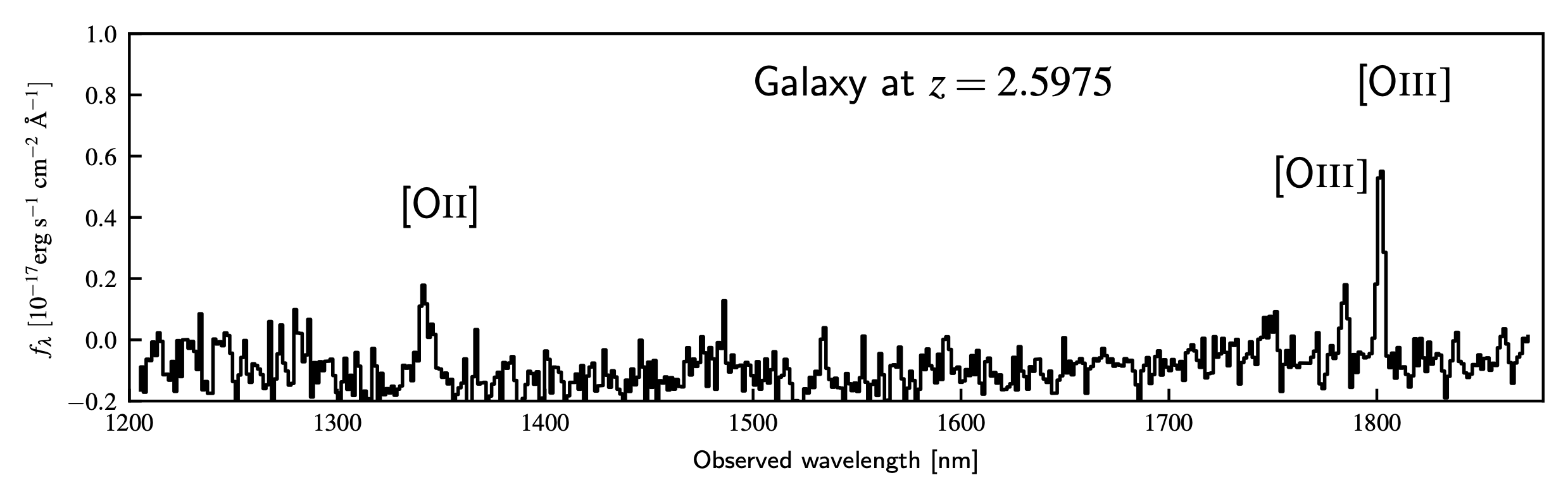}
\caption{\label{fig:gal_examples} Examples of spectra classified as galaxies at various redshifts, selected with the displayed emission lines having S/N greater than 10 }
\end{figure*}

\begin{figure*}
\centering
\includegraphics[width=18cm]{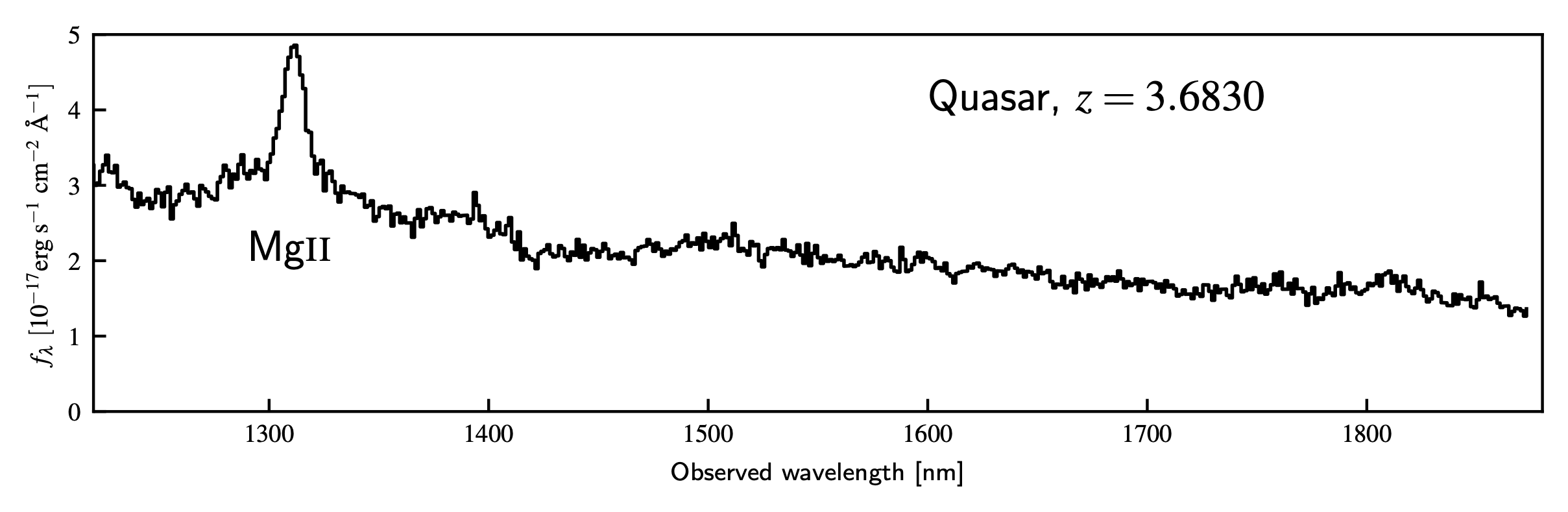}
\includegraphics[width=18cm]{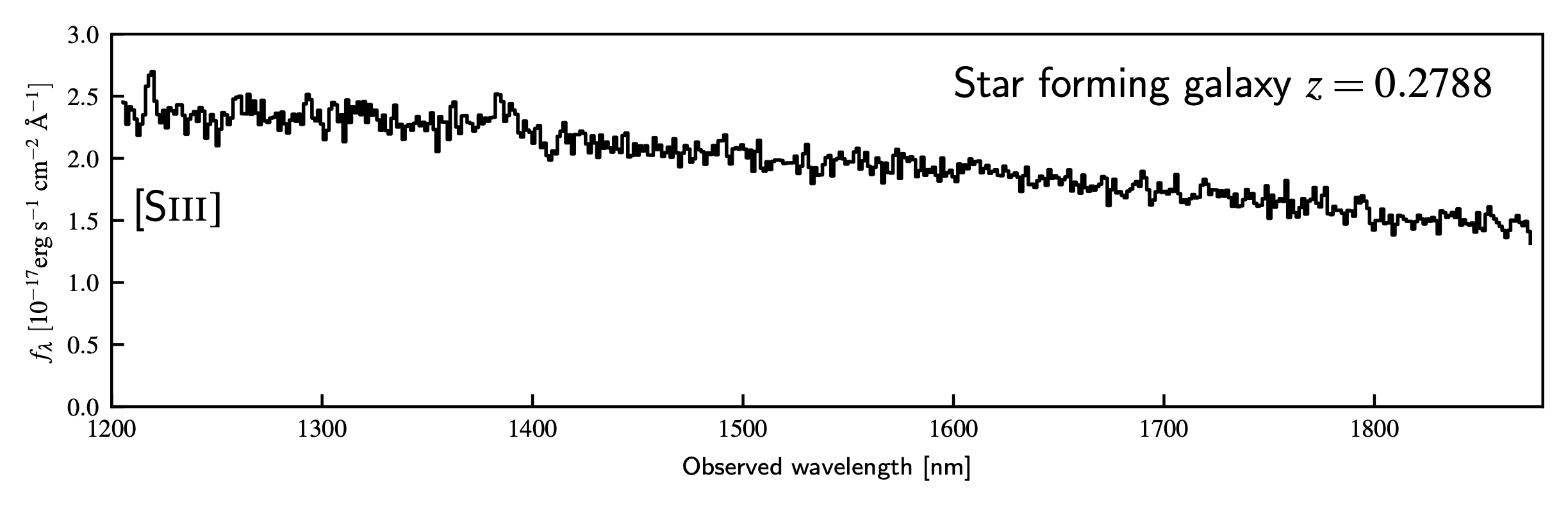}
\caption{\label{fig:qso_examples} Example spectra classified as quasars from the EWS : a $z=3$ quasar on top and a star forming galaxy on the bottom}
\end{figure*}
\begin{figure*}
\centering
\includegraphics[width=18cm]{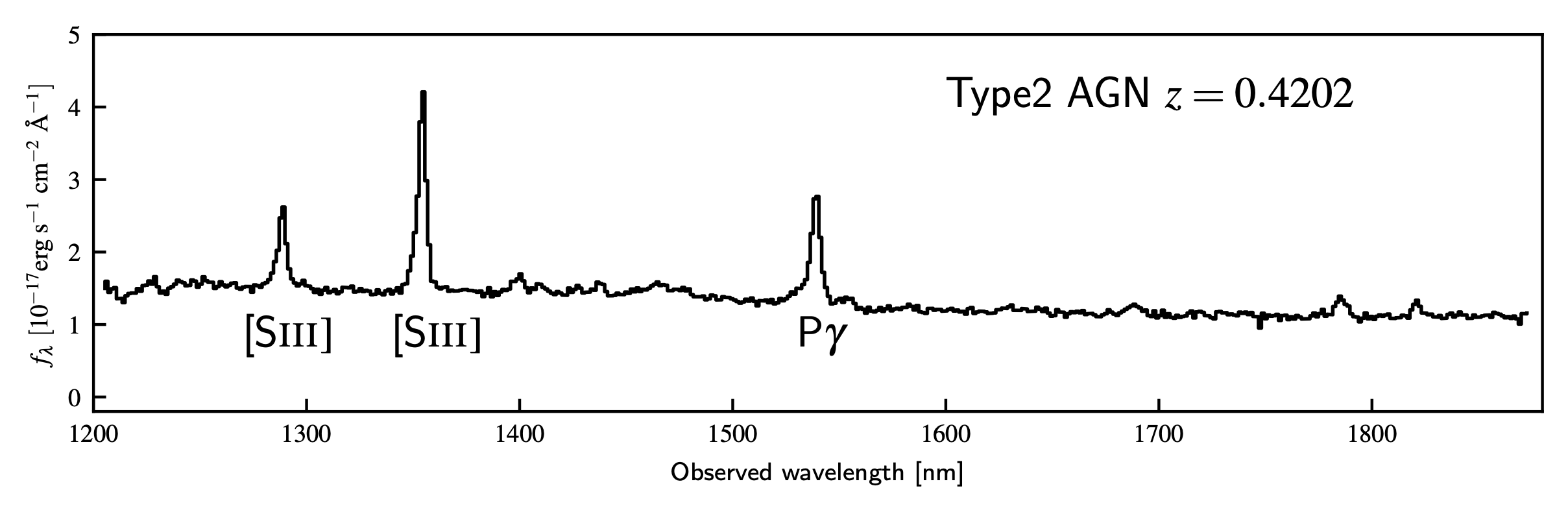}
\caption{\label{fig:AGN2} Example of an AGN2 spectrum used to qualify the line flux measurement}
\end{figure*}

\end{appendix}

\end{document}